\documentclass[12pt]{elsart}
\usepackage{graphicx}
\usepackage{amssymb}
\usepackage{amsmath}
\usepackage{amsfonts}
\usepackage{graphics}
\usepackage[breaklinks,bookmarks,colorlinks,linkcolor=darkblue,citecolor=darkblue]{hyperref}
\usepackage{color}
\usepackage{xcolor}
\usepackage{ragged2e}
\usepackage{epstopdf}
\usepackage{natbib}
\usepackage{bm}
\usepackage{txfonts} 

\definecolor{darkblue}{rgb}{0,0,.8}
\newcommand{\ket}[1]{|{#1}\rangle}
\newcommand{\bra}[1]{\langle{#1}|}

\journal{Advances in Atomic, Molecular and Optical Physics}

\begin{document}
\begin{frontmatter}
\title{Engineered Open Systems and Quantum Simulations with Atoms and Ions}


\author{Markus M\"uller}
\address{Institute for Quantum Optics and Quantum Information of the Austrian Academy of Sciences, and Institute
for Theoretical Physics, University of Innsbruck, A-6020 Innsbruck, Austria}
\address{Departamento de F\'isica Te\'orica I, Universidad Complutense, 28040 Madrid, Spain}

\author{Sebastian Diehl}
\address{Institute for Quantum Optics and Quantum Information of the Austrian Academy of Sciences, and Institute
for Theoretical Physics, University of Innsbruck, A-6020 Innsbruck, Austria}

\author{Guido Pupillo}
\address{Institute for Quantum Optics and Quantum Information of the Austrian Academy of Sciences, and Institute
for Theoretical Physics, University of Innsbruck, A-6020 Innsbruck, Austria}
\address{ISIS (UMR 7006) and IPCMS (UMR 7504), Universit\'e de Strasbourg and CNRS, Strasbourg, France}

\author{Peter Zoller}
\address{Institute for Quantum Optics and Quantum Information of the Austrian Academy of Sciences, and Institute
for Theoretical Physics, University of Innsbruck, A-6020 Innsbruck, Austria}
\date{ on 24 Feb.~2012}

\begin{abstract}
The enormous experimental progress in atomic, molecular and optical (AMO) physics during the last decades allows us nowadays to isolate single, a few or even many-body ensembles of microscopic particles, and to manipulate their quantum properties at a level of precision, which still seemed unthinkable some years ago. This versatile set of tools has enabled the development of the well-established concept of engineering of many-body Hamiltonians in various physical platforms. These available tools, however, can also be harnessed to extend the scenario of Hamiltonian engineering to a more general Liouvillian setting, which in addition to coherent dynamics also includes controlled dissipation in many-body quantum systems. Here, we review recent theoretical and experimental progress in different directions along these lines, with a particular focus on physical realizations with systems of atoms and ions. This comprises digital quantum simulations in a general open system setting, as well as engineering and understanding new classes of systems far away from thermodynamic equilibrium. In the context of digital quantum simulation, we first outline the basic concepts and illustrate them on the basis of a recent experiment with trapped ions. We also discuss theoretical work proposing an intrinsically scalable simulation architecture for spin models with high-order interactions such as Kitaev's toric code, based on Rydberg atoms stored in optical lattices. We then turn to the digital simulation of dissipative many-body dynamics, pointing out a route for the general quantum state preparation in complex spin models, and discuss a recent experiment demonstrating the basic building blocks of a full-fledged open system quantum simulator. 
In view of creating novel classes of out-of-equilibrium systems, we focus on ultracold atoms.  We point out how quantum mechanical long range order can be established via engineered dissipation, and present genuine many-body aspects of this setting: In the context of bosons, we discuss  dynamical phase transitions resulting from competing Hamiltonian and dissipative dynamics. In the context of fermions, we present a purely dissipative pairing mechanism, and show how this could pave the way for the quantum simulation of the Fermi-Hubbard model. We also propose and analyze the key properties of dissipatively targeted topological phases of matter.
\end{abstract}
\begin{keyword}
Open quantum systems, quantum simulation, atomic physics, trapped ions, quantum phase transitions, unconventional pairing mechanisms, topological phases of matter.
\end{keyword}

\end{frontmatter}

\footnote{The authors M.~M. and S.~D. contributed equally to this work.}

\tableofcontents

\section{Introduction}
\label{sec:Introduction}

The extraordinary experimental progress in AMO physics experienced during the last decades allows us nowadays to isolate one or few microscopic particles, or even many-body ensembles of them, and to manipulate, control and detect their quantum states almost perfectly. Harnessing the available tools offers unique possibilities to extend the customary idea of Hamiltonian engineering to a more general scenario, where coherent and controlled driven-dissipative dynamics appear on an equal footing. This program comprises different directions. On the one hand, the ability to control both coherent and dissipative dynamics constitutes a complete set of tools for general open-system quantum simulation, very much in the spirit of a truly universal simulator device. On the other hand, the possibility to combine coherent and dissipative dynamics opens the door to novel classes of artificial out-of-equilibrium many-body systems without immediate counterpart in condensed matter. In this work, we review theoretical and experimental progress on the quantum simulation and open-system dynamics of many-particle systems with cold atoms and trapped ions from various perspectives.

\emph{Simulation of quantum physics} on classical computers is in many cases hindered by the intrinsic complexity of many-particle quantum systems, for which the computational effort scales exponentially with the number of particles. Thus Feynman's vision was
to build a controllable quantum device which can be programmed to act as a quantum
simulator for any quantum system, and would allow one to study complex quantum systems, intractable on classical computers, from a wide plethora of research fields \citep{T_buluta-science-326-108}. 
Such a device can be built as an \emph{analog} or \emph{digital} quantum simulator, and its time evolution can represent a \emph{Hamiltonian} closed system or \emph{open system dynamics}. In \emph{analog quantum simulation} one `builds the Hamiltonian directly' by 
`always-on' tunable external control fields. Familiar examples are cold atoms
in optical lattices as analog simulators of Bose and Fermi Hubbard models \citep{lewenstein-rmp-56-135,bloch-rmp-2008,dalibard-rmp-83-1523} or Rydberg atoms \citep{saffman-rmp-82-2313} or trapped ions \citep{schneider-repprogphys-75-024401,johanning-j-phys-b-42-154009} for the simulation of spin systems. In contrast, in \emph{digital quantum simulation} the initial state of the quantum system is encoded in a register of qubits. For any many-body quantum system with few-particle interactions, the time evolution can then be efficiently approximated \citep{lloyd-science-273-1073} according to a Trotter decomposition in small, finite time steps, realized by a stroboscopic sequence of quantum gates, as familiar from quantum computing. As we will discuss below, the digital simulation approach can be applied to realize \textit{coherent}Ê as well as \textit{dissipative} many-body dynamics, in particular of open many-particle systems involving $n$-body interactions and constraints, as they naturally appear, e.g., in complex condensed matter models, quantum chemistry, high energy physics and many-body spin models of interest in the field of topological quantum information. 

In the context of \emph{engineering open many-body systems}, cold atomic gases offer a natural and versatile platform. A large part of current research in this field focuses on tailoring specific Hamiltonians, made possible by the precise control of microscopic system parameters via external fields. The resulting systems are well described as \emph{closed} quantum systems, isolated from the environment, and rest in thermodynamic equilibrium -- in close analogy to condensed matter systems. In contrast, here we will be interested in a scenario where many-body ensembles are properly viewed as \emph{open} quantum systems, much in the spirit of the setting of quantum optics and without direct condensed matter analog: A system of interest is coupled to an environment in a controlled way, and is additionally driven by external coherent fields. As anticipated above, via such reservoir engineering driven dissipation may then not only occur as a perturbation, but rather as the dominant resource of many-body dynamics. In particular, we point out that, while dissipation is usually seen as an adversary to subtle quantum mechanical correlations, in proper combination with coherent drive, it can act in exactly the opposite way -- even creating quantum mechanical order. More generally, the results presented below pinpoint the fact that the far-from-equilibrium stationary states of such driven-dissipative ensembles offer a variety of novel many-body aspects and phenomena. 

Under rather general circumstances, discussed and justified below, the dynamics of the many-particle quantum systems we are interested in here can be described by the following master equation:  \footnote{Throughout this article we set $\hbar=1$.}
\begin{align}
\label{eq:master_equation}
\partial_t \rho & = - i [H, \rho] + \mathcal{L} (\rho) 
\end{align}
for the density operator $\rho(t)$ of the many-body system \citep{zollerbook}. The \emph{coherent} part of the dynamics is described by a
Hamiltonian $H = \sum_\alpha H_\alpha$, where $H_\alpha$ act on a quasi-local subset of particles. \emph{Dissipative} time evolution is described by the Liouvillian part of the master equation,
\begin{align}
\label{eq:Lindblad_part}
\mathcal{L} (\rho) & = \sum_\beta \frac{\gamma_\beta}{2} \left( 2c_\beta \rho c_\beta^\dagger - c_\beta^\dagger c_\beta \rho
-  \rho c_\beta^\dagger c_\beta  \right),
\end{align}
where the individual terms are of Lindblad form \citep{wiseman-book} and are determined by quantum jump operators $c_\alpha$, acting on single particles or on subsets of particles, and by the respective rates $\gamma_\alpha$ at which these jump processes occur.

While there have been several comprehensive recent reviews on quantum many-body physics, quantum simulation and computation with quantum optical systems \citep{jane2003simulation,ladd-nature-464-45,Cirac12} involving atoms \citep{dalibard-rmp-83-1523,bloch-rmp-2008,lewenstein-rmp-56-135,baranov12,BlochDalibard12}, molecules \citep{njp-focus-polarmolecules},
ions \citep{blatt-nature-453-1008,schneider-repprogphys-75-024401,haffner-physrep-469-155} and photons \citep{obrien-science-318-1567}, but also solid state systems \citep{clarke-nature-453-1031,wrachtrup-jpcm-18-S807,hanson-rmp-79-1217}, we will summarize
below recent advances in these directions with a particular focus {\bf on} engineered open many-body systems and quantum simulations with atoms and ions. We note that in this review we intend, rather than providing a comprehensive overview of all recent developments in the field, to present our personal view on open-system quantum simulation, with a focus on work of the authors in Innsbruck in recent years~ \footnote{Parts of this review contain text and figure material from manuscripts by some of the authors, which have been published in other journals.}. Our emphasis is on presenting new concepts and building blocks, which we believe constitute first steps towards many-body systems far away from thermodynamic equilibrium and future large-scale many-body simulations.

\textit{Structure of this Review} -- Part \ref{sec:Digital_QS} of this review presents theoretical and experimental advances in \textit{digital} quantum simulation with trapped ions and Rydberg atoms. In Sect.~\ref{sec:concept_and_first_experiments} we outline the basic concepts of digital quantum simulation and illustrate them by discussing results of recent experiments, which demonstrate the principles of a digital quantum simulator in a trapped-ion quantum information processor \citep{lanyon_science-334-57} (Sect.~\ref{sec:concept_and_first_experiments}). Subsequently, we discuss a proposal for a scalable digital quantum simulator based on Rydberg atoms stored in optical lattices \citep{weimer-nphys-6-382}. We show how this simulation architecture based on a multi-atom Rydberg gate \citep{mueller-prl-102-170502} allows one to simulate the Hamiltonian dynamics of spin models involving coherent $n$-body interactions such as Kitaev's toric code Hamiltonian (Sect.~\ref{sec:Ryderg_simulator}). In Sect.~\ref{sec:dig_sim_opensystems} we focus on digital simulation of \textit{dissipative} many-body dynamics, which enables, e.g., the dissipative ground state preparation of the toric code via collective $n$-body dissipative processes. In this context, we discuss the corresponding reservoir-engineering techniques in the Rydberg simulator architecture, as well as recent experiments, which demonstrate the basic building blocks of an \textit{open-system} quantum simulator with trapped ions \citep{barreiro-nature-470-486}. Finally, we show how a combination of coherent and dissipative dynamics might in the future enable the simulation of more complex spin models such as a three-dimensional U(1) lattice gauge theory. Finally, in Sect.~\ref{sec:complementary_developments} we comment on the effect of gate imperfections on the simulations.

In part \ref{sec:Analog_QS_DissDyn} we turn to engineered open many-body systems of cold atoms. In Sect.~\ref{sec:Concepts} we demonstrate that quantum mechanical long-range order can be established dissipatively, and point out a route how this can be achieved via proper reservoir engineering, indeed extending the notion of quantum state engineering in cold atomic gases from the Hamiltonian to the more general Liouvillian setting \citep{diehl-natphys-4-878}. We then give accounts for further central aspects of this general setting. In Sect.~\ref{sec:Competition}, we investigate the dynamical phase diagram resulting from the competition of unitary and dissipative dynamics, and identify several intrinsic many-body phenomena, underpinning that the stationary states of such systems constitute a novel class of artificial out-of-equilibrium ensembles \citep{Diehl10a}.  In Sect.~\ref{sec:DWave}, in the context of atomic fermions we reveal a novel dissipative pairing mechanism operative in the absence of any attractive forces \citep{Diehl10b}, and point out how such systems may provide an attractive route towards quantum simulation of important condensed matter models, such as the Fermi-Hubbard model. Finally, we discuss in Sect.~\ref{sec:Topological} how engineered dissipation may pave the way towards realizing in the lab topological states of matter \citep{Diehl11}, and discuss some of their key many-body properties.

We conclude with an outlook in Sect.~\ref{sec:Conclusions_Outlook}, which summarizes present outstanding theoretical problems and challenges.


\section{Digital Quantum Simulation with Trapped Ions and Rydberg Atoms}
\label{sec:Digital_QS}


When is quantum simulation useful? As noted above, the main motivation
for quantum simulation is to solve many-body problems where classical
computers fail - or, at least, an efficient classical approach is
presently not known. Indeed remarkable classical algorithms have been
developed to solve specific problems and aspects in equilibrium and
out-of-equilibrium many body physics: examples include Monte-Carlo
techniques \citep{ceperley-rmp-67-279,prokofev-jetp-87-310}, coupled-cluster expansion \citep{shavitt-book,hammond-book}, density functional theory \citep{parr-book}, dynamical mean field theories \citep{georges-rmp-68-13}, and density matrix renormalization
group (DMRG) \citep{schollwoeck-rmp77-259,hallberg-advphys-55-477}. These techniques may fail, when one encounters, for example,
sign problems in the Monte Carlo simulation of fermionic systems, or also
in time dependent problems. An example is provided by quench
dynamics: recent optical lattice experiments \citep{Trotzky11} have studied the time
evolution after a quench, and a comparison with time-dependent DMRG
calculations revealed the difficulty of predicting the long-time evolution
due to growth of entanglement. These recent developments, enabled by the remarkable level of control achieved in analog cold-atom quantum simulators, are exciting, as they indicate for the first time possible large-scale entanglement in
many-particle dynamics, close to the heart of quantum simulation. In the following section, we will outline the complementary route to simulate the time dynamics of interacting many-particle systems by the digital, i.e. gate-based, quantum simulation approach.

\subsection{Concepts and First Experiments with Trapped Ions}
\label{sec:concept_and_first_experiments}

\subsubsection{The Digital Simulation Method}
\label{sec:Digital_Simulation_Method}

\begin{figure}
\begin{center}
\includegraphics[width=\columnwidth]{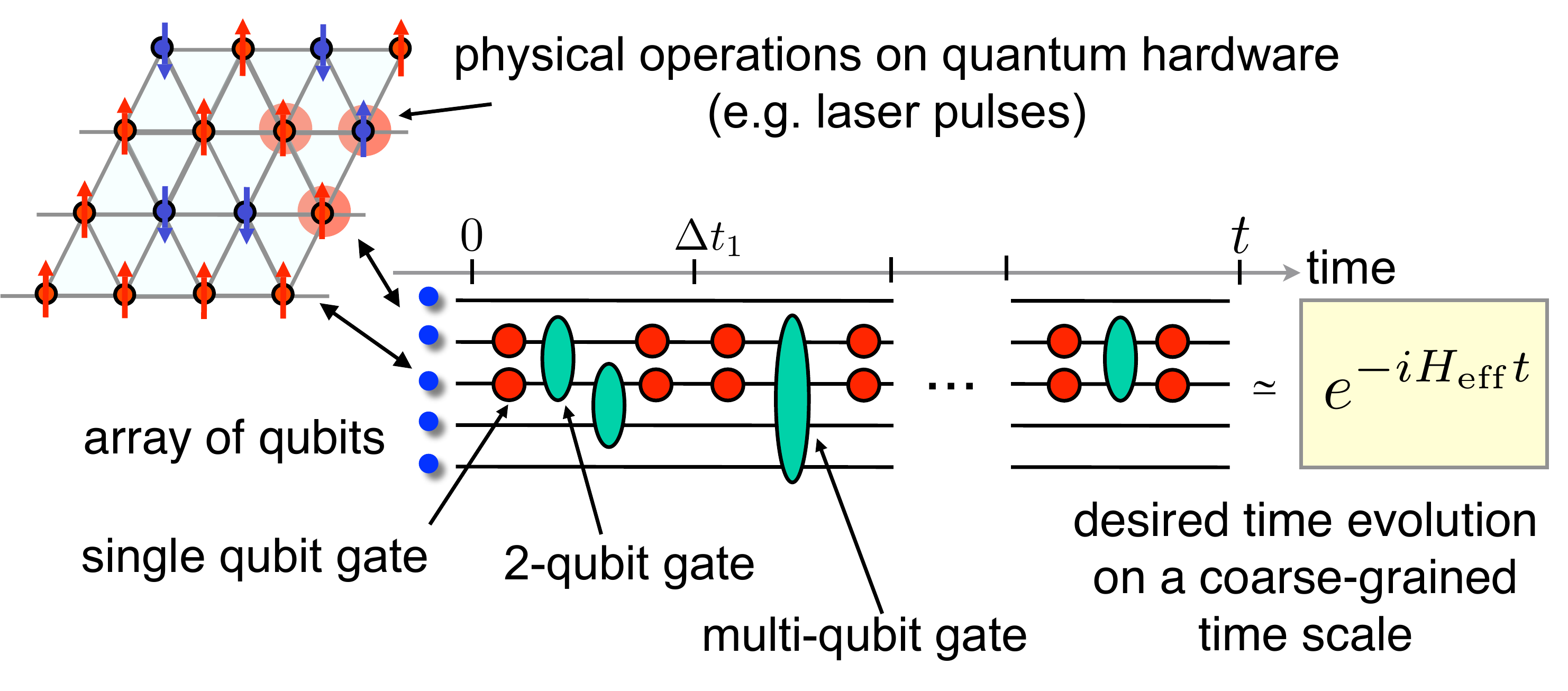}
\end{center}
\caption{(Color online) Schematics of the working principle of a digital quantum simulator: For a specific (many-body) quantum system of interest to be simulated, the initial quantum state is stored in a register of qubits, which are encoded for instance in (meta-)stable electronic states of cold atoms in optical lattices or trapped ions. Then the time evolution of the system up to a time $t$ is represented as a sequence of single- and many-qubit gates, according to a Trotter decomposition of the time evolution operator for small time steps $\Delta t$. Thus, the effective dynamics according to the desired model Hamiltonian $H_\mathrm{eff}$ arises approximately and on a coarse-grained time scale. This \textit{digital}, i.e., gate-based simulation approach is very flexible as the simulated ($n$-body) interactions can be substantially different from and more complex than the physical one- and two-body interactions, which underlie the specific simulator architecture. The concept of digital quantum simulation is not limited to purely coherent Hamiltonian dynamics, but can be extended to the simulation of dissipative dynamics, as e.g.~described by a many-body quantum master equation of the form of Eq.  (\ref{eq:master_equation}) with Liouvillian part of Eq. (\ref{eq:Lindblad_part}), and as discussed in detail in Sect.~\ref{sec:dig_sim_opensystems}.}%
\label{fig:digitalsimulator}%
\end{figure}

We start our discussion with the simulation of purely coherent dynamics generated by a possibly time-dependent many-body Hamiltonian $H(t) = \sum_\alpha H_\alpha (t)$, and proceed in Sect.~\ref{sec:dig_sim_opensystems} with a detailed discussion of the digital simulation of dissipative dynamics according to many-body master Eqs.~ (\ref{eq:Lindblad_part}). It has been shown that a digital quantum simulator can implement the unitary time evolution operator $U(t)$ generated by $H(t)$ efficiently for any local quantum system~\citep{lloyd-science-273-1073,abrams-prl-79-2586,ortiz-pra-64-022319}, i.e., where the individual terms $H_\alpha$ are quasi-local. This means that they operate on a finite number of particles, due to interaction strengths that fall off with distance, for example. In this case it is possible to divide the simulation time $t$ into small time steps $\Delta t = t/n$ and to implement the time evolution through a Trotter expansion of the propagator, $U(t) \simeq \prod_{m=1}^{n} \exp(-i H(m \Delta t) \Delta t)$. The key idea of the Trotter expansion is to approximate each propagator for a small time step according to the full Hamiltonian $H(t)$ by a product of evolution operators for each quasi-local term, $\exp(- i H(m \Delta t) \Delta t) \simeq \prod_{\alpha} \exp(- i H_\alpha(m \Delta t) \Delta t)$. In a digital quantum simulator each of the quasi-local propagators $\exp(- i H_\alpha(m \Delta t) \Delta t)$ can be efficiently approximated by (or in many cases exactly decomposed into) a fixed number of operations from a universal set of gates \citep{lloyd-prl-75-346,kitaev-russmathsurv-52-1191,nielsen-book}. As a consequence, the evolution is approximated by a stroboscopic sequence of many small time steps of dynamics due to the quasi-local interactions present in the system. The desired global time evolution according to the full many-body Hamiltonian, $\dot{\rho} = - i [H(t), \rho]$ (see coherent part of Eq.~(\ref{eq:master_equation})) emerges as an effective, coarse-grained description of the dynamics, as sketched in Fig.~\ref{fig:digitalsimulator}. For a finite number of time steps $n$, errors from possible non-commutativity of the quasi-local terms in the Hamiltonian, $[H_{\alpha}, H_{\alpha'}] \neq 0$, are bounded \citep{nielsen-book,berry-comm-mat-phys-270-359,bravyi-prl-101-070503} and can be reduced by resorting to shorter time steps $\Delta t$ or higher-order Trotter expansions \citep{suzuki-pla-165-387}. 

We note that it had been recognized early-on that dissipative dynamics can be efficiently simulated by carrying out unitary dynamics on an enlarged Hilbert space \citep{lloyd-science-273-1073}, such that efficient simulation of Hamiltonian dynamics is in principle sufficient to also realize open-system dynamics. In Sect.~\ref{sec:dig_sim_opensystems}, we will discuss an alternative approach for digital simulation of dissipative dynamics, which combines unitary operations and dissipative elements (in a Markovian setting). Recently, explicit error bounds for dissipative Trotter dynamics according to many-body master Eqs. (\ref{eq:Lindblad_part}) have been derived in~\citet{kliesch-prl-107-120501}. 

\begin{figure}
\begin{center}
\includegraphics[width=\columnwidth]{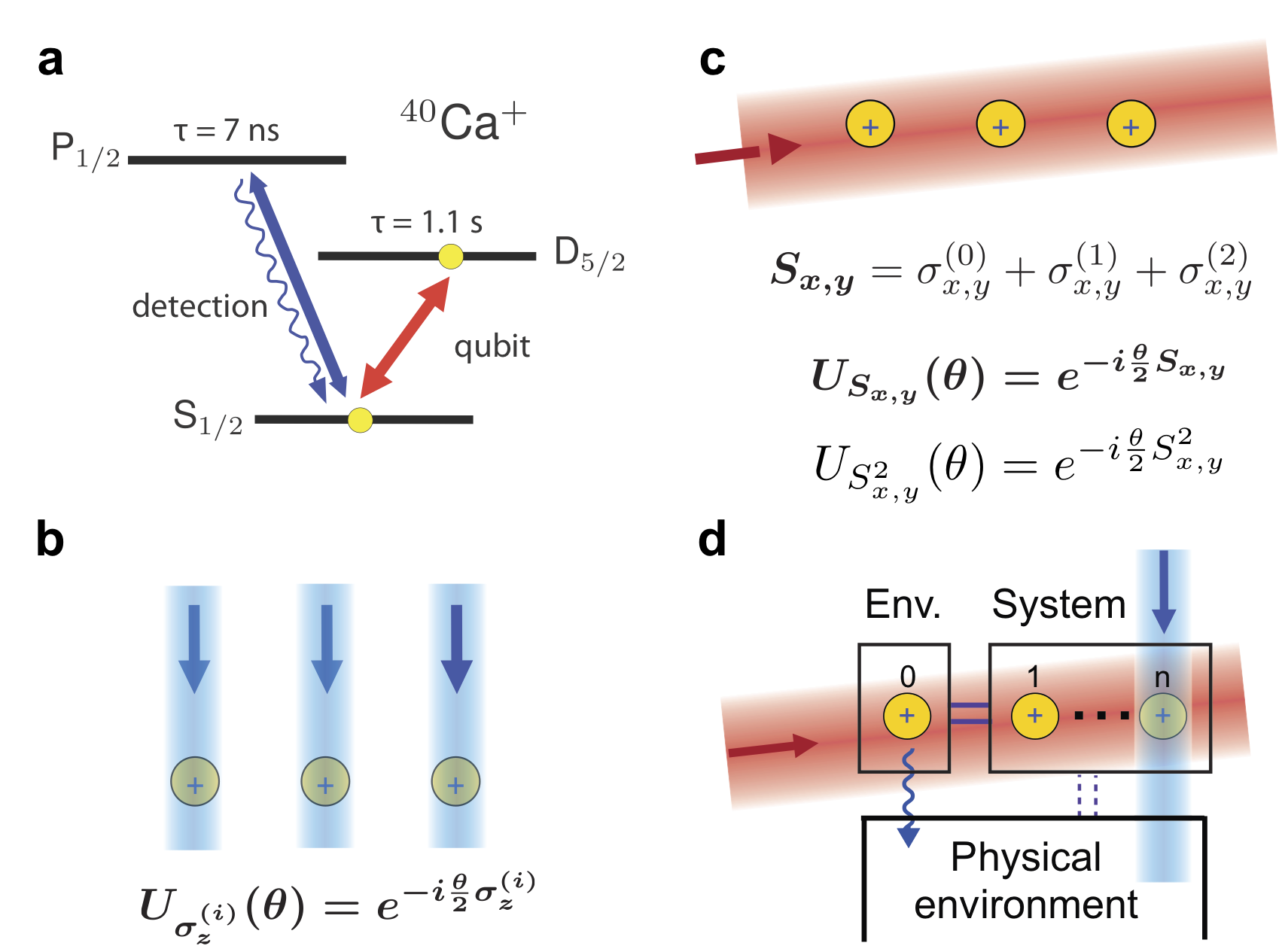}
\end{center}
\caption{(Color online) Toolbox of quantum operations in the Innsbruck ion trap quantum computer. a) Simplified level scheme of laser-cooled $^{40}$Ca$^+$ ions stored in a linear Paul trap: Long-lived internal electronic states $\ket{D} = \ket{0}$ and $\ket{S} = \ket{1}$ represent the qubit, while short-lived transitions are used for read-out of the quantum state of the qubit using a fluorescence measurement technique. b) The universal set of gates is formed by addressed single-qubit z-rotations and c) collective x- and y-rotations as well collective entangling operations $U_{S_{x,y}^2}$, as suggested by \cite{molmer-prl-82-1835}. d) For the simulation of open-system dynamics (see Sect.~\ref{sec:dig_sim_opensystems}) the string of ions can be divided into system qubits S (ions 1 through n) and an ``environment" qubit E. Coherent gate operations on S and E, combined with a controllable dissipative mechanism involving spontaneous emission of a photon from the environment ion via an addressed optical pumping technique \citep{science-schindler-332-1059}, allow one to tailor the coupling of the system qubits to an artificial environment (see ~\cite{barreiro-nature-470-486} for experimental details). This should be contrasted to the residual, detrimental coupling of the system (and environment) ions to their physical environment. Figure adapted from~\cite{barreiro-nature-470-486}.
}%
\label{fig:ion_qs_techniques}%
\end{figure}%

\subsubsection{Coherent Digital Simulation with Trapped Ions}
\label{sec:coh_sim_ions}

A recent experiment  carried out on a small-scale trapped ion quantum computer \citep{lanyon_science-334-57} has explored and demonstrated in the laboratory the various aspects of digital Hamiltonian quantum simulation. In a series of digital quantum simulations according to different types of interacting quantum spin models the performance of the digital simulation approach for systems of increasing complexity in the interactions and increasing system sizes was quantitatively studied. The experiments, whose main aspects we will briefly summarize in this section, have been enabled by remarkable progress in the implementation of individual gate operations (see Fig.~\ref{fig:ion_qs_techniques} for details on the experimental simulation toolbox). In particular, multi-ion entangling gates have been realized with fidelities higher than 99\% for two ions \citep{benhelm-nphys-4-463, roos-njp-10-013002}, and for up to 14 qubits \citep{monz-prl-106-130506}.

\begin{figure}
 \begin{center}
        \vspace{-2mm}
        \includegraphics[width=0.95\columnwidth]{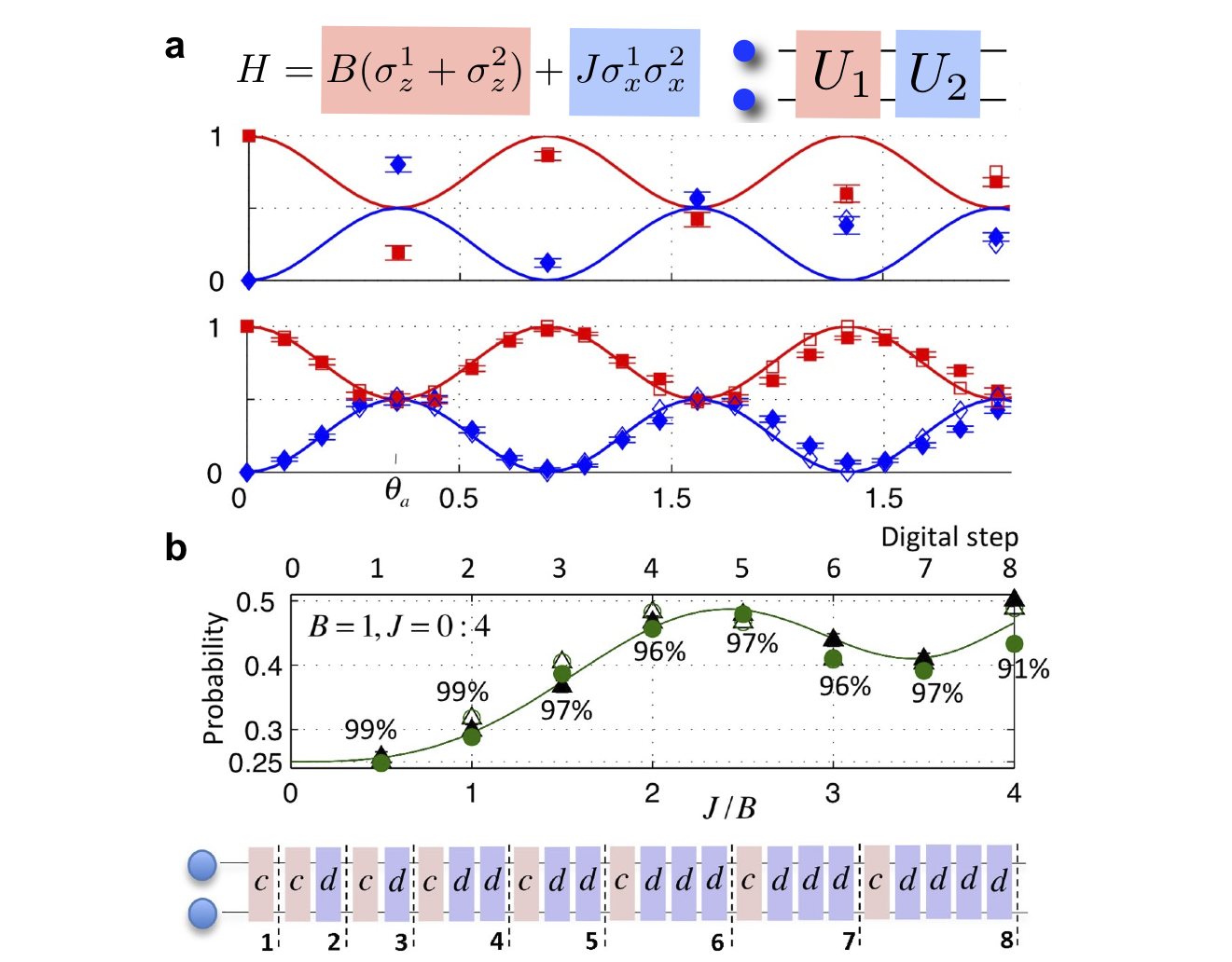}
        \end{center}
\caption{(Color online) Digital Hamiltonian simulation with trapped ions. This figure and the following one present some basic concepts of digital Hamiltonian simulation, and illustrate them with examples from a recent experiment with trapped ions \citep{lanyon_science-334-57}, where the digital approach was used to simulate various interacting quantum spin models of different complexity in the interactions and different system sizes. The simulations were realized using the toolbox of available coherent gates specified in Fig.~\ref{fig:ion_qs_techniques}. a) \textit{Time-independent Hamiltonian simulation.} Dynamics of the initial state $\ket{\uparrow\uparrow}$ under a time-independent two-spin Ising Hamiltonian with $J=2B$: As expected, the simulated dynamics according to a first-order Trotter decomposition converge closer to the exact dynamics as the digital resolution is increased, i.e.~the size of the individual time steps is decreased. It is convenient to introduce a dimensionless Hamiltonian $\tilde{H}$, i.e.~$H{=}E\tilde{H}$ such that $U{=}e^{-i\tilde{H}E \Delta t}$ and the evolution is quantified by a unitless phase $\theta = E \Delta t$. Each single digital step is given by $U_1 U_2 = U_{S_x^2}(\theta_a/n) \,U_{\sigma_z^{(1,2)}}(\theta_a/n)$ with $\theta_a = \pi/2\sqrt{2}$ and $n=1$ and $n=4$ (finer Trotter resolution). 
(Labeling: Lines: exact dynamics. Unfilled shapes: ideal digitised (Trotter decomposition). Filled shapes: experimental data. $\color{red}{\blacksquare}$$\uparrow\uparrow$, $\color{blue}{\Diamondblack}$$\downarrow\downarrow$). b) \textit{Time-dependent Hamiltonian dynamics.} Time evolution under a two-spin Ising Hamiltonian, where the spin-spin interaction strength $J$ increases linearly from 0 to 4$B$ during a total evolution given by $\theta_t{=}\pi/2$. In the experiment, the continuous dynamics is approximated using a sequence of 24 gates, with $c{=}U_{\sigma_z^{(1,2)}}(\pi/8)$, $d{=}U_{S_x^2}(\pi/16)$. The increase of $J$ over time is reflected by an increase in the number of $d$-blocks per Trotter step. The observed oscillation in population expectation values (measured in the $\sigma_x$-basis) is a diabatic effect due to the finite speed in ramping up the interaction term $H_\mathrm{int}$ ($\color[rgb]{0, 0.498, 0}{\medbullet}$$\rightarrow\rightarrow_x$, $\blacktriangle$$\leftarrow\leftarrow_x$). Percentages: fidelities between measured and exact states with uncertainties less than 2\%. Figure reprinted with
permission from~\citet{lanyon_science-334-57}. Copyright 2011 by MacMillan.
}
\label{QuantumSimulation1}
\end{figure}

\begin{figure}
 \begin{center}
        \vspace{-2mm}
        \includegraphics[width=0.95\columnwidth]{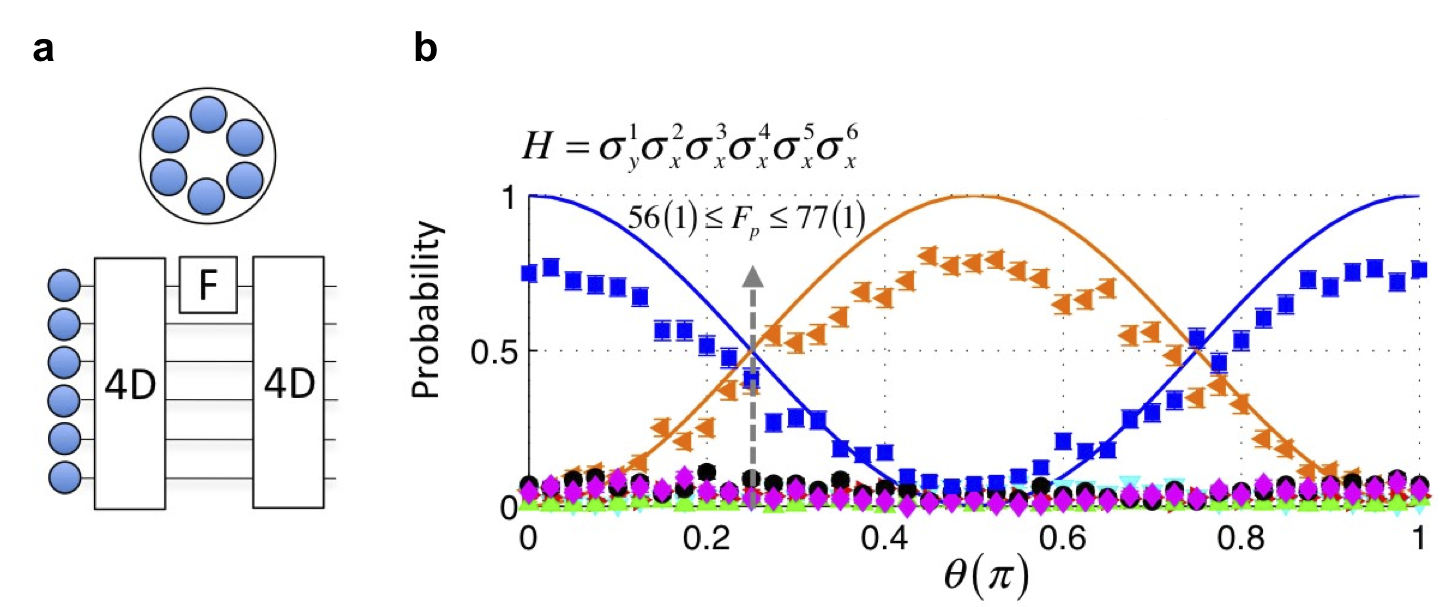}
        \end{center}
 \caption{(Color online) Continuation of digital Hamiltonian simulation with trapped ions.
\textit{Simulation of $n$-body interactions.} a) In a digital simulation, $n$-body spin interactions (with $n>2$) are usually realized by quantum circuits involving $2n$ two-qubit C-NOT gates \citep{nielsen-book}. However, the availability of high-fidelity, collective entangling gates acting on $n$ ions allows one to bundle the effect of such series of two-qubit gates and thus to realize, e.g., six-body interactions by a highly compact, experimentally efficient quantum circuit involving two six-ion gates ($4D{=}U_{S_x^2}(\pi/4)$), interspersed with one single-ion z-rotation ($F {=}U_{\sigma_z^{(1)}}(2\theta)$). The strength of the six-body interaction is controlled by the phase $\theta$ in the single-qubit rotation (see Ref.~\cite{mueller-njp-13-085007} for theoretical details).
b) Experimentally observed dynamics induced by a six-body spin interaction, which directly couples the states $\ket{\uparrow\uparrow\uparrow\uparrow\uparrow\uparrow}$ and $\ket{\downarrow\downarrow\downarrow\downarrow\downarrow\downarrow}$, periodically producing a maximally entangled GHZ state. Lines: exact dynamics. Filled shapes: experimental data. The quantitative characterization and assessment of errors of such multi-qubit building blocks is a non-trivial task, as standard quantum process tomography is impractical for more than 3 qubits. The inequality bounds the quantum process fidelity $F_p$ at $\theta{=} 0.25$ -- see online material of ~\cite{lanyon_science-334-57} for details on the employed technique. Figure reprinted with permission from~\citet{lanyon_science-334-57}. Copyright 2011 by MacMillan.
}
 \label{QuantumSimulation2}
\end{figure}

\textit{Exploring Trotter dynamics with two spins} -- To illustrate the Trotter simulation method, the conceptually most simple example of an Ising system of two interacting spin-1/2 particles as an elementary building block of larger and more complex spin models was studied: The Hamiltonian is given by the sum of two non-commuting terms, $H = H_\mathrm{int} + H_\mathrm{magn}$, where
$H_\mathrm{int} = J \sigma_x^1 \sigma_x^2$ describew a spin-spin interaction, and $H_\mathrm{magn}$ the coupling to an effective, transverse magnetic field $H_\mathrm{magn} = B (\sigma_z^1 + \sigma_z^2)$. This was one of the first systems to be simulated with trapped ions following an analog approach \citep{friedenauer-nphys-4-757,porras-prl-92-207901}. The experiments \citep{lanyon_science-334-57} studied the two-spin dynamics both for the time-independent Ising-Hamiltonian (see Fig.~~\ref{QuantumSimulation1}(a)), as well as for the time-dependent case where the interaction term $H_\mathrm{int}$ was linearly ramped up in time (see Fig.~\ref{QuantumSimulation1}(b)). The time evolution was realized by a first-order Trotter decomposition, where the propagators for small time steps according to the two Hamiltonian terms were decomposed into sequences of experimentally available single- and two-qubit gates. 

\textit{Simulation of larger systems and $n$-body interactions} -- Experiments with up to six ions \citep{lanyon_science-334-57} showed that the digital approach allows arbitrary interaction distributions for larger interacting spin systems to be programmed. For instance, it is possible to implement spatially inhomogeneous distributions of interaction strengths and to simulate $n$-body interaction terms, with $n>2$, in a non-perturbative way (see Fig.~\ref{QuantumSimulation2}). Many-body spin interactions of this kind are an important ingredient in the simulation of systems with strict symmetry requirements~\citep{kassal-annrevphyschem-62-185}. Furthermore, they appear in the context of many-body quantum systems exhibiting topological order \citep{nayak-rmp-80-1083} and in the context of topological quantum computing and memories \citep{kitaev-annalsphys-303-2,dennis-j-mat-phys-43-4452}. In Sect.~\ref{sec:toric_code} we will discuss in more detail Kitaev's toric model \citep{kitaev-annalsphys-303-2} as an example for a complex spin model involving four-body spin interaction terms. Engineering of three-body interactions in analog quantum simulators has been suggested for trapped ions \citep{bermudez-pra-79-060303} and polar molecules \citep{buechler-natphys-3-726}; however, it is in general very difficult to achieve dominant, higher-order interactions of substantial strength via analog quantum simulation techniques. Fig.~\ref{QuantumSimulation2}(b) shows the digital simulation of  time evolution according to a six-spin many-body interaction, where each Trotter time step was experimentally realized by a highly compact quantum circuit involving two collective six-ion entangling gates as essential resource \citep{lanyon_science-334-57}.

In view of these remarkable experimental advances and the demonstrated flexibility and control achieved so far, two major remaining challenges are (i) the quantum simulation of \textit{open-system} quantum dynamics according to many-body master equations of the form (\ref{eq:master_equation}) and (ii) to scale up the simulations from a few qubits to larger system sizes. Regarding the latter aspect, we will in the next section leave the trapped ions for a moment and switch to another physical platform, where we will discuss an \textit{a priori} scalable, digital simulation architecture based on Rydberg atoms stored in optical lattices or magnetic trap arrays. In Sect.~\ref{sec:dig_sim_opensystems} we will then extend the discussion to open many-particle quantum systems and describe how to simulate complex \textit{dissipative} many-body dynamics. In this context we will come back to trapped ions, where recently the building blocks of an open-system quantum simulator have been successfully implemented \citep{barreiro-nature-470-486}.

\subsection{Scalable Quantum Simulation with Rydberg Atoms}
\label{sec:Ryderg_simulator}
Laser excited Rydberg atoms \citep{gallagher-book} offer unique possibilities for quantum information processing and the study of strongly correlated many-body dynamics. Atoms excited to high-lying Rydberg states interact via strong and long-range dipole-dipole or Van der Waals forces \citep{gallagher-book} over distances of several $\mu$m, which are internal state-dependent and can be up to 12 orders of magnitude stronger than interactions between ground state atoms at a comparable distance \citep{saffman-rmp-82-2313}. Electronic level shifts associated with these interactions can be used to block transitions of more than one Rydberg excitation in mesoscopic atomic ensembles. This ``dipole blockade'' \citep{jaksch-prl-85-2208,lukin-prl-87-037901} mechanism underlies the formation of ``superatoms'' in atomic gases with a single Rydberg excitation shared by many atoms within a blockade radius. This effect gives rise to strongly correlated, dominantly coherent many-body dynamics \citep{raitzsch-prl-100-013002}, which has been explored in recent years both experimentally \citep{tong-prl-93063001, singer-prl-93-163001, cubel-pra-72-023405, vogt-prl-97-083003, mohapatra-prl-98-113003, heidemann-prl99-163601, reetz-lamour-prl-100-253001} and theoretically \citep{pohl-prl-104-043002,weimer-prl-101-250601,olmos-prl-103-185302,sun-njp-10-045032,honer-prl-105-160404}. In the context of quantum information processing, it has been recognized that these strong, switchable interactions between pairs of atoms potentially provides fast and addressable two-qubit entangling operations \citep{jaksch-prl-85-2208,lukin-prl-87-037901} or effective spin-spin interactions \citep{lesanovsky-prl-106-025301,pohl-prl-104-043002,schachenmayer-njp-12-103044,weimer-prl-101-250601}; recent theoretical proposals have extended Rydberg-based protocols towards a single-step, high-fidelity entanglement of a mesoscopic number of atoms \citep{moller-prl-100-170504,mueller-prl-102-170502}. Remarkably, the basic building blocks of Rydberg-based quantum information processing have been demonstrated recently in the laboratory by several groups, which observed the dipole blockade between a pair of neutral Rydberg atoms stored in optical tweezers \citep{urban-natphys-5-110,gaetan-natphys-5-115}. Here, the Rydberg blockade was used as a mechanism to create two-atom entanglement \citep{wilk-prl-104-010502} and to realize the first neutral atom two-qubit C-NOT gate \citep{isenhower-prl-104-010503}. 

\begin{figure}
\begin{center}
\includegraphics[width=0.6\columnwidth]{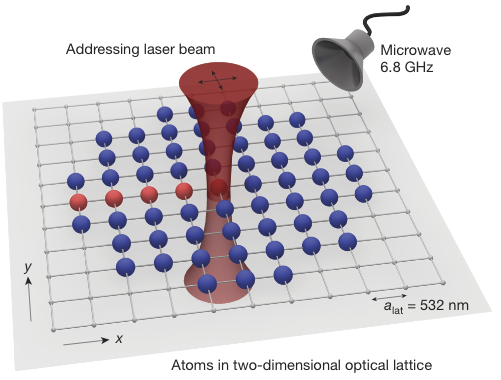}
\includegraphics[width=0.35\columnwidth]{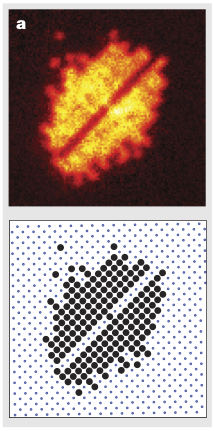}
\end{center}
\caption{(Color online) Single-site addressing of atoms in an optical lattice. The left part shows a schematics of atoms loaded into a square optical lattice, where they form a Mott insulator state with one atom per lattice site. Atoms residing on individual lattice sites in the $x-y$ plane can be optically addressed with an off-resonant laser beam, which can be focused to individual sites by means of a high-aperture microscope objective. The upper part of (a) shows an experimentally obtained fluorescence image of a Mott insulator site with one atom per site, where a subset of atoms (diagonal of the image) has been transferred from an internal state $\ket{0}$ to $\ket{1}$ by means of the single-site addressed beam. Before fluorescence detection, the atoms in $\ket{1}$ are removed from the lattice by a resonant laser pulse. The bottom part shows the reconstructed atom number distribution (see ~\citet{sherson-nature-467-68} for details on the reconstruction algorithm), where filled black circles correspond to single atoms and dots indicate the position of the lattice sites.  Figure adapted with
permission from~\citet{weitenberg-nature-471-319}. Copyright 2011 by MacMillan.
}
\label{fig:SingleSiteAddressability}%
\end{figure}

On the other hand, cold atoms stored in optical lattices or magnetic trap arrays offer a versatile platform for \textit{a priori} scalable quantum information processing and quantum simulation \citep{jaksch-prl-81-3108, jaksch-ann-phys-315-52, greiner-nature-415-39,bloch-rmp-2008,lewenstein-rmp-56-135,dalibard-rmp-83-1523}. In particular, in sufficiently deep lattices, where tunneling between neighboring lattice sites is suppressed, single atoms can be loaded and kept effectively frozen at each lattice site, with long-lived atomic ground states representing qubits or effective spin degrees of freedom. Working with large-spacing lattices, with inter-site distances of the order of a few $\mu$m \citep{nelson-natphys-3-556,whitlock-njp-11-023021} allows single-site addressing with laser light, and thus individual manipulation and readout of atomic spins. Very recently, several groups have achieved single-site addressing in optical lattices \citep{bakr-nature-462-74,sherson-nature-467-68,bakr-science-329-547} and manipulation of individual spins in this setup \citep{weitenberg-nature-471-319} (see Fig.~\ref{fig:SingleSiteAddressability}).

As we will discuss below, given these achievements and the future integration of techniques for coherent laser excitation of Rydberg atoms in addressable (optical) lattice setups~\citep{viteau-prl-107-060402,anderson-prl-107-263001}, in principle all essential ingredients seem to exist already in the laboratory to build a scalable, digital quantum simulator based on cold Rydberg atoms \citep{weimer-nphys-6-382}. 

Before specifying in more detail the concrete physical architecture of the Rydberg quantum simulator proposed in ~\citet{weimer-nphys-6-382}, we will in the next section discuss a specific many-body spin model of interest: Kitaev's toric code \citep{kitaev-annalsphys-303-2}. This model represents a paradigmatic example of a large class of spin models, which have in the last years attracted great interest in the context of topological quantum information processing and as strongly interacting many-body quantum systems exhibiting topological order \citep{nayak-rmp-80-1083,wen-book}. This example illustrates the generic challenges and goals of a quantum simulation of complex many-body models, which are to be addressed in a concrete physical implementation of a quantum simulator. The realization of a more complex setup of a three-dimensional $U(1)$ lattice gauge theory giving rise to a spin liquid phase will be discussed below in Sect.~\ref{sec:digital_simulation_toolbox_and_LGT}.

\subsubsection{Paradigmatic Example: Simulation of Kitaev's toric Code Hamiltonian}
\label{sec:toric_code}

\begin{figure}
\begin{center}
\includegraphics[width=\columnwidth]{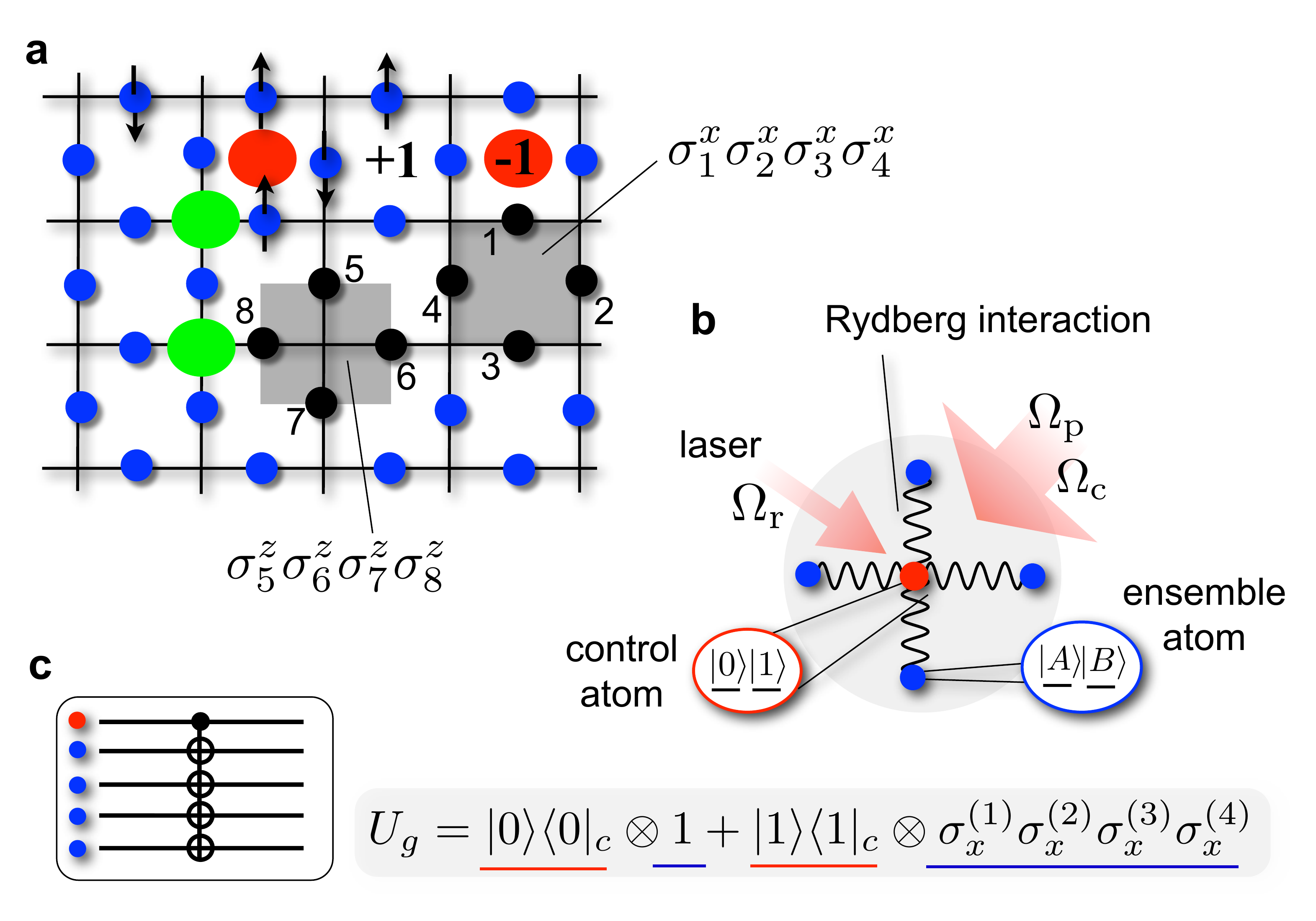}
\end{center}
\caption{(Color online) Schematics of the Rydberg quantum simulator architecture and a multi-atom C-NOT$^N$ Rydberg gate as its principal building block. a) The Rydberg quantum simulator \citep{weimer-nphys-6-382} is particularly suited for the simulation of coherent and dissipative dynamics of complex quantum spin models involving $n$-body interactions and constraints. A paradigmatic example is Kitaev's toric code Hamiltonian \citep{kitaev-annalsphys-303-2}, where spins are located on the edges of a two-dimensional square lattice and interact via four-body plaquette or vertex interactions. The model exhibits two types of localized quasi-particle excitations (depicted as red and green dots), which exhibit Abelian anyonic statistics under braiding, i.e.~when they are winded around each other. b) A mesoscopic multi-atom Rydberg gate \citep{mueller-prl-102-170502} applied to subsets of four spins around plaquettes and vertices, and additional control atoms, which are located at the centers of the plaquettes and on the vertices of the lattice, allows one to efficiently realize such many-body plaquette and vertex interactions. Here, controllable strong and long-range Rydberg interactions mediate effective four-body interactions among the system spins. By a combination of the multi-qubit C-NOT gate shown in (c) with optical pumping on the auxiliary control atoms, it is possible to engineer dissipative $n$-body processes. This many-body reservoir engineering can be used to realize cooling dynamics, which leads, e.g.,  to the dissipative ground state preparation of Kitaev's toric code Hamiltonian.
}%
\label{fig:RydbergSimulator}%
\end{figure}%

Kitaev's toric code is a paradigmatic, exactly solvable model, out of a large class
of spin models, which have recently attracted a lot of interest in the context of studies on
topological order and quantum computation. It considers a two-dimensional
setup, where spins are located on the edges of a square lattice
\citep{kitaev-annalsphys-303-2}. The Hamiltonian $H=-E_{0}\left(\sum_{p}A_{p}+\sum_{s}B_{s}\right)$
is a sum of mutually commuting stabilizer operators $A_{p}=\prod_{i\in p}\sigma_{i}^{x}$
and $B_{s}=\prod_{i\in s}\sigma_{i}^{z}$, which describe four-body
interactions between spins located around plaquettes ($A_{p}$) and
vertices ($B_{s}$) of the square lattice (see Fig. \ref{fig:RydbergSimulator}a). All $A_p$ and $B_s$ stabilizer operators mutually commute, thus the ground state  of the Hamiltonian is a simultaneous eigenstate of all stabilizer
operators $A_{p}$ and $B_{s}$ with eigenvalues $+1$,  and gives rise to a topological phase:
the ground state degeneracy depends on the boundary conditions and topology of the setup, and
the elementary excitations exhibit Abelian anyonic statistics under braiding, i.e.~when they are winded around each other.
The toric code shows two types of localized excitations corresponding to $-1$ eigenstates of each
stabilizer $A_p$ (``magnetic charge'', filled red dots in Fig.~\ref{fig:RydbergSimulator}a) and $B_p$ (``electric charge'', filled green dots).

\begin{figure}
\begin{center}
\includegraphics[width=\columnwidth]{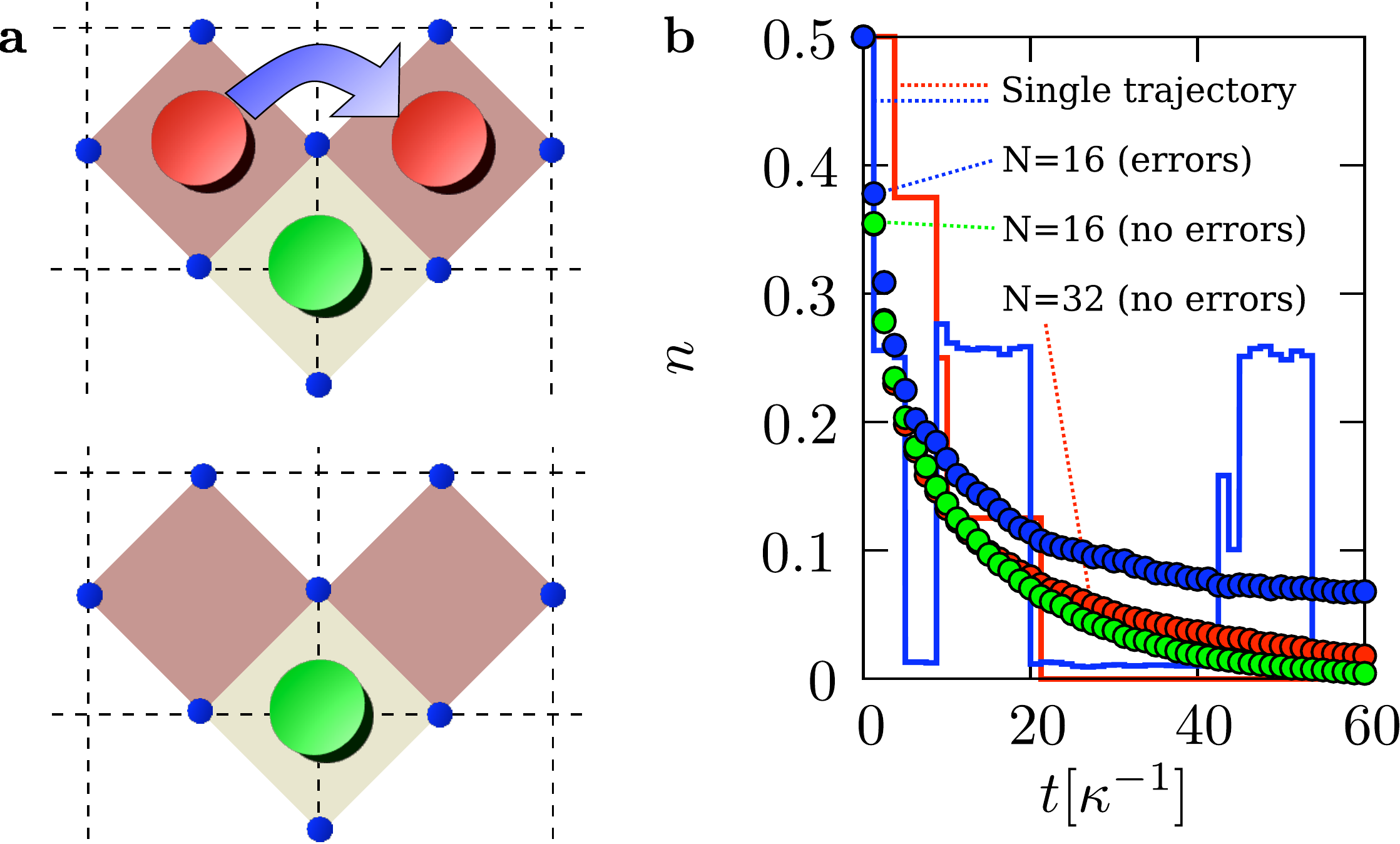}
\end{center}
\caption{(Color online) Cooling of Kitaev's toric code: a) A dissipative time step incoherently moves
one anyonic excitation (red dot) on top of a second anyon
located on a neighboring plaquette, annihilating each other and
thus lowering the internal energy of the system. The anyon of the other type (an ``electric charge'', filled green dot located on a vertex of the lattice) remains unaffected by this cooling step. b) Numerical simulation
of the cooling for $N$ lattice sites (periodic boundary conditions).
Single trajectories for the anyon density $n$ over time are shown as solid lines. Filled
circles represent averages over 1000 trajectories. The initial state for the simulations is the
fully polarized, experimentally easily accessible state of all spins down. For perfect gates in the digital quantum simulation discussed in detail in Sect.~\ref{sec:stabilizer_pumping}, the energy of the system reaches the ground state energy in the long
time limit, while for imperfect gates heating events can occur (blue solid line) and a finite
density of anyons $n$ remains present (blue circles). The characteristic time scale
$\kappa^{-1}$ for cooling is set by (i) the gate parameters in the quantum circuit decomposition discussed below (see Sect.~\ref{sec:stabilizer_pumping} and (ii) by the duration for the implementation of the underlying quantum gates. Figure reprinted with
permission from~\citet{weimer-nphys-6-382}. Copyright 2010 by MacMillan.
}%
\label{fig:toriccodecooling}%
\end{figure}

\indent In addition to the toric code Hamiltonian, one can formulate a dissipative many-body dynamics, which {}``cools'' into the ground state manifold of the many-body Hamiltonian. Such dissipative time evolution is provided by a Liouvillian (\ref{eq:Lindblad_part}) with quantum jump operators,
\begin{equation}
c_{p}=\frac{1}{2}\sigma_{i}^{z} (1-A_{p} ),\hspace{10pt}c_{s}=\frac{1}{2}\sigma_{j}^{x} (1-B_{s}),
\label{eq:Kitaev_jump_operators}
\end{equation}
with $i\in p$ and $j\in s$, which act on four spins located around plaquettes $p$ and vertices $s$, respectively. In Sect.~\ref{sec:stabilizer_pumping} we will discuss in detail how these four-body quantum jump operators can be physically implemented in the Rydberg simulator architecture of ~\citet{weimer-nphys-6-382}. The jump operators are readily understood as operators which {}``pump''
from $-1$ into $+1$ eigenstates of the stabilizer operators: the part $(1-A_{p})/2$ of $c_{p}$ is a projector onto the eigenspace of
$A_p$ with $-1$ eigenvalue (an excited state with a ``magnetic charge'' present), while all states in the $+1$ eigenspace are dark states.
The subsequent  spin flip $\sigma_{i}^{z}$ transfers the excitation to the neighboring plaquette.
The jump operators then give rise to a random walk of anyonic excitations on the lattice, and whenever two excitations of the
same type meet they are annihilated, resulting in a cooling process, see Fig.~\ref{fig:toriccodecooling}a. Similar arguments apply to the jump operators $c_{s}.$ Efficient cooling is achieved by alternating the index $i$ of the spin, which is flipped.\\
\indent Our choice of the jump operator follows the idea of reservoir engineering
of interacting many-body systems as discussed in~\citet{diehl-natphys-4-878,kraus-pra-78-042307} and in Sect.~\ref{sec:Analog_QS_DissDyn}. In contrast to alternative schemes for measurement based state preparation
\citep{aguado-prl-101-260501}, here, the cooling is part of the time evolution
of the system. These ideas can be readily generalized to more complex stabilizer
states and to setups in higher dimensions, as in, e.g., the color
codes developed in~\citet{bombin-prl-97-180501,bombin-pra-76-012305}, and the simulation of a three-dimensional U(1) lattice gauge theory, which will be discussed in Sect.~\ref{sec:digital_simulation_toolbox_and_LGT}. \\
\indent In conclusion, the main challenge in the quantum simulation of coherent Hamiltonian dynamics and dissipative ground state preparation of many-body spin models such as Kitaev's toric code Hamiltonian lies in (i) the realization of strong $n$-body interactions, and (ii) the ability to tailor multi-particle couplings of the many-body system to a reservoir, such that the dissipative dynamics gives rise to ground state cooling, as described by a many-body master Eq. (\ref{eq:Lindblad_part}) with many-body quantum jump operators of Eq.(\ref{eq:Kitaev_jump_operators}).

\subsubsection{A Mesoscopic Rydberg Gate}
Let us now turn to the physical implementation of the digital Rydberg simulator setup suggested in~\citet{weimer-nphys-6-382}. A key ingredient of the proposed architecture are additional auxiliary qubit atoms in the lattice, which play a two-fold role: First, they control and \emph{mediate} effective $n$-body spin interactions among a subset of $n$ system spins residing in
their neighborhood in the lattice, as e.g.~the four-body plaquette and vertex interactions of Kitaev's toric code Hamiltonian discussed above. In the proposed scheme this is achieved efficiently
making use of single-site addressability and a parallelized multi-qubit gate,
which is based on a combination of strong and long-range Rydberg interactions
and electromagnetically induced transparency (EIT) and is schematically shown in Fig.~\ref{fig:RydbergSimulator}b. This gate has been suggested and analyzed in~\citet{mueller-prl-102-170502}.  As it plays a central role in the simulation architecture, we will briefly and on a qualitative level review its main features here. Second, the auxiliary atoms can be optically pumped, thereby providing a dissipative element, which in combination with Rydberg interactions results in effective collective dissipative dynamics of a
set of spins located in the vicinity of the auxiliary particle. This enables, e.g., the simulation of dissipative dynamics for ground state cooling of Kitaev's toric code and related models.

\textit{Setup of the Rydberg gate} -- The envisioned setup is illustrated in Fig.~\ref{fig:RydbergSimulator}b. A control atom and a
mesoscopic ensemble of, say, four atoms are stored in separate trapping potentials,
e.g. in two dipole traps as in~\citet{wilk-prl-104-010502,isenhower-prl-104-010503} or in neighboring lattice sites of a
(large-spacing) optical lattices or magnetic trap array \citep{whitlock-njp-11-023021}. The multi-qubit gate exploits state-dependent Rydberg interactions and realizes a controlled-NOT$^N$ (CNOT$^N$) gate, which is defined by
\begin{equation}
\label{eq:CNOT}
U_g = \left|0\right\rangle \!\left\langle 0\right|_{c}\mathop{\otimes}_{i=1}^{N} 1_i+\left|1\right\rangle \!\left\langle 1\right|_{c}\mathop{\otimes}_{i=1}^{N}\sigma^x_i.
\end{equation}
Depending on the state of the control qubit -- the state of all $N$ target qubits is left unchanged or flipped. Here, $|0\rangle$, $|1\rangle$ and $|A\rangle$, $|B\rangle$
denote long-lived ground states of the control and ensemble atoms,
respectively (see Fig.~\ref{fig:RydbergSimulator}b), and $\sigma_i^x \ket{A}_i = \ket{B}_i$ and $\sigma_i^x \ket{B}_i = \ket{A}_i$.

The basic elements of the gate of Eq. (\ref{eq:CNOT}) are: (i) the control atom can be individually addressed
and laser excited to a Rydberg state conditional to its internal state,
thus (ii) turning on or off the strong long-range Rydberg-Rydberg
interactions of the control with ensemble atoms, which (iii) via EIT-type
interference suppresses or allows the transfer of all ensemble
atoms from $|A\rangle$ or $|B\rangle$ conditional to the state of
the control atom. It does not necessarily require individual
addressing of the ensemble atoms, in contrast to a possible implementation of the gate (\ref{eq:CNOT}) by a sequence of $N$ two-qubit C-NOT gates.

\begin{figure}
\begin{center}
\includegraphics[width=\columnwidth]{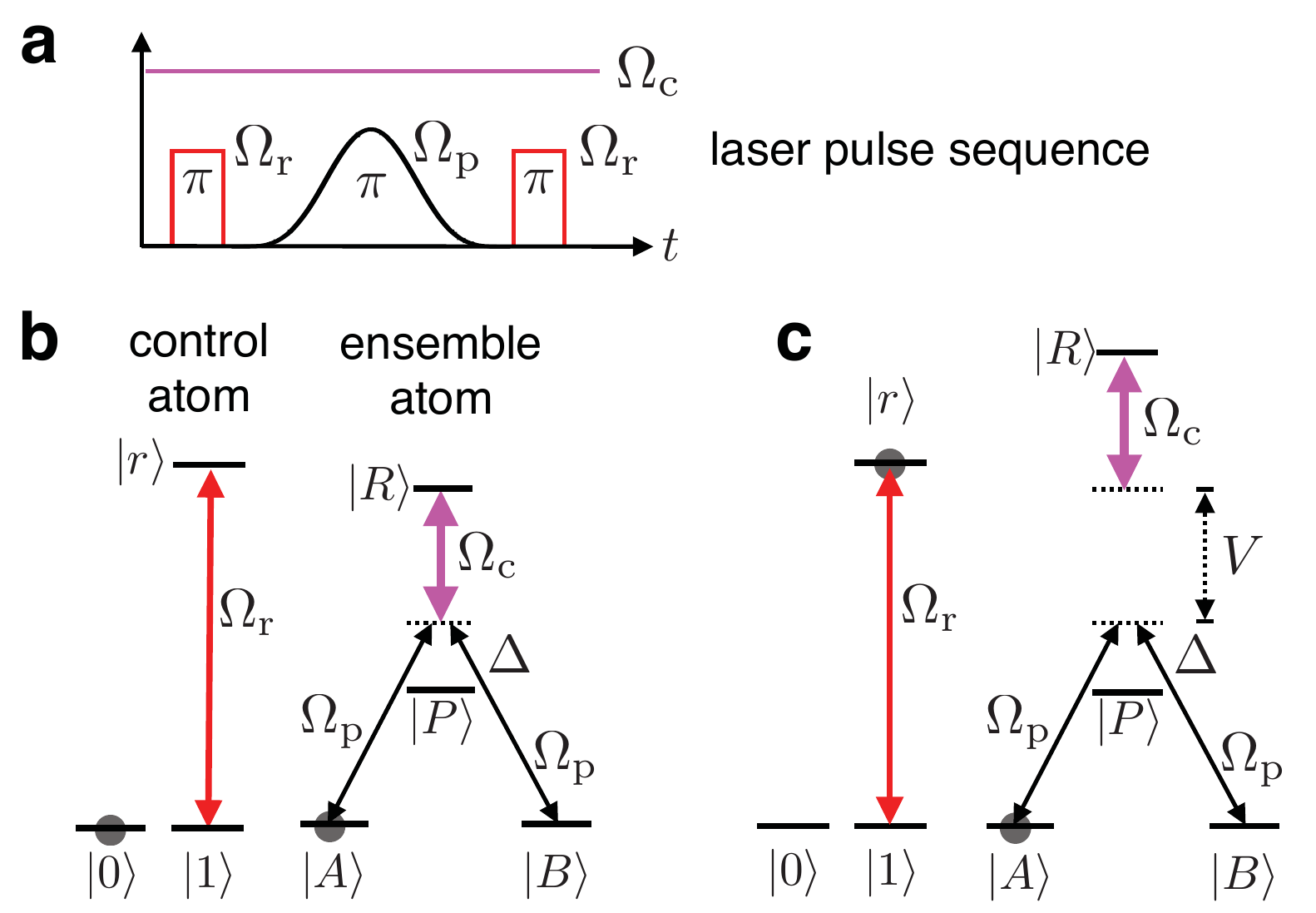}
\end{center}
\caption{(Color online) Mesoscopic Rydberg gate. a) Sequence of laser pulses (not to scale). b)
Electronic level structure of the control and ensemble atoms. The
ground state $\left|1\right\rangle $ is resonantly coupled to the
Rydberg state $\left|r\right\rangle $. The states $\left|A\right\rangle $
and $\left|B\right\rangle $ are off-resonantly coupled (detuning
$\Delta$, Rabi frequency $\Omega_{\mathrm{p}}$) to $\left|P\right\rangle $.
A strong laser with Rabi frequency $\Omega_{\mathrm{c}}\gg\Omega_{\mathrm{p}}$
couples the Rydberg level $\left|R\right\rangle $ to $\left|P\right\rangle $
such that $\left|R\right\rangle $ is in two-photon resonance with
$\left|A\right\rangle $ and $\left|B\right\rangle $. In this situation
(known as EIT) Raman transfer from $\left|A\right\rangle $ to $\left|B\right\rangle $
is inhibited. c) With the control atom excited to $\left|r\right\rangle $
the two-photon resonance condition is lifted as the level $\left|R\right\rangle $
is shifted due to the interaction energy $V$ between the Rydberg
states, thereby enabling off-resonant Raman transfer from $\left|A\right\rangle $
to $\left|B\right\rangle $. Figure adapted from~\citet{mueller-prl-102-170502}
}%
\label{fig:RydbergGate}%
\end{figure}%

\textit{Implementation of the gate operation} -- For the physical realization of the operation (\ref{eq:CNOT}), an auxiliary Rydberg level $|r\rangle$ of the control atom is used, which is resonantly coupled to $|1\rangle$ by a laser with (two-photon) Rabi frequency $\Omega_{\mathrm{r}}$ (see Fig.~\ref{fig:RydbergGate}). For the ensemble atoms the two stable ground states $|A\rangle$ and $|B\rangle$ are coupled far off-resonantly in a $\Lambda$-configuration with Rabi frequency $\Omega_{\mathrm{p}}$ and detuning $\Delta$ to a low-lying, intermediate state $|P\rangle$ (e.g.~$5^{2}P_{3/2}$ in case of $^{87}$Rb). A second laser with Rabi frequency $\Omega_{\mathrm{c}}$ ($\Delta \gg \Omega_{c} > \Omega_{p}$) couples $|P\rangle$ to a Rydberg state $|R\rangle$ of the ensemble atoms, such that the two ground states $|A\rangle$ and $|B\rangle$ are in two-photon resonance with $|R\rangle$, as depicted in Fig.~\ref{fig:RydbergGate}b.

The conditional, coherent transfer of population between the ground states of the ensemble atoms, as required for the C-NOT operation, is then achieved by a sequence of three laser pulses (shown in Fig.~\ref{fig:RydbergGate}a): (i) a short $\pi$-pulse on the control atom,
(ii) a smooth Raman $\pi$-pulse $\Omega_{\mathrm{p}}(t)$ with $\int_{0}^{T}\mathrm{d}t\,\Omega_{\mathrm{p}}^{2}(t)/(2\Delta)=\pi$
acting on all ensemble atoms, and (iii) finally a second $\pi$-pulse on the control atom. The effect of this pulse sequence can be understood by distinguishing the two possible cases of (a) blocked transfer (for the control atom initially residing in the logical state $\ket{0}$) and
(b) enabled transfer (with the control atom initially in $\ket{1}$).

\textit{(a) Blocked population transfer:} For the control atom initially residing in the logical state $\ket{0}$ the first pulse has no effect. In the regime $\Omega_\mathrm{p} \ll \Omega_\mathrm{c}$, the laser configuration of the ensemble atoms realizes an EIT scenario~\citep{fleischhauer-rmp-77-633}, where the strong always-on ``control'' laser field $\Omega_\mathrm{c}$ suppresses via destructive interference coupling of the ``probe'' laser $\Omega_\mathrm{p}$ to the intermediate state $\ket{P}$ and thus also the second-order Raman coupling. This also effectively inhibits population transfer between $\ket{A}$ and $\ket{B}$. As a consequence, the Raman pulses $\Omega_\mathrm{p}$ are ineffective (as well as the second $\pi$-pulse applied to the control atom in $\ket{1}$). The state of the ensemble atoms remains unchanged, thereby realizing the first logical half of the gate (\ref{eq:CNOT}).

\textit{(b) Enabled population transfer:} If the control atom initially resides in $\ket{1}$, it is excited to the Rydberg state $\ket{r}$ by the first pulse. Due to strong repulsive Rydberg interactions $V>0$ between the control atom in $\ket{r}$ and ensemble atoms in $\ket{R}$, the Rydberg level of the ensemble atom is now shifted by the energy $V$ (see Fig.~\ref{fig:RydbergGate}c), despite the fact that the Rydberg state $\ket{R}$ of the ensemble atoms is not populated. This interaction-induced energy shift lifts the two-photon resonance condition, which underlies the EIT scenario and is crucial to block the Raman transfer between $|A \rangle$ and $|B\rangle$. Now, the Raman lasers couple off-resonantly to $|P\rangle$ and the coherent population transfer between $|A\rangle$ and $|B\rangle$ takes place.

A quantitative analysis of the gate performance in~\citet{mueller-prl-102-170502} shows that the effect of the relevant error sources such as radiative decay from the $\ket{P}$ and the Rydberg states and possible mechanical effects are negligible for realistic atomic and laser parameters. Remarkably, undesired destructive many-body effects originating from undesired, but possibly strong Rydberg interactions between the ensemble atoms can be effectively suppressed and minimized in the limit $\Omega_\mathrm{p} \ll \Omega_\mathrm{c}$. As a consequence, the gate also works reliably and with high fidelity for a moderate number of ensemble atoms separated by up to a few microns, it is robust with respect to inhomogeneous inter-particle distances and varying interaction strengths and can be carried out on a microsecond timescale \citep{mueller-prl-102-170502}.

\subsubsection{Simulation of Coherent Many-Body Interactions}

\begin{figure}
\begin{center}
\includegraphics[width=0.9\columnwidth]{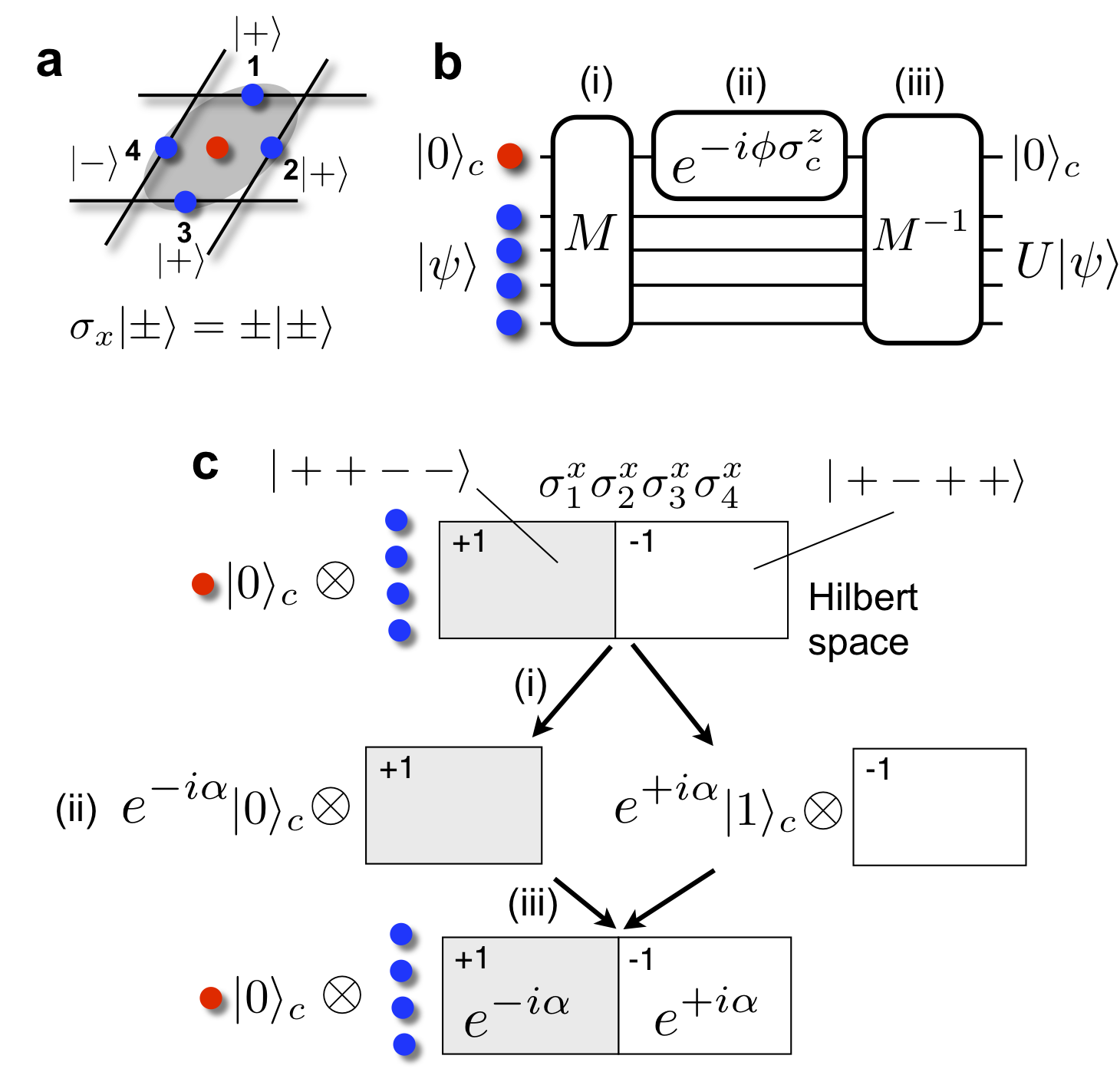}
\end{center}
\caption{(Color online) Simulation of coherent $n$-body interactions. a) Kitaev plaquette term corresponding to four-body interactions $A_p = \prod_{i=1}^4\sigma_{i}^{x}$. b) Three-step gate sequence, which implements desired time evolution $U=\exp(-i\phi A_{p})$ of the four system spins, mediated by an auxiliary control qubit. c) The gate sequence $M$ coherently maps
the information, whether the system spins reside in a $+1$ (e.g.~$\ket{++--}$) or $-1$ eigenstate (e.g.~$\ket{+-++}$) of the many-body interaction $A_p$ onto the internal state $|0\rangle_{c}$ and $|1\rangle_{c}$ of the control qubit. The mapping is given by $M = \exp(+i\pi\sigma_c^{y}/4) \, U_{g}\, \exp(-i\pi\sigma_c^{y}/4)$, i.e., up to single-qubit y-rotations of the control qubit, by the multi-atom Rydberg gate of Eq.~(\ref{eq:CNOT}). After the mapping, a single-qubit z-rotation of the control qubit $\exp\left(-i\phi\sigma_{c}^{z}\right)$ effectively imprints a phase $\exp(\mp i \phi)$ on all $\pm 1$ eigenstates of $A_p$.  After the mapping $M$ is reversed, the control qubit returns to $|0\rangle_{c}$ and thus factors out from the dynamics of the system spins, which have evolved according to $U$.
}%
\label{fig:circuits}%
\end{figure}

The many-qubit Rydberg gate \citep{mueller-prl-102-170502} discussed in the previous section is the key building block of the Rydberg quantum simulator architecture \citep{weimer-nphys-6-382}. Using an auxiliary qubit located at the center of a four-atom plaquette allows one to efficiently simulate coherent $n$-body interactions such as the four-body spin plaquette interactions
$A_{p}=\prod_{i}\sigma_{i}^{x}$ appearing in Kitaev's toric code Hamiltonian (Fig.~\ref{fig:circuits}). The general approach
consists of a sequence of three coherent steps, as depicted in Fig.~\ref{fig:circuits}b: 
(i) First, a gate sequence $M$ is performed, which coherently encodes
the information whether the four system spins are in a $+1$ or $-1$ eigenstate of $A_{p}$ in the
two internal states of the auxiliary control qubit (see Fig.~\ref{fig:circuits}c). (ii) In a second step,
a single qubit-gate operation, which depends on the internal state of the
control qubit, is applied. Due to the previous mapping this manipulation of
the control qubit is equivalent to manipulating the subspaces
with fixed eigenvalues $\pm1$ of $A_{p}$. Thus, effectively, the application of a single-qubit gate $\exp\left(-i\phi\sigma_{c}^{z}\right)$ on the control qubit imprints a phase shift $\exp(\mp i \phi)$ on all $\pm 1$ eigenstates of the stabilizer operator $A_p$. (iii) Finally, the mapping $M$ is reversed, and the control qubit returns to its initial state $\ket{0}$. Consequently, at the end of the sequence, the auxiliary qubit effectively factors out from the dynamics of the four system spins, which in turn have evolved according to the desired time evolution
\begin{equation}
U=\exp(-i\phi A_{p})=M^{-1}\exp\left(-i\phi\sigma_{c}^{z}\right)M.
\end{equation}
Note that the essential resource for one time step consists of two applications of the mesoscopic Rydberg gate $U_g$, which up to local rotations realizes the mappings $M$ and $M^{-1}$.  In contrast, a standard implementation via two-qubit C-NOT gates would correspond to eight entangling operations \citep{nielsen-book}.

For small phase imprints $\phi\ll1$ the mapping reduces to the standard equation for coherent time
evolution according to the master equation $\partial_{t}\rho=-i E_{0} [ A_{p},\rho ] + o(\phi^{2})$
and thus implements the propagator for a small Trotter time step according to the four-body spin interaction $A_{p}$ on one plaquette. The above scheme for the implementation of the many-body interaction
$A_{p}$ can be naturally extended to arbitrary many-body interactions between the system spins
surrounding the control atom, as e.g., the $B_{p}$ interaction terms in the toric
code. Gate operations on single system spins allow
to transform $\sigma_{i}^{x}$ into $\sigma_{i}^{y}$ and $\sigma_{i}^{z}$, in accordance with previous proposals for digital simulation of spin Hamiltonians \citep{sorensen-prl-83-2274},
while selecting only certain spins to participate in the many-body gate
via local addressability gives rise to the identity operator for the non-participating spins.

The associated energy scale of the many-body interactions becomes $E_{0}=\phi/\tau$
with $\tau$ the physical time needed for the implementation of all gates, which are required for a single time step according to the many-body Hamiltonian on the whole lattice. Note that in principle many of these operations at sufficiently distant areas of the lattice can be done in parallel, for instance by using super-lattices \citep{lee-prl-99-020402,folling-nature-448-1029} for the application of the required laser pulses. In this case the energy scale $E_0$ becomes independent of the lattice size, and is essentially only limited by the fast micro-second time scale of the Rydberg gates, potentially allowing for characteristic energy scales $E_0$ on the order of 10-100 kHz \citep{weimer-nphys-6-382,weimer-quantuminfprocess-10-885}.

\subsection{Digital Simulation of Open-System Dynamics}
\label{sec:dig_sim_opensystems}

In the previous sections, we have focused on the principles and physical examples of digital simulation of \textit{coherent} many-body interactions. Let us now extend the discussion to the digital simulation of \textit{dissipative} many-body dynamics. The dynamics of an open quantum system $\mathrm{S}$ coupled to an environment
$\mathrm{E}$ can be described by the unitary transformation $\rho_{SE}\mapsto
U\rho_{SE}U^{\dagger}$, with $\rho_{SE}$ the joint density matrix of the
composite system $\mathrm{S+E}$. Thus, the reduced density operator of the
system will evolve as $\rho =\mathrm{Tr}_{E} ( U\rho_{SE}U^{\dagger}) $.  The
time evolution of the system can also be described by a completely positive
Kraus map \citep{nielsen-book}
\begin{equation}
\label{eq:Kraus_map}
\rho \mapsto \mathcal{E}(\rho) =  \sum_k E_k \rho E_k^\dagger,
\end{equation}
where $\rho$ denotes the reduced density operator of the system, $\{E_k\}$ is a set of operation elements satisfying $\sum_k E_k^\dagger E_k = 1$, and we assume an initially uncorrelated system and environment. For the case of a closed system, decoupled from the environment, the map of Eq. (\ref{eq:Kraus_map}) reduces to $\rho \mapsto U \rho U^\dagger$ with $U$ the unitary time evolution operator of the system. The Markovian limit of the general quantum operation (\ref{eq:Kraus_map}) for the \emph{coherent} and \emph{dissipative} dynamics of a many-particle system is given by the many-body master Eq. (\ref{eq:master_equation}) discussed above.

Control of both coherent and dissipative dynamics is then achieved by finding
corresponding sequences of maps specified by sets of
operation elements $\{E_k\}$ and engineering these sequences in the laboratory.
In particular, for the example of dissipative quantum state preparation,
pumping to an entangled state $|\psi\rangle$ reduces to implementing
appropriate sequences of dissipative maps. These maps are chosen to drive the
system to the desired target state irrespective of its initial state. The
resulting dynamics have then the pure state $|\psi\rangle$ as the unique
attractor, $\rho \mapsto|\psi\rangle\langle\psi|$. In quantum optics and
atomic physics, techniques of optical pumping and laser cooling are
successfully used for the dissipative preparation of quantum states, although
on a \textit{single-particle} level. The engineering of dissipative maps for
the preparation of entangled states can be seen as a generalization of this
concept of pumping and cooling in driven dissipative systems to a
\textit{many-particle} context. For a discussion of Kraus map engineering from a control-theoretical viewpoint see also the
literature \citep{lloyd-pra-65-010101,wu-jpa-40-5681,bolognani-IEEE-55-2721,verstraete-nphys-5-633} and the discussion on \textit{open} -- vs. \textit{closed-loop} simulation scenarios at the end of Sect.~\ref{sec:stabilizer_pumping}. To be concrete, here we focus on dissipative preparation of stabilizer states, which represent a large family of entangled
states, including graph states and error-correcting
codes~\citep{steane-prl-77-793,calderbank-pra-54-1098}.
Similar ideas for dissipative preparation of correlated quantum phases are discussed in Sect.~\ref{sec:Analog_QS_DissDyn} in the context of \textit{analog} many-body quantum simulation in cold bosonic and fermionic atomic systems.

\subsubsection{Bell State Pumping}
\label{sec:Bell_state_pumping}
\begin{figure}
\begin{center}
\includegraphics[width=0.7\columnwidth]{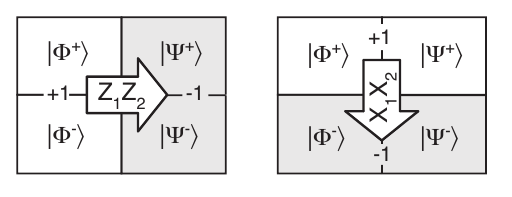}
\includegraphics[width=\columnwidth]{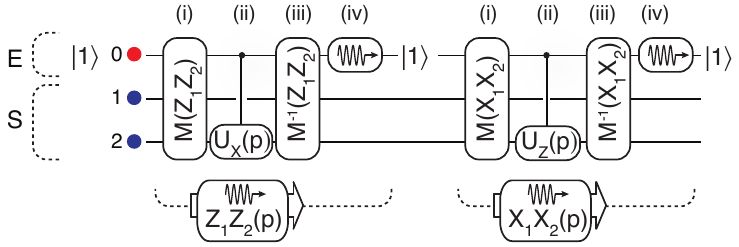}
\end{center}
\caption{(Color online) Bell state pumping $\rho \mapsto|\Psi^-\rangle\langle\Psi^-|$. Upper part: Pumping dynamics in Hilbert space, realized by two dissipative maps, under which two system qubits are irreversibly transferred from the +1 into
the -1 eigenspaces of $Z_1Z_2$ and $X_1X_2$. Lower part:  Schematics of the circuit decomposition of the two dissipative maps into unitary operations (i) - (iii), acting on the two system qubits S and an ancilla qubit playing the role of an environment E, followed by a dissipative reset (iv) of the ancilla. See main text for details.  Figure adapted from~\citet{barreiro-nature-470-486}
}%
\label{fig:BSC_circuit}%
\end{figure}

Before discussing the dissipative preparation of many-body phases such as ground state cooling of Kitaev's toric code Hamiltonian, we start by outlining the concept of dissipative Kraus map engineering for the simplest
non-trivial example of ``cooling'' a system of two qubits into a Bell
state. The Hilbert space of two qubits is spanned by the four Bell states
defined as $\ket{\Phi^{\pm}} = \frac{1}{\sqrt{2}} (\ket{00} \pm \ket{11})$ and
$\ket{\Psi^{\pm}} = \frac{1}{\sqrt{2}} (\ket{01} \pm \ket{10})$. Here,
$\ket{0}$ and $\ket{1}$ denote the computational basis of each qubit, and we
use the short-hand notation $\ket{00} = \ket{0}_1 \ket{0}_2$, for example.
These maximally entangled states are stabilizer states: the Bell state
$\ket{\Phi^+}$, for instance, is said to be \textit{stabilized} by the two
stabilizer operators $Z_1Z_2$ and $X_1X_2$, where $X$ and $Z$ denote the usual
Pauli matrices, as it is the only two-qubit state being an eigenstate of
eigenvalue +1 of these two commuting observables,
i.e.~$Z_1Z_2\ket{\Phi^+}=\ket{\Phi^+}$ and $X_1X_2\ket{\Phi^+}=\ket{\Phi^+}$.
In fact, each of the four Bell states is uniquely determined as an eigenstate
with eigenvalues $\pm1$ with respect to $Z_1Z_2$ and $X_1X_2$. The key idea of
cooling is that we can achieve dissipative dynamics which pump the system into
a particular Bell state, for example
$\rho \mapsto|\Psi^-\rangle\langle\Psi^-|$, by constructing two dissipative
maps, under which the two qubits are irreversibly transfered from the +1 into
the -1 eigenspaces of $Z_1Z_2$ and $X_1X_2$, as sketched in the upper part of Fig.~\ref{fig:BSC_circuit}.
The dissipative maps are engineered with the aid of an ancilla ``environment"
qubit~\citep{lloyd-pra-65-010101,duer-pra-78-052325} and a quantum circuit of
coherent and dissipative operations.

\textit{Kraus maps for Bell state pumping} -- For $Z_1Z_2$, the dissipative map which induces pumping into the -1 eigenspace is given by $\rho \mapsto \mathcal{E}(\rho) = E_1 \rho E^\dagger_1 + E_2 \rho E^\dagger_2$ with
\begin{align}
\label{eq:KrausOperators}
E_{1} & = \sqrt{p} \, X_2 \frac{1}{2} \left(1 + Z_1Z_2 \right), \nonumber \\
E_{2} & = \frac{1}{2} \left( 1 - Z_1Z_2 \right) +  \sqrt{1-p} \,\frac{1}{2} \left( 1 + Z_1Z_2 \right).
\end{align}
The map's action as a uni-directional pumping process can
be seen as follows: since the operation element $E_1$ contains the
projector $\frac{1}{2} ( 1 + Z_1Z_2)$ onto the +1 eigenspace of
$Z_1Z_2$, the spin flip $X_2$ can then convert +1 into -1 eigenstates
of $Z_1Z_2$, e.g.,~$\ket{\Phi^+} \mapsto \ket{\Psi^+}$. In contrast,
the -1 eigenspace of $Z_1Z_2$ is left invariant.
The cooling dynamics are
determined by the probability of pumping from the +1 into the -1 stabilizer
eigenspaces, which can be directly controlled by varying the parameters in the
employed gate operations (see below). For pumping with unit probability ($p=1$), the two
qubits reach the target Bell state --- regardless of their initial state ---
after only one cooling cycle, i.e.,~by a single application of each of the two
maps. In contrast, in the limit $p\ll1$, the repeated application of this map generates dynamics according to a master
equation (\ref{eq:Lindblad_part}) with Lindblad quantum jump operator $c=\frac{1}{2}X_2(1-Z_1Z_2)$.

The map is implemented by a quantum circuit of three
unitary operations (i)-(iii) and a dissipative step (iv), acting on two system qubits S and an ancilla which plays the role of
the environment E (see lower part of Fig.~\ref{fig:BSC_circuit}):
 (i)~Information about whether the system is in the +1 or -1 eigenspace
of $Z_1Z_2$ is mapped by $M(Z_1Z_2)$ onto the logical states $\ket{0}$ and
$\ket{1}$ of the ancilla (initially in $\ket{1}$): (ii)~A controlled gate $C(p)$ converts +1 into -1 eigenstates by
flipping the state of the second qubit with probability $p$, where
\begin{equation}
C(p) = \ket{0}\bra{0}_\mathrm{0} \otimes U_{X_2}(p) +
\ket{1}\bra{1}_\mathrm{0} \otimes {\bf 1}.\nonumber
\end{equation}
Here, $U_{X_2}(p) = \exp (i \alpha X_2)$ and $p = \sin^2 \alpha$ controls the pumping probability. (iii)~The initial mapping is inverted by $M^{-1}(Z_1Z_2)$. At this
stage, in general, the ancilla and system qubits are entangled. (iv) The ancilla is dissipatively reset to $\ket{1}$, which allows to carry away entropy to ``cool'' the two system qubits. The second map for pumping into the -1 eigenspace of $X_1X_2$ is obtained from interchanging the roles of $X$ and $Z$ above.

\begin{figure}
\begin{center}
\includegraphics[width=\columnwidth]{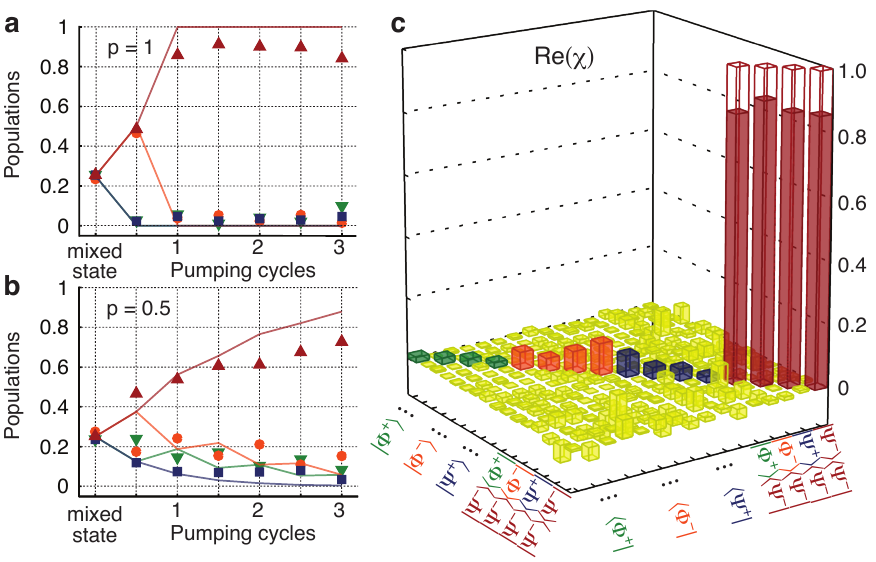}
\end{center}
\caption{(Color online) Experimental Bell state pumping. Evolution of
  the Bell-state populations $|\Phi^+\rangle$ (down triangles),
  $|\Phi^-\rangle$ (circles), $|\Psi^+\rangle$ (squares) and $|\Psi^-\rangle$
  (up triangles). a) Pumping process of an initially mixed state with probability $p=1$ into the target Bell state $|\Psi^-\rangle$.
  Regardless of experimental imperfections, the target state population is preserved under the repeated application of further cooling
cycles and reaches up to 91(1)\% after 1.5 cycles (ideally 100\%). b) In a second experiment towards the simulation of master-equation dynamics, the probability is set at $p=0.5$ to probe probabilistic cooling dynamics. In this case the target state is
approached asymptotically. After cooling the system for 3 cycles with $p=0.5$, up to 73(1)\% of the initially mixed population
cools into the target state (ideally 88\%). Error bars, not shown, are smaller than 2\% ($1\sigma$). c) In order to completely
characterize the Bell-state cooling process, a quantum process
tomography was performed~\citep{nielsen-book}. As an example, the figure shows the reconstructed process matrix $\chi$ (real part) for deterministic pumping with $p=1$, displayed in the Bell-state basis, describing the deterministic pumping of the two ions after 1.5 cycles. The reconstructed process matrix has a Jamiolkowski process fidelity~\citep{gilchrist-pra-71-062310} of 0.870(7) with the ideal
dissipative process $\rho \mapsto\ket{\Psi^-}\bra{\Psi^-}$. The ideal process mapping any input state into the state
  $|\Psi^-\rangle$ has as non-zero elements only the four transparent bars
  shown. Figure adapted from ~\citet{barreiro-nature-470-486}.
}%
\label{fig:Bellstatepumping_ions}%
\end{figure}

\textit{Experimental Bell state pumping} -- The described dynamics of ``Bell state pumping'' has been explored experimentally with three ions encoding the two system qubits and the ancilla qubit \citep{barreiro-nature-470-486} (see Fig.~\ref{fig:ion_qs_techniques}d). The unitary steps (i)-(iii) have been decomposed into the available set of coherent gate operations as shown in Fig.~\ref{fig:ion_qs_techniques}b and c. The dissipative reset of the ancilla qubit (iv) to its initial state $\ket{1}$ is realized by an addressed optical pumping technique, which leaves the quantum state of the system qubits unaffected \citep{science-schindler-332-1059}. The experimental results of various cycles of deterministic ($p=1$) and probabilistic ($p=0.5$) Bell state pumping are shown and discussed in Fig.~\ref{fig:Bellstatepumping_ions}.

\subsubsection{Stabilizer Pumping and Ground State Cooling of the Toric Code Hamiltonian}
\label{sec:stabilizer_pumping}

\begin{figure}[t]
\begin{center}
\includegraphics[width=0.9\columnwidth]{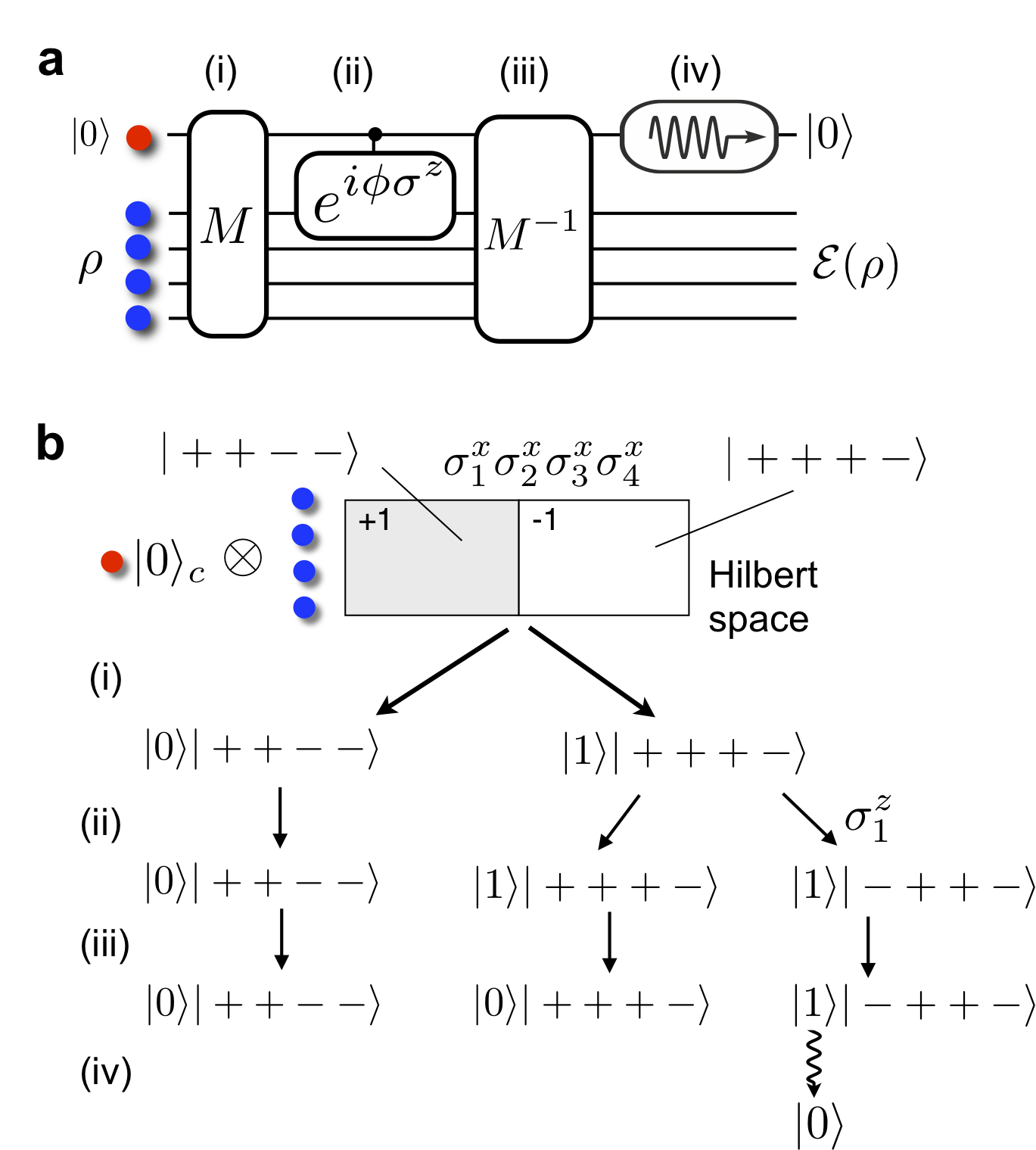}
\end{center}
\caption{(Color online) In a) Quantum circuit for the simulation of dissipative $n$-body interactions. In b) In a first step (i) the information whether the system spins reside in a +1 or -1 eigenstate of $A_{p}$ is coherently mapped onto the logical states of the auxiliary qubit - in direct analogy to the simulation of coherent $n$-body interaction discussed above. (ii) Subsequently a two-qubit gate $U_{\rm Z,i}(\phi) = \left|0\right\rangle \!\left\langle 0\right|_{c}\otimes{\bf 1}+\left|1\right\rangle \!\left\langle 1\right|_{c}\otimes\Sigma$ with $\Sigma = \exp(i \phi \sigma_i^z)$ is applied. 
The ``low-energy" $+1$ eigenstates of $A_p$ are not affected by $U_{\rm Z,i}$ as they have been mapped onto $\ket{0}_c$ in step (i). In contrast -- with probability $p=\sin^2\phi$ -- the two-qubit gate induces a spin flip on the $i$-th system spin, if the system spins are in ``high-energy" $-1$ eigenstates of $A_p$. (iii) The mapping $M$ is reversed and (iv) finally, the auxiliary control qubit is incoherently reinitialized in state $|0\rangle_{c}$ by optical pumping. Controlling the angle $\phi$ in the quantum circuit allows one to realize either probabilistic cooling ($\phi \ll \pi/2$) described by a master equation with four-spin jump operators $c_p$ as given in Eq.~(\ref{eq:Kitaev_jump_operators}) or deterministic cooling ($\phi = \pi/2$) as described by a discrete Kraus map of Eq.~(\ref{eq:Kraus_map}).
}%
\label{fig:diss_circuits}%
\end{figure}

The engineering of dissipative maps can be readily generalized to
systems of more qubits. In particular, in the Rydberg simulator architecture \citep{weimer-nphys-6-382} a combination of coherent multi-atom Rydberg gates $U_g$ (Eq.~(\ref{eq:CNOT})) with optical pumping of ancillary control atoms allows one to implement collective dissipative many-particle dynamics in an \textit{a priori} scalable system. As an example, we outline the engineering of dissipative dynamics for ground state cooling of Kitaev's toric code according to the plaquette and vertex four-body quantum jump operators given in Eq.~(\ref{eq:Kitaev_jump_operators}). In direct analogy to the quantum circuit for Bell state pumping discussed in the previous section, four-qubit stabilizer pumping for a single plaquette is realized by a sequence of three unitary steps (shown in Fig.~\ref{fig:diss_circuits}a), which are applied to the four system spins and the ancilla atom located at the center of the corresponding plaquette, followed by (iv) a dissipative reset of the ancilla qubit to its initial state.

To this purpose, as for the simulation of coherent many-body dynamics (i) one first applies the mapping $M$ (as specified in detail in the caption of Fig.~\ref{fig:circuits}) to coherently encode the information, whether the four system spins are in a +1 or -1 eigenstate of the stabilizer $A_p$ in the logical states of the auxiliary qubit, as schematically shown in Fig.~\ref{fig:diss_circuits}b. (ii) Subsequently, a controlled spin flip onto one of the four system spins is applied, which converts a -1 (``high-energy") into a +1 (``low-energy") eigenstate of $A_p$, with a certain, tunable probability determined by a phase $\phi$ (see Fig.~\ref{fig:diss_circuits}a). (iii) After reversing the mapping $M$, the auxiliary qubit remains in the state $|1\rangle_{c}$, if one of the system spins has been flipped in the previous step (ii). Thus, (iv) finally addressed optical pumping resets the auxiliary ion from $|1\rangle_{c}$ to its initial state $|0\rangle_{c}$, thereby guaranteeing that the auxiliary qubit factors out from the system dynamics and is ``refreshed" for subsequent simulation steps.

For small phases $\phi$ (and thus small probabilities for pumping from the -1 into +1 subspace of $A_p$ in one step) and under a repeated application of this dissipative map, the density matrix $\rho$ of the spin system evolves according to the Lindblad master Eq. (\ref{eq:Lindblad_part}) with the jump operators $c_{p}$ given in Eq.~(\ref{eq:Kitaev_jump_operators}) and the cooling rate $\kappa=\phi^{2}/\tau$. Note, that the cooling also works for large phases $\phi$; in fact, the most efficient dissipative state preparation
is achieved with $\phi=\pi/2$, i.e.~for deterministic pumping where an anyonic excitation, if it is present on the plaquette under consideration, is moved to a neighboring plaquette with unit probability. If this dynamics is applied to all plaquettes of the lattice, it leads, as discussed above and illustrated in Fig.~\ref{fig:toriccodecooling}, to a dissipative random walk and pairwise annihilation of anyonic quasi-particle excitations, and thus in the long-time limit to a cooling of the system into the ground-state manifold.

\begin{figure}
\begin{center}
\includegraphics[width=14 cm]{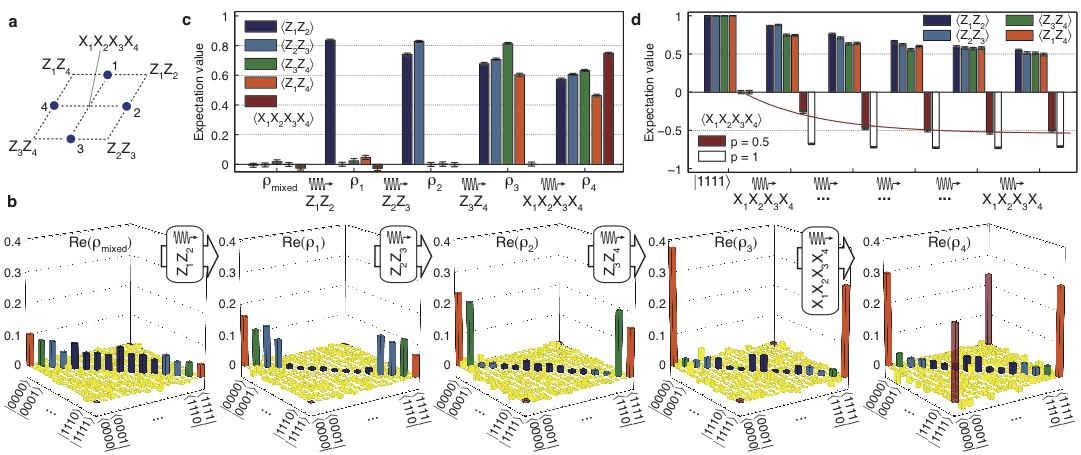}
\end{center}
\caption{(Color online) Experimental four-qubit stabilizer pumping, which can be regarded as dissipative ground state preparation of one plaquette of Kitaev's toric code \citep{kitaev-annalsphys-303-2}. a) Schematic of the four system qubits to be
  cooled into the GHZ state $(\ket{0000} + \ket{1111})/\sqrt{2}$, which is uniquely characterized as the simultaneous eigenstate with eigenvalue +1 of the shown stabilizers. b) Reconstructed density matrices (real part) of the
  initial mixed state $\rho_\text{mixed}$ and subsequent states
  $\rho_{1,2,3,4}$ after sequentially pumping the stabilizers $Z_1Z_2$,
  $Z_2Z_3$, $Z_3Z_4$ and $X_1X_2X_3X_4$.  Populations in the initial mixed
  state with qubits $i$ and $j$ antiparallel, or in the -1 eigenspace of the
  $Z_iZ_j$ stabilizer, disappear after pumping this stabilizer into the +1
  eigenspace.  For example, populations in dark blue disappear after
  $Z_1Z_2$-stabilizer pumping. A final pumping of the stabilizer
  $X_1X_2X_3X_4$ builds up the coherence between $\ket{0000}$ and $\ket{1111}$,
  shown as red bars in the density matrix of $\rho_4$. The reconstructed density matrices for the initial and subsequent states arising
in each step have a fidelity, or state overlap~\citep{josza-jmo-41-2315}, with
the expected states of \{79(2),89(1),79.7(7),70.0(7),55.8(4)\}\%. c) Measured
  expectation values of the relevant stabilizers; ideally, non-zero expectation
  values have a value of +1. d) Evolution of the measured expectation
  values of the relevant stabilizers for repetitively pumping an initial state
  $|1111\rangle$ with probability $p=0.5$ into the -1 eigenspace of the
  stabilizer $X_1X_2X_3X_4$.  The incremental cooling is evident by the red
  line fitted to the pumped stabilizer expectation value. The evolution of the
  expectation value $\langle X_1X_2X_3X_4\rangle$ for deterministic cooling
  ($p=1$) is also shown.  The observed decay of $\langle Z_iZ_j\rangle$ is due
  to imperfections and detrimental to the pumping process. Error bars in (c) and (d), $\pm1\sigma$. Figure reprinted with
permission from~\citet{barreiro-nature-470-486}. Copyright 2011 by MacMillan.
}%
\label{fig:plaquettepumping_ions}%
\end{figure}%

\textit{Many-body stabilizer pumping with trapped ions} -- Whereas for the described Rydberg simulator setup, all required components are not yet available in a single laboratory, \citet{barreiro-nature-470-486} demonstrated the described four-qubit stabilizer pumping in a proof-of-principle experiment with 5 trapped ions. Specifically, pumping dynamics into a four-qubit Greenberger-Horne-Zeilinger (GHZ) state
$(\ket{0000} + \ket{1111})/\sqrt{2}$ was realized. This state can be regarded as the ground state of a minimal instance of Kitaev's toric code, consisting of a single square plaquette, as sketched in Fig.~\ref{fig:plaquettepumping_ions}a. The state is uniquely characterized as
the simultaneous eigenstate of the four stabilizers $Z_1Z_2$, $Z_2Z_3$,
$Z_3Z_4$ and $X_1X_2X_3X_4$, all with eigenvalue +1. Therefore, cooling dynamics into the GHZ
state are realized by four consecutive dissipative steps, each pumping the
system into the +1 eigenspaces of the four stabilizers (Fig.~\ref{fig:plaquettepumping_ions}b-d).  In a system of 4+1
ions encoding the four system spins and an ancillary qubit, such cooling dynamics has been realized in analogy with the Bell-state pumping discussed in Sect.~\ref{sec:Bell_state_pumping}. Here, however, the experimental complexity is considerably larger, as the circuit decomposition of one cooling cycle involves 16 five-ion entangling M\o lmer-S\o rensen gates, 20 collective and 34
single-qubit rotations; further details in~\citet{barreiro-nature-470-486}.

\textit{Open- vs. closed-loop control scenarios} -- In the discussed examples of engineering of dissipative dynamics for Bell-state and four-qubit stabilizer pumping  the available quantum resources were used by coupling the system qubits to an ancilla qubit by a universal set of gates. Such set was constituted by entangling multi-ion MS gates in combination with single-ion rotations \citep{barreiro-nature-470-486}, or the the mesoscopic Rydberg gate \citep{mueller-prl-102-170502} in combination with single-atom gates in the Rydberg simulator architecture \citep{weimer-nphys-6-382}. The engineered environment was here represented by ancilla ions or Rydberg atoms, undergoing optical pumping by dissipative coupling to the vacuum modes of the radiation field. Note that in the described scenario, the ancilla qubit remains unobserved, representing an \textit{open-loop} dynamics. For such \textit{open} quantum systems, though, it was noted in~\citet{bacon-pra-64-062302,lloyd-pra-65-010101} that using a single ancilla qubit the most general Markovian open-system dynamics cannot be obtained with a finite set of non-unitary open-loop transformations. However, such a universal
dynamical control can be achieved through repeated application of coherent
control operations and measurement of the auxiliary qubit, followed by
classical feedback operations onto the system. In the trapped-ion experiments in~\citet{barreiro-nature-470-486} the simulation toolbox was complemented by the demonstration of a quantum-non-demolition (QND) measurement of a four-qubit stabilizer operator via an auxiliary qubit. In combination with classical feedback \citep{riebe-nphys-4-839}, such QND readout operations provide the basis for such \textit{closed-loop} dynamics.

Furthermore, in the context of quantum error correction, QND measurements of stabilizer operators constitute a crucial ingredient for the realization of quantum error-correcting codes \citep{steane-prl-77-793,calderbank-pra-54-1098}. Such readout operations correspond to error syndrome measurements, and the obtained information can be classically processed and used to detect and correct errors \citep{dennis-j-mat-phys-43-4452}. For instance, in~\citet{mueller-njp-13-085007} it is explicitly worked out how minimal instances of complete topogical quantum error correcting codes \citep{bombin-prl-97-180501} can be realized with the currently available toolbox for open-system quantum simulation with trapped ions \citep{barreiro-nature-470-486}.

\subsubsection{Digital Simulation of a U(1) Lattice Gauge Theory}
\label{sec:digital_simulation_toolbox_and_LGT}

The above analysis for the coherent simulation and ground state cooling of Kitaev's toric code can be extended to a large class of interesting models. In~\citet{weimer-quantuminfprocess-10-885} it is discussed how the digital Rydberg simulator architecture enables the simulation of Heisenberg-like spin models, and in principle also fermionic Hubbard models, by mapping lattice fermions to a spin Hamiltonian involving many-body interactions that can be realized in the Rydberg simulator.

\textit{Three-dimensional U(1) lattice gauge theory} -- The toric code is the ground state of the frustration-free, exactly solvable toric code Hamiltonian involving four-qubit plaquette and vertex interactions \citep{kitaev-annalsphys-303-2}. It belongs to the class of stabilizer states and exhibits Abelian topological order. It is also possible to provide (digital) simulation protocols for the simulation of coherent many-body dynamics and ground state preparation of more complex spin models. In~\citet{weimer-nphys-6-382} such a protocol was developed for the example of a three-dimensional $U(1)$-lattice gauge theory \citep{kogut-rmp-51-659} and it was shown how to achieve dissipative ground state preparation also for such a complex system. Such models have attracted interest in the search for `exotic' phases  and spin liquids \citep{moessner-prl-86-1881,motrunich-prl-89277004,hermele-prb-69-064404,levin-rmp-77-871,levin-prb-71-045110}.

\begin{figure}
\begin{center}
\includegraphics[width=10 cm]{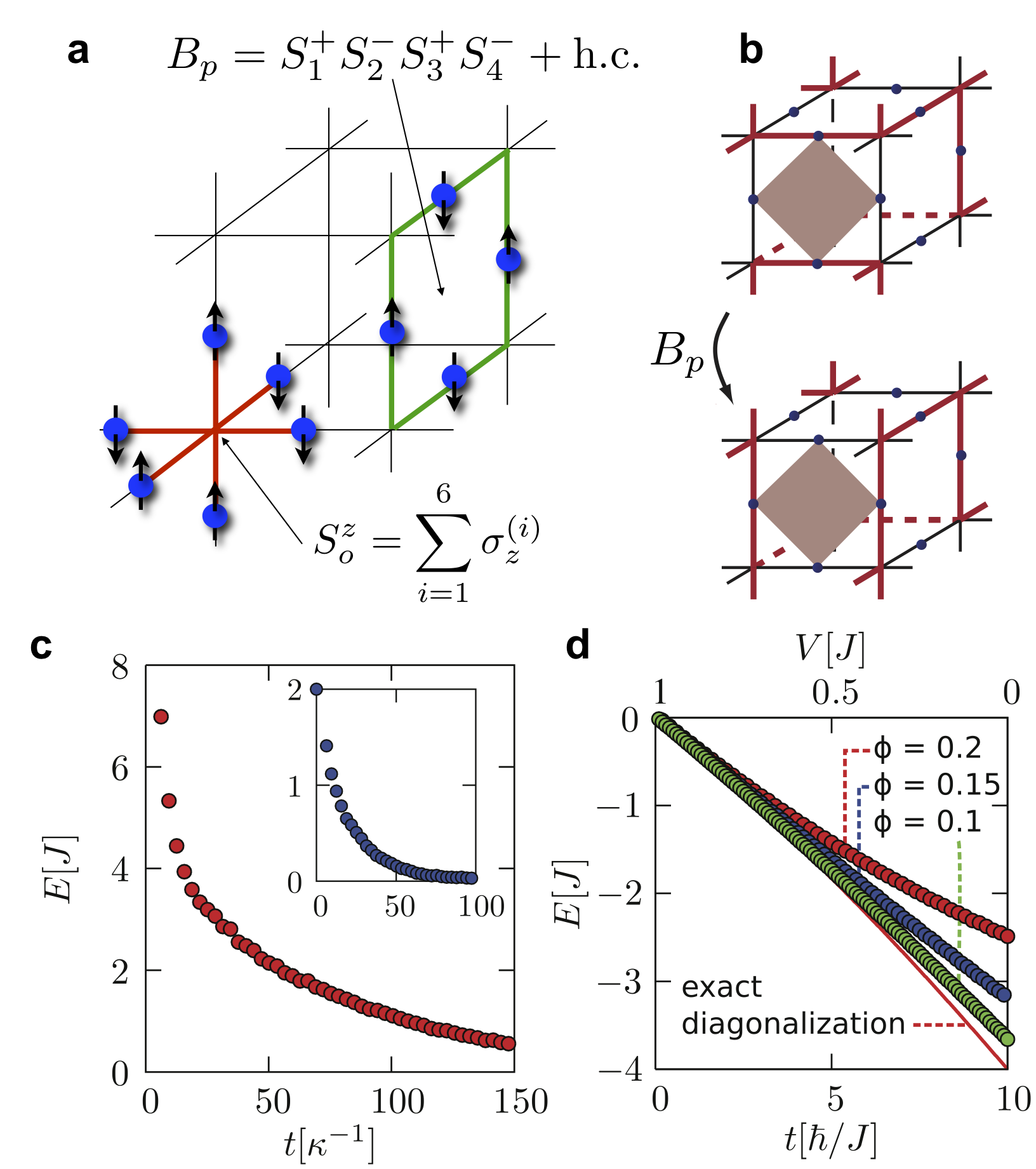}
\end{center}
\caption{(Color online) Simulation of a three-dimensional U(1) lattice gauge theory. a) Spins located on the edges of a cubic lattice interact via a six-spin low-energy constraint term $S^{z}_{o}$ (indicated by red links), which imposes the condition of an equal number of three up- and three down-spins on each octahedron, and via a four-spin ring-exchange plaquette interaction $B_p$ (green links) -- see Eq.~(\ref{eq:U1LGT}) and details in the main text. b) In the language of dimer coverings, the ring exchange terms $B_p$ coherently convert flippable plaquettes from one configuration into another. c) Numerical simulation of the cooling into the ground state at the Rokhsar-Kivelson point $V=J$ for a system of 4 unit cells (12 spins): The cooling into the low-energy subspace defined by the octahedra constraints can be realized in analogy to the cooling of the toric code \citep{weimer-nphys-6-382}; alternatively one can directly start the protocol in an initial (classical) state, which satisfies all constraints. The inset shows the cooling into the equal-weight superposition of all dimer coverings starting from an initial state which already satisfies the $S_o^z$-constraints. d) Coherent time evolution from the Rokhsar-Kivelson point with a linear ramp of the Rokhsar-Kivelson term $V(t) = (1- tJ /10)$: the solid line denotes the exact ground state energy, while dots represent the digital time evolution during an adiabatic ramp for different phases $\phi$ during each time step \citep{weimer-nphys-6-382}. The differences arise from Trotter expansion errors due to non-commutative terms in the Hamiltonian (\ref{eq:U1LGT}). Parts (b) - (d) of the figure adapted from~\citet{weimer-nphys-6-382}]
}%
\label{fig:U1LGT}%
\end{figure}%

Fig.~\ref{fig:U1LGT}a shows the setup of the U(1) lattice gauge theory. Spins are located on the edges of a three-dimensional cubic lattice and interact via the many-body Hamiltonian
\begin{equation}
H =U \sum_{o}\left( S^{z}_{o}\right)^{2}   -  J\sum_{p}B_{p} + V \: N_{\rm RK}.
\label{eq:U1LGT}
\end{equation}
The first term with $S^{z}_{o} = \sum_{k \in o} \sigma_z^{(k)}$ describes pairwise two-body interactions of six spins located at the corners of octahedra, located around the vertices of the square lattice (see the spins connected by red lines in Fig.~\ref{fig:U1LGT}a). The inequality $U \gg |V| , |J|$ defines a low-energy sector of the theory, which consists of spin configurations with an equal number of three up and three down spins,  i.e., states with vanishing total spin $S^{z}_{o}$ on each octahedron. The second term describes a ring-exchange interaction $B_{p}=S_{1}^{+}S_{2}^{-}S_{3}^{+}S_{4}^{-}+S_{1}^{-}S_{2}^{+}S_{3}^{-}S_{4}^{+}$ of four spins located around each plaquette of the lattice (see green plaquette in Fig.~\ref{fig:U1LGT}a); here $S_{i}^{\pm}=(\sigma_{i}^{x}\pm i\sigma_{i}^{y})/2$. This interaction flips the state of four plaquette spins with alternating spin orientation, e.g., $B_p |0101\rangle_p = |1010\rangle_p$, and leaves other states unchanged, e.g., $B_p |1001\rangle_p = 0$. Note that while the ring-exchange interaction term commutes with the $S^z_o$ spin constraint terms, ring-exchange terms on neighboring plaquettes do not commute.

The last term of the Hamiltonian of Eq.~(\ref{eq:U1LGT}) counts the total number of flippable plaquettes $N_{\rm RK} = \sum_{p} B_{p}^2$. It is introduced since at the so-called Rokhsar-Kivelson point with $J=V$, the system becomes exactly solvable \citep{rokshar-prl-61-2376}. If one identifies each spin up with a ``dimer'' on a link of the lattice, all states satisfying the low-energy constraint of vanishing $S^{z}_{o}$ on all octahedra can be viewed as an ``allowed'' dimer covering with three dimers meeting at each site of the cubic lattice. Fig.~\ref{fig:U1LGT}b shows how the $B_p$ ring exchange interaction term flips one dimer covering into another. Within this dimer description, the ground state at the Rokhsar-Kivelson point is given by the condensation of the dimer coverings \citep{levin-prb-71-045110}, i.e., the equal weight superposition of all allowed dimer coverings. It has been suggested that in the non-solvable parameter regime $0\leq V\leq J$ of interest the ground state of the system is determined by a spin liquid smoothly connected to the Rokhsar-Kivelson point \citep{hermele-prb-69-064404}.

\textit{Simulation protocol} -- To reach the $0\leq V\leq J$ phase of interest, the idea is to (i) implement dissipative dynamics, which first cools the system at the Rokshar-Kivelson point ($J=V$) into the ground state given by the symmetric superposition of dimer coverings, and (ii) subsequently to slowly decrease the strength of the Rokshar-Kivelson term $V \: N_{\rm RK}$ in the Hamiltonian (\ref{eq:U1LGT}) such that the ground state is adiabatically transformed into the quantum phase of interest:

(i) If one starts in some initial state, which satisfies the $S^z_o$ on all octahedra, the condensation of the dimer coverings can be achieved by dissipative dynamics according to plaquette jump operators
\begin{equation}
\label{eq:c_p_jump_ops}
  c_{p} = \frac{1}{2}\sigma_{i}^{z}\left[1-B_{p}\right]B_{p}.
\end{equation}
The jump operator $c_p$ has by construction two dark states, which are the $0$ and $+1$ eigenstates of $B_p$. The $0$ eigenstates
correspond to a non-flippable plaquette (e.g.~$c_p |1001\rangle_p = 0$), while the $+1$ eigenstate is the equal-weight superposition of the
original dimer covering and the dimer covering obtained by flipping the plaquette, $|1010\rangle_p + |0101\rangle_p$.
Finally, the jump operator $c_p$ transforms the third eigenstate with eigenvalue $-1$ into the $+1$ eigenstate. As a consequence, as Fig.~\ref{fig:U1LGT}c illustrates, under this dynamics acting on all plaquettes of the cubic lattice, for long times the system asymptotically approaches the ground state consisting of the symmetric superposition of all allowed dimer coverings.

(ii) Subsequently, this ground state is transformed adiabatically into the phase at $0\leq V\leq J$ by slowly ramping down the Rokshar-Kivelson term. Such adiabatic passage can be realized by decomposing the coherent dynamics according to the Hamiltonian with the time-dependent Rokshar-Kivelson term $V (t) \: N_{\rm RK}$ into small Trotter time steps (conceptually similar to the simulation of two-spin time-dependent Trotter dynamics discussed in Sect.~\ref{sec:coh_sim_ions}). The different curves in Fig.~\ref{fig:U1LGT}d indicate deviations of the simulated adiabatic passage from the exact dynamics due to Trotter errors originating from the non-commutativity of terms in the Hamiltonian (\ref{eq:U1LGT}).

The Hamiltonian terms (\ref{eq:U1LGT}) and quantum jump operators (\ref{eq:c_p_jump_ops}) for the simulation of the U(1) lattice gauge theory are more complex than the ones for ground state cooling and Hamiltonian dynamics according to the toric code Hamiltonian. However, in the Rydberg simulator architecture they can also be implemented by combinations of many-atom Rydberg gates and optical pumping of ancilla qubits, which are located on the plaquettes and corners of the qubit lattice; see~\citet{weimer-nphys-6-382} for details and explicit circuit decompositions.

\subsection{The Effect of Gate Imperfections on Digital Quantum Simulation}
\label{sec:complementary_developments}
Imperfect gate operations in the quantum circuits which are used to implement coherent and dissipative steps of time evolution according to discrete Kraus maps (\ref{eq:Kraus_map}) or many-body master equations (\ref{eq:master_equation}) lead to deviations of the actually realized system from the envisioned dynamics. In the simulation of many-body dynamics for a given time $t$ via a Trotter decomposition this leads in practice to a trade-off: On the one hand, the number of simulation steps $n$ according to small time intervals $t/n$ should be chosen large, in order to keep the effect of Trotterization errors originating from non-commuting terms small. On the other hand, the practical implementation of each time step has a certain cost in terms of resources and is associated to a certain experimental error, which favors the decomposition of the simulated time dynamics into a not too large number of steps.

Small imperfections typically provide in leading order small perturbations for the simulated Hamiltonian dynamics and weak additional dissipative terms. The specific form is strongly dependent on the particular implementation platform and its dominant error sources; see the analysis in~\citet{duer-pra-78-052325} for a general discussion. For the Rydberg quantum simulator architecture \citep{weimer-nphys-6-382} the influence of errors in the multi-atom Rydberg gate \citep{mueller-prl-102-170502} on the simulation of Kitaev's toric code Hamiltonian and ground state cooling in this model has been analyzed: Fig.~\ref{fig:toriccodecooling}b shows that in the presence of small gate imperfections the desired cooling into the ground state of the model is accompanied by weak, unwanted heating processes, such that in the long-time limit a finite anyon density remains present in the many-body system. Such effects have also been observed experimentally in the dissipative state preparation of a minimal system of one plaquette of the toric code with trapped ions \citep{barreiro-nature-470-486}, as discussed in Sect.~\ref{sec:stabilizer_pumping}: Fig.~\ref{fig:plaquettepumping_ions}d shows that under repeated pumping into the -1 eigenspace of the four-qubit-stabilizer $X_1X_2X_3X_4$, the expectation value of the two-qubit stabilizers $Z_iZ_j$, which should ideally be unaffected by the $X_1X_2X_3X_4$-pumping and should remain at their initial value of +1, undergo a decay. This detrimental effect can be interpreted as ``heating processes'' due to experimental imperfections in the underlying quantum circuits; see also~\citet{mueller-njp-13-085007} where a theoretical modeling of these errors is discussed. However, the thermodynamic properties (quantum phases) and dynamical behaviour of a strongly interacting many-body system are in general robust to small perturbations in the Hamiltonian; e.g., the stability of the toric code for small magnetic fields has recently been demonstrated \citep{vidal-prb-79-033109}. Consequently, small imperfections in the implementation of the gate operations leading to deviations from the ideal simulated dynamics are expected to be tolerable.


\section{Engineered Open Systems with Cold Atoms}
\label{sec:Analog_QS_DissDyn}




As anticipated in Sect.~\ref{sec:Introduction}, here we will be interested in a scenario where many-body ensembles of cold atoms are properly viewed as \emph{open} quantum systems, in a setting familiar from quantum optics: A system of interest is coupled to an environment, giving rise to dissipative processes, and is additionally driven by external coherent fields. This creates a non-equilibrium many-body setting without immediate counterpart in condensed matter systems. In particular, in the first part of this section, we point out how the conspiracy of laser drive and dissipation can give rise to off-diagonal long-range order, a trademark of macroscopic quantum phenomena. We also argue how this can be achieved via proper reservoir engineering, in this way fully extending the notion of quantum state engineering from the Hamiltonian to the more general Liouvillian setting, where controlled dissipation is included.

In the following parts of this section, we will give accounts for further central aspects of this general setting. In the context of atomic bosons, we point out in which sense these systems indeed constitute a novel class of artificial out-of-equilibrium many-body systems, by analyzing a stationary state phase diagram resulting from competing unitary and dissipative dynamics. In the context of atomic fermions, we present a dissipative pairing mechanism which builds on a conspiracy of Pauli blocking and dissipative phase locking, based on which we argue that such systems may provide an attractive route towards quantum simulation of important condensed matter models, such as the Fermi-Hubbard model. We then explore the possibilities of dissipatively realizing topological phases in the lab, and elaborate on the specific many-body properties of such dissipatively stabilized states of matter.

The results presented here highlight the fact that the \emph{stationary states} of such driven-dissipative ensembles, representing flux equilibrium states far from thermodynamic equilibrium, feature interesting many-body aspects. This places these systems in strong contrast to the dynamical non-equilibrium phenomena which are currently actively investigated in closed systems in the cold atom context, focusing on thermalization \citep{PhysRevA.72.063604,PhysRevLett.100.030602,Rigol2008,Wenger06,Hoffer07,Trotzky11} and quench dynamics \citep{PhysRevLett.96.136801,PhysRevLett.98.180601,GreinerM02,Sadler06}.

\subsection{Long-Range Order via Dissipation }
\label{sec:Concepts}

\subsubsection{Driven-Dissipative BEC}
\label{sec:DissBEC}

\begin{figure}[tbp]
\begin{center}
\includegraphics[width=.9\columnwidth]{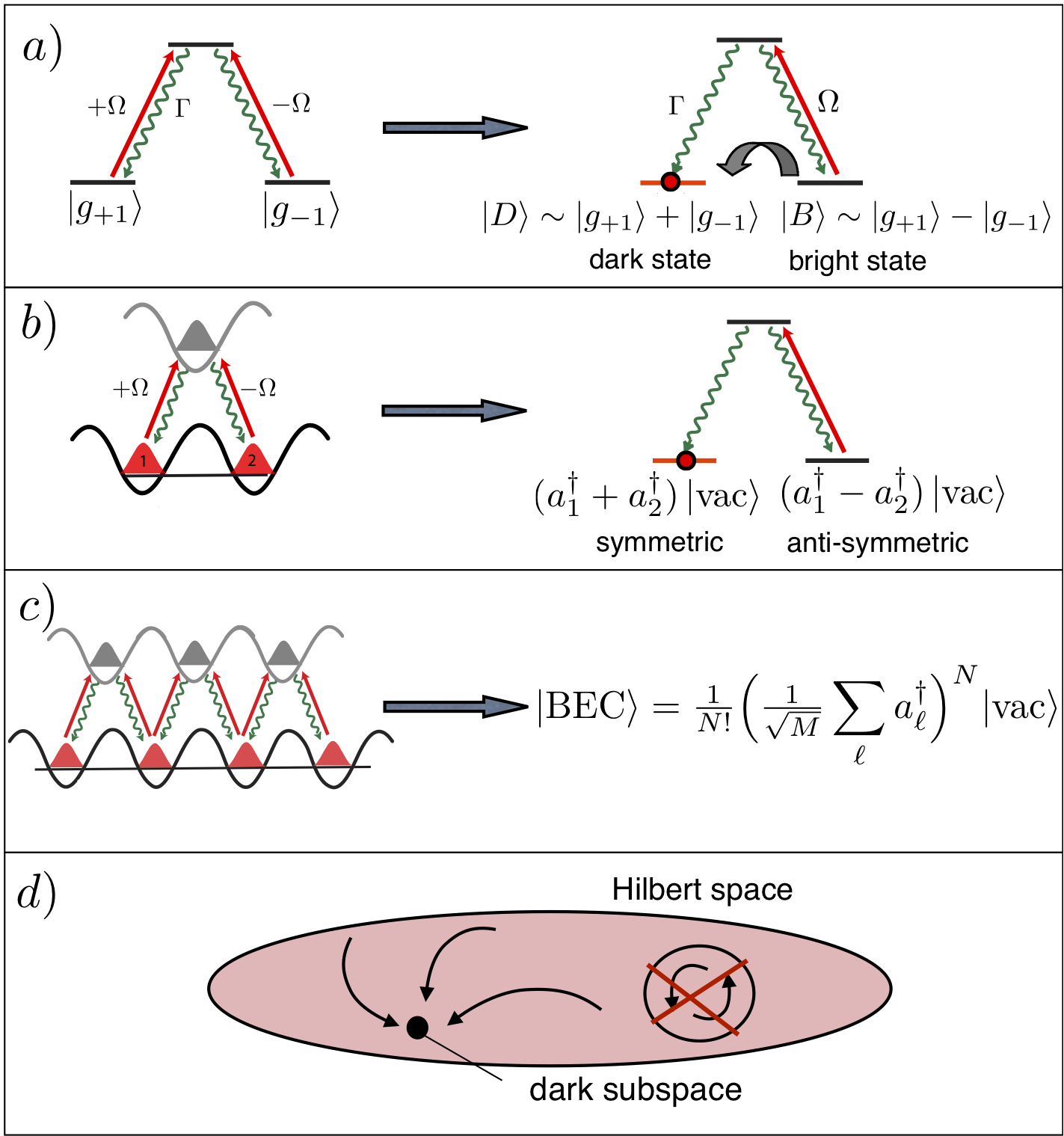}
\end{center}
\caption{(Color online) Dark states in many-body systems from an analogy with optical pumping: a) A coherently driven and spontaneously decaying atomic $\Lambda$-system with metastable excited state has the symmetric superposition of the degenerate ground states as a dark state for antisymmetric driving. b) The internal degrees of freedom are replaced by external degrees of freedom, such as the sites of an optical superlattice, with the same consequences once antisymmetric driving and spontaneous emission are properly engineered. c) The unit cell is naturally cloned in a translation invariant lattice setting. The symmetric phase locking on each pair of sites generates coherence over the whole system, corresponding to a fixed number BEC. d) Uniqueness: If the dark subspace consists of one dark state only, and no subspace exists which is left invariant under the set of jump operators, the many-body density matrix converges to the dark state irrespective of its initial condition.}
\label{DarkState}
\end{figure}

\emph{Qualitative picture: Dark states in single- and many-particle systems} -- For long times, a system density matrix governed by Eq. \eqref{eq:master_equation} will approach a flux equilibrium stationary state, $\rho(t)\rightarrow \rho_{ss}$, in the presence of dissipation, which generically is a mixed state. However, under suitable circumstances the stationary state can be a pure state, $\rho_{ss}=\left|D\right>\left<D\right|$. In the language of quantum optics, such states $\left|D\right>$ are called \emph{dark states}. A familiar example on the level of single particles is optical pumping or dark state laser cooling to subrecoil temperatures \citep{aspect88,kasevich92}, illustrated in Fig. \ref{DarkState}a: By coherently coupling two degenerate levels to an auxiliary excited state with antisymmetric Rabi frequencies $\pm\Omega$, from which spontaneous emission leading back to the ground states occurs symmetrically, a dark state is given by the symmetric superposition of the ground states. For sufficient detuning, it is then clear that the population will entirely end up in this dark state decoupled from the light field. In our setting, we replace the \emph{internal} degrees of freedom of an atom by \emph{external}, motional degrees of freedom, realized schematically by an optical potential configuration with an intermediate site on the link between degenerate ground states, cf. Fig. \ref{DarkState}b. Below we will discuss how to realize the relevant driving and decay processes. Clearly, the same arguments then lead to a phase locked, symmetric superposition dark state as above, i.e. $(a_1^\dag + a_2^\dag)|\text{vac}\rangle$ in a second quantization notation. However, two generalizations follow immediately: First, the levels (lattice sites) can be populated with bosonic degrees of freedom, i.e. there is no limit on the occupation number. Second, and most natural in an optical lattice context, the ``dark state unit cell'' can be cloned in a translation invariant way to give a complete lattice setting, in one or higher dimensions.  The key ingredient is antisymmetric drive of each pair of sites, and the spontaneous decay back to the lower states, as depicted in Fig. \ref{DarkState}c. The phase is then locked on each two adjacent sites, such that eventually only the symmetric superposition over the whole lattice persists. This is the only state not being recycled into the dissipative evolution. This state is nothing but a Bose-Einstein condensate (BEC) with a fixed but arbitrary particle number $N$, which for a one-dimensional geometry with $M$ sites depicted in Fig. \ref{DarkState}c reads
\begin{eqnarray}\label{BECState}
\vert \mathrm{BEC}\rangle_N = \tfrac{1}{\sqrt{N!}} \big(\tfrac{1}{\sqrt{M}}\sum_i a_i^\dag \big)^N\vert \mathrm{vac}\rangle =  \tfrac{1}{\sqrt{N!}}a_{\mathbf{q}=0}^{\dag \,N} \vert \mathrm{vac}\rangle .
\end{eqnarray}
In consequence, quantum mechanical long range order is built up from \emph{quasilocal, number conserving} dissipative operations. The system density matrix is purified, in that a zero entropy state is reached from an arbitrary initial density matrix, as will be discussed next.

\emph{Driven-dissipative BEC as unique stationary state} --  Here we make the above intuitive picture more precise by discussing the Lindblad jump operators which drive into the BEC state Eq. \eqref{BECState}. In a slight generalization, we consider the dynamics of $N$ bosonic atoms on a  $d$-dimensional lattice with spacing $a$ and $M^{d}$ lattice sites, and lattice vectors $\mathbf{e}_{\lambda}$. For simplicity, we first address the purely dissipative case of Eq. \eqref{eq:master_equation} and set $H=0$. The goal is then achieved by choosing the jump operators Eq. \eqref{eq:Lindblad_part} as
\begin{equation}
c_{\beta}\equiv c_{ij}=\left(a_{i}^{\dagger}+a_{j}^{\dagger}\right)\left(a_{i}-a_{j}\right)\label{eq:jump},
\end{equation}
 acting between each pair of adjacent lattice sites $\beta\equiv\langle i,j\rangle$ with an overall dissipative rate $\kappa_\beta \equiv \kappa_{ij} = \kappa$. Because the annihilation part of the normal ordered operators $c_\beta$ commute with the generator of the BEC state $\sum_i a_i^\dag$, we have
 \begin{eqnarray} \label{DSCond}
 (a_{i}-a_{j})\left\vert \mathrm{BEC}\right\rangle =0\,\forall\,\langle i,j\rangle,
 \end{eqnarray}
making this state indeed a many-body dark state (or dissipative zero mode) of the Liouville operator defined with jump operators Eq. \eqref{eq:jump}.

From the explicit form of the jump operators, we see that the key for obtaining a dark state with long range order is a coupling to the bath which involves a current or discrete gradient operator between two adjacent lattice sites. The temporally local jump operator $c_{ij}$ describes a pumping process, where the annihilation part $a_{i}-a_{j}$ removes any anti-symmetric (out-phase) superposition on each pair of sites $\langle i,j\rangle$, while $a_{i}^{\dagger}+a_{j}^{\dagger}$ recycles the atoms into the symmetric (in-phase) state. As anticipated above, this process can thus be interpreted as a dissipative locking of the atomic phases of every two adjacent lattice sites, in turn resulting into a global phase locking characteristic of a condensate.

We also note from Eq. \eqref{DSCond} that the dark state property of $\left\vert \mathrm{BEC}\right\rangle $ is mainly determined by the annihilation part of the jump operator. In fact, any linear combination of $a_{i}^{\dagger}, a_{j}^{\dagger}$ of recycling operators will work, except for a hermitian $c_{ij}$, i.e. for the combination $a^\dag_{i}-a^\dag_{j}$. In this case, the dissipative dynamics would result in dephasing instead of pumping into the dark state. This case is then qualitatively similar to the generic situation in atomic physics. There, a bath typically couples to the atomic density with jump operators $n_{i}=a_{i}^{\dagger}a_{i}$, as in the case of decoherence due to spontaneous emission in an optical lattice \citep{PhysRevA.82.063605}, or for collisional interactions.

We now discuss the uniqueness of the stationary dark state. The following two requirements have to be fulfilled to ensure uniqueness (in the absence of Hamiltonian dynamics) \citep{baumgartner07,kraus-pra-78-042307}:\\
(i) The dark subspace is one-dimensional, i.e. there is exactly one normalized dark state $\left|D\right>$, for which
\begin{eqnarray}\label{eq:DaSt}
c_{\beta}\left|D\right>=0\,\,\, \forall\beta.
\end{eqnarray}
(ii) No stationary solutions other than the dark state exist.

In the above example, so far we have only argued that the BEC state is \emph{a} dark state. However, it is easily seen that no other dark states are present, since the non-hermitean creation and annihilation operators can only have eigenvalue zero on an $N$-particle Hilbert space. In particular, the creation part $a^\dag_i+ a^\dag_j$ never has a zero eigenvalue, as it acts on an $N-1$ particle Hilbert space. We can therefore focus on the annihilation part alone, where the Fourier transform $\sum_\lambda(1 - e^{\mathrm i \mathbf q \mathbf{e}_\lambda })a_\mathbf{q}$  reveals indeed exactly one zero mode at $\mathbf{q} =0$. As to (ii), uniqueness of the dark state as a stationary state is guaranteed if there is no other subspace of the system Hilbert space which is left invariant under the action of the operators $c_{\beta}$ \citep{baumgartner07,kraus-pra-78-042307}. This can be shown explicitly for the example above \citep{kraus-pra-78-042307}. More generally, it can be proved that for any given pure state there will be a master equation so that this state becomes the unique stationary state. Uniqueness is a key property: under this circumstance, the system will be attracted to the dark state for arbitrary initial density matrix, as illustrated in Fig. \ref{DarkState}d. These statements remain true for a Hamiltonian dynamics that is compatible with the Lindblad dynamics, in the sense of the dark state being an eigenstate of the Hamiltonian, $H\left|D\right>=E\left|D\right>$. One example is the addition of a purely kinetic Hamiltonian, since $H_{0}\left\vert \mathrm{BEC}\right\rangle =N\epsilon_{\mathbf{q}=0}\left\vert \mathrm{BEC}\right\rangle $, where $\epsilon_{\mathbf{q}}=2J\sum_{\lambda}\sin^{2}\mathbf{q}\mathbf{e}_{\lambda}/2$ is the single particle Bloch energy for quasimomentum $\mathbf{q}$.

Finally, we remark that as a consequence of the symmetry of global phase rotations exerted by $e^{\mathrm i \varphi \hat N}$ on the set of jump operators (i.e. $[c_\beta , \hat N]=0 $ $\forall \beta$, where $\hat N = \sum_i a_i^\dag a_i$ is the total particle number operator), which is present microscopically, any breaking of this symmetry must occur spontaneously. This gives room for concept of spontaneous symmetry breaking to be applicable in the thermodynamic limit for such driven-dissipative systems.

\subsubsection{Implementation with Cold Atoms}
\label{sec:Implementation}

Before sketching an explicit implementation scheme of the above dynamics, we point out that the existence of a microscopic scale, where a description of the system in terms of a temporally local evolution equation is possible, is far from obvious in a many-body context. In fact, in usual  condensed matter settings, typical baths have arbitrarily low energies which can be exchanged with a given many-body system of interest, giving rise to temporally non-local memory kernels in the description of environmental effects.
Instead, the validity of the master equation rests on the Born-Markov approximation with system-bath coupling in rotating wave approximation. This means that the bath is gapped in a condensed matter language. For typical quantum optics settings, these approximations are excellent  because the (optical) system frequencies providing for the gap are much larger than the decay rates. Below we argue how to mimic such a situation in an optical lattice context. At the same time, this setting makes clear the need for external driving in order to provide the energy necessary to access the decaying energy levels. The validity of this combination of approximations then fully extends the scope of microscopic control in cold atom systems from unitary to combined unitary-dissipative dynamics.

\begin{figure}[tbp]
\begin{center}
\hspace{-.2cm}\includegraphics[width=1.0\columnwidth]{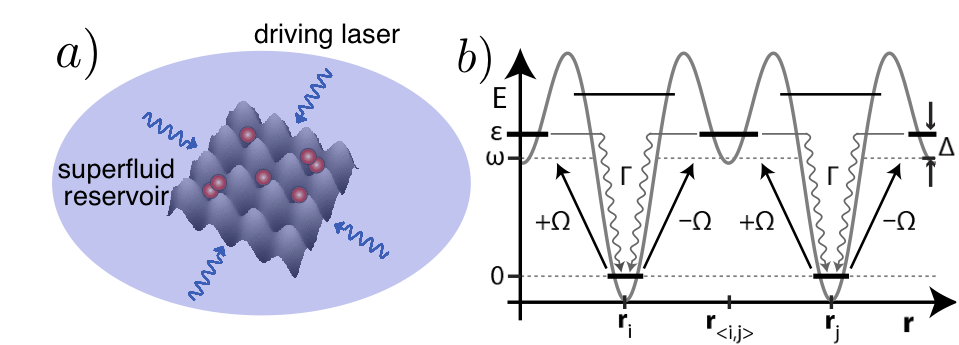}
\end{center}
\caption{(Color online) Cold atom implementation of a driven-dissipative condensate: a) A coherently driven lattice gas is immersed in a surrounding condensate. b) Schematic realization of the effective dissipative process in an optical superlattice, which provides for excited states gapped by $\varepsilon$ and localized on the links  of neighboring lattice sites $\langle ij\rangle$: A Raman laser couples the  ground- and excited bands with effective Rabi-frequency $\Omega$ and detuning $\Delta=\omega-\varepsilon$ from the inter-band transition. Only the antisymmetric component of atoms on neighboring lattice sites is excited to the upper band due to the spatial modulation of the Raman-laser. The inter-band decay with a rate $\Gamma$ back to the lower band is obtained via the emission of Bogoliubov quasiparticle excitations into the surrounding BEC. Figure adapted from \cite{diehl-natphys-4-878}.}
\label{Implementation}
\end{figure}

A concrete possible implementation in systems of cold bosonic atoms $a$ builds on the immersion of a coherently driven optical lattice system into a large BEC of atoms $b$ \citep{Griessner06}, cf. Fig. \ref{Implementation}a. In order to realize the key $\Lambda$-configuration, we consider a superlattice setting as illustrated in one-dimensional geometry in Fig. \ref{DarkState}a, with an additional auxiliary lattice site on each of the links. The optical lattice corresponding to a single link is shown in Fig. \ref{Implementation}b, where the $\Lambda$-system is implemented with the two Wannier functions of lattice sites 1 and 2 representing two ground states, and the auxiliary state in the middle representing an excited state. In order to achieve the annihilation part of the jump operator, we drive this three-level system by Raman transitions from the two ground to the excited states with Rabi frequencies $\Omega$ and $-\Omega$, respectively. This could be realized in a translation invariant way for the whole lattice by, e.g., a commensurate ratio of lattice and Raman laser wavelengths, $\lambda_{\text{Raman}}= 2\lambda_\text{latt} $, which would guarantee the relative sign via a $\pi$-phase shift for the Rabi frequency. In the next step, the dissipation needs to be introduced. To this end, the coherently driven system is placed into a large BEC reservoir. This condensate interacts in the form of a conventional s-wave contact potential with interspecies scattering length  $a_{ab}$ with the lattice atoms $a$, and acts as a bath of Bogoliubov excitations. Such a coupling provides an efficient mechanism for decay of atoms $a$ from the excited to the lower Bloch band by emission of Bogoliubov quasiparticles. This replaces photon emission in a conventional quantum optics situation. The conspiracy of coherent drive and dissipation explained here also gives rise to the physical picture of the coherence of the driving laser beam being imprinted onto the matter system -- any deviation from the above commensurability condition would be reflected in a length scale in the driven-dissipative BEC. We note however, that the ratio of wavelengths can be controlled with high precision in state-of-the-art experiments.

In the presence of a large condensate, linearization of the system-bath interaction around the bath condensate expectation value, together with the harmonic bath of Bogoliubov excitations, realizes the generic system-bath setting of quantum optics. In particular, a key element is the presence of the largest energy scale provided by the Hubbard band separation $\epsilon$ (cf. Fig. \ref{Implementation}b), ensuring the validity of Born-Markov and rotating wave approximations. This in turn leads to a temporally local master equation description. As long as this scale exceeds the bath temperature $ \epsilon\gg T_\text{BEC}$, the occupation of modes at these energies is negligible and the BEC thus acts as an \emph{effective zero temperature reservoir}. At the same time, the role of coherent driving with energy $\omega$ in order to bridge the energy separation of the two bands becomes apparent. The fact that energy is constantly pumped into the system in our driven-dissipative non-equilibrium setting highlights the fact that our setting can indeed realize states of zero entropy, or in practice an entropy substantially lower than the surrounding reservoir gas, without conflicting with the second law of thermodynamics.

If we further specialize to the limit of  weak driving $\Omega \ll \Delta$, where $\Delta = \omega -\epsilon$ is a detuning from the upper Hubbard band, adiabatic elimination of the excited Bloch band results in a master equation generated by jump operators of the type \eqref{eq:jump}. In this case, on the full lattice, the laser excitation to the upper band $\sim (a_{i}-a_{j})$ for each pair of sites is followed by immediate return of the atoms into the lowest band, which generically happens in a symmetric fashion such as $\sim(a^\dag_{i}+a^\dag_{j})$, in this way realizing jump operators of the form of Eq. \eqref{eq:jump}. Details of the return process, however, depend on the Bogoliubov excitation wavelength in the bath: For  wavelength $\lambda_{\mathrm b}$ larger or smaller than the optical lattice spacing $a$, spontaneous emission is either correlated or uncorrelated. However complicated, the existence of a dark state in the present case is guaranteed by $(a_{i}-a_{j})\left\vert \mathrm{BEC}\right\rangle =0$, a property which follows from the laser excitation step alone. We will therefore concentrate below on the jump operators defined in Eq. \eqref{eq:jump}.

Finally, we emphasize that the basic concept for the dissipative generation of long-range order in many-particle systems can be explored in very different physical platforms beyond the cold atom context, offering additional opportunities for  implementations. For example, microcavity arrays have been identified as promising candidates for  the realization of the above dynamics with state-of-the-art technology \citep{Marcos12}, where the bosonic degrees of freedom are realized by microwave cavity photons. The auxiliary system is there realized by two interacting superconducting qubits, which are placed between two neighboring microwave resonators. The symmetric and antisymmetric superposition modes of the resonators are coupled to the qubit system and the dissipative step is realized naturally via spontaneous decay of the latter.

In an even broader context, also different kinds of intrinsically quantum mechanical correlations, such as entanglement, can be targeted dissipatively. Examples have been discussed in trapped ion systems above. In addition, in a recent breakthrough experiment entanglement has been generated dissipatively between two macroscopic spin ensembles \citep{PhysRevLett.107.080503,PhysRevA.83.052312}. On the theory side, creation of atomic entanglement has been proposed in the context of optical cavities \citep{PhysRevLett.106.090502}, and the generation of squeezed states of matter has been investigated for the case of macroscopic two-mode boson ensembles \citep{Watanabe11}. Furthermore, dissipation has been proposed as a means to purify many-body Fock states as defect-free registers for quantum computing with cold atoms~\citep{Pupillo2004,Brennen2005}, as well as to enforce three-body constraints in Hamiltonian dynamics~\citep{Daley2009,Kantian2009,PhysRevLett.104.165301,PhysRevLett.104.096803}. Recent landmark experiments have used it to build strong correlations in, and thus to stabilize, a metastable weakly interacting molecular gas in one-dimension~\citep{Syassen-2008,Porto-2008}.

So far we have discussed the proof-of-principle for the concept of state engineering in many-particle systems by tailored dissipation in the conceptually simplest example, the driven-dissipative BEC. In the following subsections, we will review different research directions which address many-body aspects in such systems, where dissipation acts as a dominant resource of dynamics:
In the context of bosonic systems, we present a dynamical phase transition resulting from the competition of the engineered Liouville- with a Hamiltonian dynamics, defining a novel class of interacting non-equilibrium many-body systems with interesting stationary states. The phase transition is seen to share features of both quantum and classical phase transitions, and we identify an intriguing phase where global phase rotation and translation symmetry are simultaneously broken spontaneously.
In the context of atomic fermions, we discuss a dissipative pairing mechanism, which is operative in the absence of attractive forces and allows us to target states of arbitrary symmetry, such as d-wave paired states in two dimensions. Beyond the identification of this new far-from-equilibrium pairing mechanism, this makes dissipative state engineering potentially relevant for the experimental efforts towards the quantum simulation of the two-dimensional Fermi-Hubbard model, where the ground state is believed to have pairing with d-wave symmetry away from half filling.
Finally, we show how dissipation engineering can be used in order to reach fermionic states with topological order dissipatively. While so far topological phases have been exclusively discussed in a Hamiltonian context, we develop here a dissipative counterpart for such phases. We discuss the associated phenomena resulting when such systems are suitably constrained in space, such as the emergence of unpaired Majorana edge modes.

\subsection{Competition of Unitary and Dissipative Dynamics in Bosonic Systems}
\label{sec:Competition}

\emph{Motivation} -- In a Hamiltonian ground state context, a quantum phase transition results from the competition of two non-commuting parts of a microscopic Hamiltonian $H=H_{1}+gH_{2}$, if the ground states for $g\ll1$ and $g\gg1$ have different symmetries  \citep{SachdevBook}. A critical value $g_{c}$ then separates two distinct quantum phases described by pure states, while in thermodynamic equilibrium for finite temperature this defines a quantum critical region around $g_{c}$ in a $T$ vs.~$g$ phase diagram. Classical phase transitions may occur for fixed parameter $g$ by increasing the temperature, and can be viewed as resulting from the competition of the specific ground state stabilized by the Hamiltonian vs. the completely mixed structureless infinite temperature state.
In contrast, here we study a non-equilibrium situation, in which there is a competition between a Hamiltonian and a dissipative dynamics. We extract the complete steady state phase diagram, revealing that the resulting transitions share features of quantum phase transitions, in that they are interaction driven, and classical ones, in that the ordered phase directly terminates into a strongly mixed state.
It contains an extended region where global phase rotation and translation symmetry are both broken spontaneously, as a consequence of a subtle renormalization effect on the complex excitation spectrum of the low-lying modes. In addition, we study the dynamical critical behavior in the long-time limit of the combined unitary and dissipative evolution.

Those aspects underpin the fact that the driven-dissipative systems investigated here add a new class of non-equilibrium stationary states to those which have been studied so far. One prominent example is certainly electron systems in condensed matter, which are exposed to a bias voltage \citep{KL2009}. In this context, also characteristic many-body behavior such as the effect of non-equilibrium conditions on quantum critical points has been investigated \citep{MTKM2006}. Further routes of driving many-body systems out of thermodynamic equilibrium are discussed in the context of exciton-polariton Bose-Einstein condensates \citep{SnokeBook,Kasp2006}, or more recently in driven noisy systems of trapped ions or dipolar atomic gases \citep{TDGA2010,DallaTorre11}.

\subsubsection{Dynamical Phase Transition}
\label{sec:DynamicalPhaseTransition}

\emph{Model and Analogy to Equilibrium Quantum Phase Transition} -- We now extend the purely dissipative  dynamics leading to a BEC state determined by Eq. \eqref{eq:jump} by the generic Hamiltonian in optical lattice systems, the Bose-Hubbard Hamiltonian:
\begin{eqnarray}
\partial_{t}\rho&=&  -i[H,\rho]+{\cal L}[\rho],\\\nonumber
H&=&-J\sum_{\langle \ell,\ell'\rangle}b_{\ell}^{\dagger}b_{\ell'} - \mu\sum_{\ell}\hat{n}_{\ell} +\frac{1}{2}U\sum_{\ell}\hat{n}_{\ell}(\hat{n}_{\ell}-1)~.
\end{eqnarray}
This Hamiltonian is defined with the parameters $J$, the hopping amplitude, and $U$, the onsite interaction strength; $\hat{n}_{\ell}=b_{\ell}^{\dagger}b_{\ell}$ is the number operator for site $\ell$. Its ground state physics provides a seminal example for a quantum phase transition in the cold atom context \citep{PhysRevB.40.546,PhysRevLett.81.3108,GreinerBloch02,bloch-rmp-2008}: For a given chemical potential $\mu$, which in equilibrium fixes the mean particle density $n$, the critical coupling strength $g_c = (U/Jz)_{c}$ separates a superfluid regime $Jz \gg U$ from a Mott insulator regime $Jz \ll U$ ($z$ is the lattice coordination number).

As indicated above, here in contrast we are interested in the competition of Hamiltonian vs. dissipative dynamics. As indicated above, the hopping $J$ is a compatible energy scale, in the sense that a purely kinetic Hamiltonian has the dissipatively targeted $|\mathrm{BEC}\rangle$ as an eigenstate. On the other hand, the onsite interaction $U$ counteracts the off-diagonal order and thus leads to a competition with dissipation of strength $\kappa$. This provides a nonequilibrium analog to the generic purely Hamiltonian equilibrium scenario, in which $g = U/\kappa z$ plays the role of a competition parameter -- a dominant dissipation $g\ll1$ supports a condensed steady state, whereas dominant interaction $g\gg1$ results in a diagonal density matrix.

A yet different kind of dynamical phase transitions, which result from the competition between different terms of the dissipative Liouvillian, have been anticipated in \cite{verstraete-nphys-5-633}, and discussed in more detail in \cite{Eisert10} and \cite{Hoening11}, where in particular the key aspect of criticality in terms of diverging length and time scales has been established.
Furthermore, our scenario is in a sense dual to the dissipative quantum phase transition of a single particle on a lattice coupled to a long wavelength heat bath, known to undergo a transition from diffusive to localized behavior upon \emph{increasing} dissipation strength \citep{PhysRevLett.51.1506,PhysRevLett.56.2303,PhysRevB.36.3651,PhysRevB.35.7256}.

\emph{Theoretical approach} -- The absence of standard concepts for thermodynamic equilibrium, such as the existence of a free energy and associated variational principles, makes it necessary to argue directly on the level of the equation of motion (EOM) for the density operator, resp. on the associated full set of correlation functions. This is in general a formidable task, even numerically intractable in the thermodynamic limit in which we are here interested.
For this reason, we have developed a generalized Gutzwiller mean field approximation scheme, which captures the physics in the two well-understood limiting cases $g\ll1,g\gg1$, and otherwise provides an interpolation scheme. It is implemented by a product ansatz $\rho = \bigotimes_{\ell}\rho_{\ell}$ for the full density matrix, such that the reduced local density operators $\rho_{\ell} = \mathrm{Tr}_{\ne \ell}\,\rho$ are obtained by tracing out all but the $\ell$th site. Compared to the standard bosonic Gutzwiller procedure for the Bose-Hubbard model at zero temperature, where the factorization is implemented for the wave function, it allows for the description of mixed state density matrices. It treats the onsite physics exactly, and drops the (connected) spatial correlations, such that it can be expected to be valid in sufficiently high dimensions. The equation of motion for the reduced density operator reads
\begin{equation}\label{eq:redmasterequation}
\partial_{t}\rho_{\ell}  = -i [h_{\ell},\rho_{\ell}] +{\cal L}_{\ell}[\rho_{\ell}]~,
\end{equation}
where the local mean field Hamiltonian and Liouvillian are given by
\begin{eqnarray}
h_{\ell} &=& - J \sum_{\langle \ell' | \ell \rangle} (\langle b_{\ell'} \rangle b_{\ell}^{\dag} + \langle b_{\ell'}^{\dag}\rangle b_{\ell} ) -\mu \hat{n}_{\ell} +\frac{1}{2} U \hat{n}_{\ell}(\hat{n}_{\ell} - 1),\nonumber\\\nonumber
{\cal L}_{\ell}[\rho_{\ell}] &=& \kappa \sum_{\langle \ell'| \ell \rangle} \sum_{r,s=1}^{4} \Gamma_{\ell'}^{rs}[2 A_{\ell}^{r} \rho_{\ell} A_{\ell}^{s\dag} - A_{\ell}^{s\dag} A_{\ell}^{r} \rho_{\ell} - \rho_{\ell} A_{\ell}^{s\dag} A_{\ell}^{r}].\\
\end{eqnarray}
$h_\ell$ is in accord with the form of the standard Gutzwiller approach. The addition of the chemical potential $\mu$ to the Hamiltonian $h_{\ell}$ does not change the dynamics, because the model conserves the average particle filling $n = \sum_{\ell} \langle \hat{n}_{\ell} \rangle / M^d$. The freedom to fix the chemical potential is necessary to solve the equation $\partial_{t} \rho_{ss} = 0$ for the steady state of the system \citep{Diehl10a,Tomadin11}. The Liouvillian is constructed with the operator valued vector ${\bf A}_\ell = (1, b_{\ell}^{\dag}, b_{\ell}, \hat{n}_{\ell})$, and the correlation matrix $\Gamma_{\ell'}^{r,s} = \sigma^{r} \sigma^{s} {\rm Tr}_{\ell} A_{\ell}^{(5-s)\dag} A_{\ell}^{(5-r)}$, with $\sigma = (-1,-1,1,1)$. Note that the correlation matrix is $\rho$-dependent -- this makes the mean field master equation effectively \emph{nonlinear} in $\rho$. Such a feature is well-known in mean field approximations, e.g. in the Gross-Pitaevski equation, where an $N$-body quantum-mechanical linear Schr\"odinger equation is approximated by a non-linear classical field equation.

The information encoded in Eq. \eqref{eq:redmasterequation} can equivalently be stored in the full set of correlation functions, resulting in an \emph{a priori} infinite hierarchy of nonlinear coupled equations of motion for the set spanned by the normal ordered expressions $\langle b_\ell^nb_\ell^m\rangle$ for $n,m\in \mathbf{N}$ and all lattice sites $\ell$. This formulation is advantageous in the low density limit $n\ll1$, where we have identified a power counting showing that a closed (nonlinear) subset of six correlation functions $(\psi_\ell=\langle b_\ell\rangle$, $\langle b_{\ell}^{2} \rangle$, $\langle b_{\ell}^{\dag} b^{2} \rangle, c.c.)$, decouples from the infinite hierarchy. For technical reasons, it is sometimes favorable to study the equivalent set of seven connected correlation functions, $(\psi_\ell,\langle \delta b_{\ell}^\dag\delta b_\ell \rangle $, $\langle \delta b_{\ell}^{2} \rangle$, $\langle \delta b_{\ell}^{\dag} \delta b^{2}_\ell \rangle, c.c.)$, where $\delta b_\ell = b_\ell -\psi_\ell$. This allows to obtain a number of results analytically in this limit, such as the condensate fraction as a function of interaction strength in the homogeneous limit, and the complete shape of the phase diagram.

\emph{Basic picture for the dynamical quantum phase transition} -- To better understand the phase transition, we consider the limiting cases of vanishing and dominant interaction. For $U=0$, the spontaneous breaking of the $U(1)$ phase symmetry is reflected by an exact steady state solution in terms of a homogeneous coherent state $\rho^{\rm (c)}_{\ell} = |\Psi\rangle_\ell\langle\Psi|$, with $|\Psi\rangle_{\ell} = \exp (- n/2) \sum_{m} [(n e^{i \theta})^{m/2}) / \sqrt{m!}]|m\rangle_{\ell}$ for any $\ell$, together with the choice $\mu = - Jz$.
The effect of  a finite interaction $U$ is best understood using a rotating frame transformation on Eq. \eqref{eq:redmasterequation}, $\hat{V}(U) = \exp[i U \hat{n}_{\ell}(\hat{n}_{\ell} -1)t]$.
While the interaction term is then removed from the Heisenberg commutator, the annihilation operators become $\hat{V} b_{\ell} \hat{V}^{-1} = \sum_{m} \exp(i m U t) |m\rangle_{\ell}\langle m| b_{\ell}$. $U$ therefore rotates the phase of each Fock states differently, thus dephasing the coherent state $\rho_{\ell}^{(c)}$. In consequence, off-diagonal order will be completely suppressed for sufficiently large $U$ and the density matrix takes a diagonal form. Under the assumption of diagonality, the master equation reduces to a rate equation
\begin{eqnarray}
\partial_{t} \rho_{\ell} &=& \kappa [(n+1) (2 b_{\ell} \rho_{\ell} b_{\ell}^\dag - \{b_{\ell}^\dag b_{\ell}, \rho_{\ell}\}) \\\nonumber
&& + n (2 b_{\ell}^\dag \rho_{\ell} b_{\ell} - \{b_{\ell} b_{\ell}^\dag , \rho_{\ell}\})].
\end{eqnarray}
This is the equation for bosons coupled to a thermal reservoir with thermal occupation $n$, with thermal state solution $\rho_{\ell;m,k}^{\rm (t)} = n^m / (n+1)^{m+1}\delta_{m,k}$, where $m,k$ are the Fock space indices of the $\ell$th site.
At this point two comments are in order. First, in contrast to the standard case of an external heat reservoir, the terms $n$, $n+1$ are \emph{intrinsic} quantities, meaning that the strongly interacting system provides its own effective heat bath. Second, from the solution we note the absence of any distinct commensurability effects for integer particle number densities, contrasting the Mott scenario at zero temperature. This can be traced back to the fact that in the latter case, the suppression of off-diagonal order is additionally constrained by the purity of the state, such that (at least on the mean field level) the diagonal pure Mott state is the only possible choice. The driven-dissipative system has no such constraint on the purity of the state.

\subsubsection{Critical Behavior in Time}
\label{sec:Critical}

\begin{figure}[tbp]
\begin{center}
\includegraphics[width=.9\columnwidth]{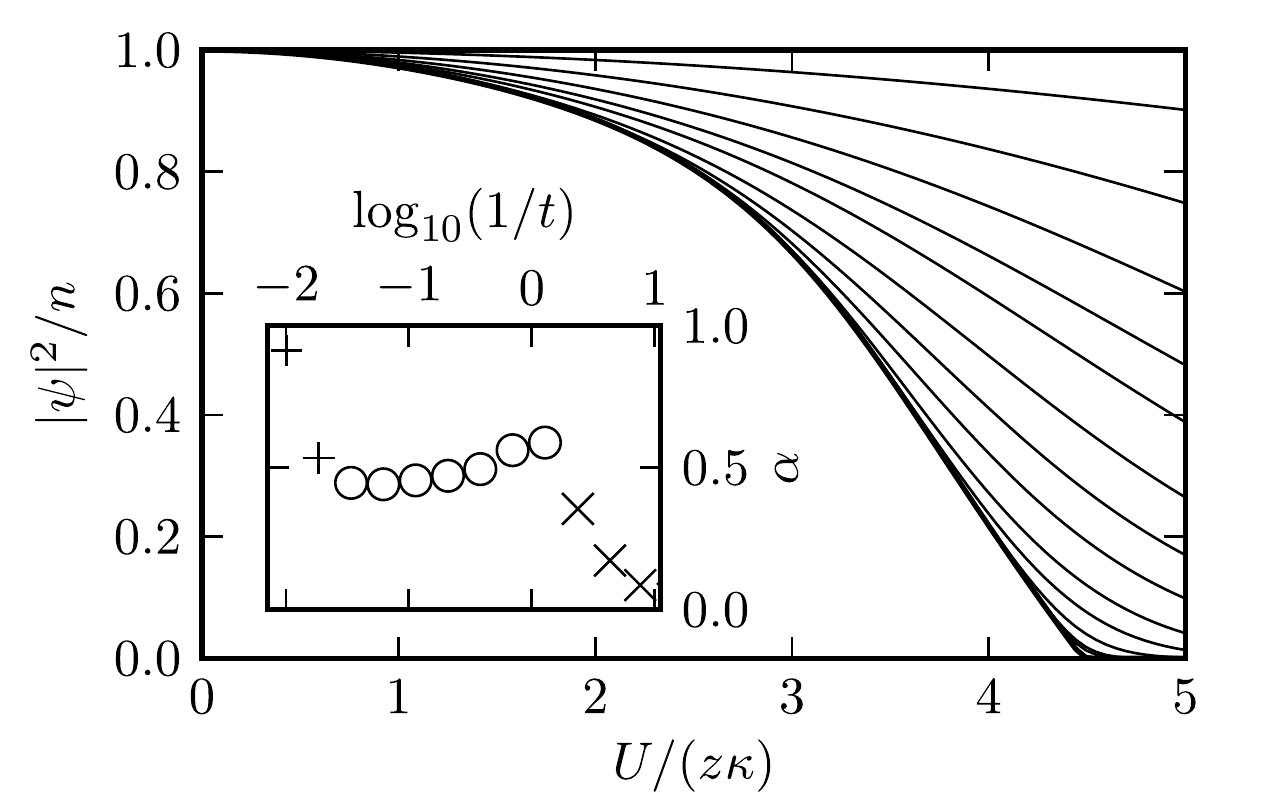}
\end{center}
\caption{Dynamical phase transition: Relaxation dynamics of the condensate fraction to the stationary state from an initial fully condensed state as a function of the interaction strength $U$, for $J = 1.5\,\kappa, n=1$. Each line corresponds to a stroboscopic snapshot.
Inset: Near critical evolution reflected by the time-evolution of the logarithmic derivative of the order parameter $\psi(t)$, for $J=0$, $n=1$, and $U\lesssim U_{\rm c}$.
The early exponential decay (tilted crosses) of the initial fully-condensed state is followed by a scaling regime (empty circles) with exponent $\alpha \simeq 0.5$. The final exponential runaway (vertical crosses) indicates a small deviation from the critical point. Figure reprinted with
permission from~\cite{Diehl10a}. Copyright 2010 by MacMillan.
}
\label{PhaseTransition}
\end{figure}

Fig.~\ref{PhaseTransition} shows stroboscopically the approach to the steady state in the homogeneous limit as a function of interaction strength. In particular, we note the expression of a non-analyticity as $t\to \infty$, characteristic of a second order phase transition. In the low density limit, the steady state condensate fraction can be obtained analytically and reads
\begin{equation}\label{eq:Depletion}
\frac{|\psi_0|^{2}}{n} = 1 - \frac{2 u^2 \left(1+(j + u)^2\right)}{1+u^2 + j( 8 u +6 j \left(1+2 u^2\right) +24 j^2 u + 8 j^3)}~,
\end{equation}
with dimensionless variables $u = U/(4\kappa z)$, $j = J/(4\kappa)$. The boundary between the thermal and the condensed phase with varying $J,n$ is shown in Fig.~\ref{PhaseDiagram} with solid lines.

On general grounds, one expects a critical slowing down at the phase transition point when approaching it in time at the critical interaction strength. More precisely, the order parameter evolution of the generic form $|\psi| \sim \exp (- m^2 t)/t^\alpha$ should have a vanishing mass or gap term $m^2$ (real part of the lowest eigenvalue), leading to a polynomial evolution. The associated scaling of the order parameter is reflected in the plateau regime in the inset of Fig.~\ref{PhaseTransition}, which sets in after an initial transient and is followed by an exponential runaway for a slight deviation from the exact critical point. In the low density limit, it is possible to extract the associated dynamical critical exponent: At criticality, the order parameter evolution is seen to be governed by a cubic dissipative nonlinearity $\sim |\psi|^3$, implying solutions $|\psi| \simeq 1 / (4 \sqrt{\kappa t})$ with exponent $\alpha =1/2$. This is a mean field result and not indicative of the precise universality class of the system, governed by anomalous critical exponents. This issue is currently under investigation in a Keldysh path integral approach. Nevertheless, already the above result highlights that in our dynamical system, criticality could be monitored directly as a function of time, e.g. by stroboscopically measuring the condensate fraction.

\subsubsection{Dynamical Instability and Spontaneous Translation Symmetry Breaking}
\label{sec:DynInstab}

\begin{figure}[tbp]
\begin{center}
\includegraphics[width=.9\columnwidth]{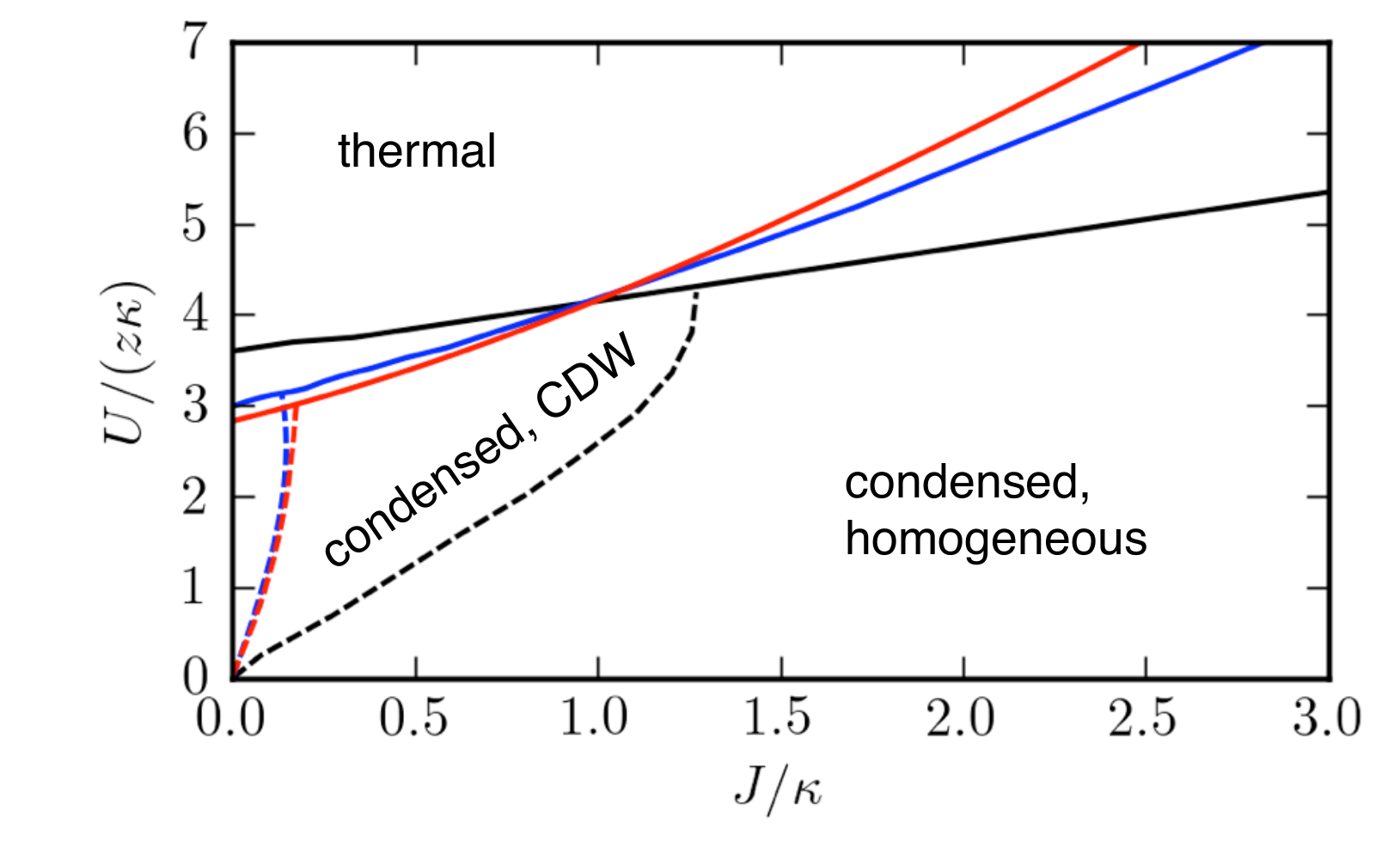}
\end{center}
\caption{(Color online) Stationary state phase diagram for different regimes of density: $n=1$ (black), $n=0.1$ (red: analytical low density limit calculation; blue: numerical low density calculation). The coincidence of analytical and numerical results is enhanced as $n\to0$. All regimes of density exhibit the same qualitative features with the three phases discussed in the text. Figure adapted from \cite{Diehl10a}.  }
\label{PhaseDiagram}
\end{figure}

An intriguing feature of the non-equilibrium stationary state phase diagram is an extended region in parameter space, where both the symmetries of phase rotations and translations are broken spontaneously, in this sense defining a supersolid phase. This state is characterized by a density modulation which is incommensurate with the lattice spacing. As illustrated in the phase diagram Fig.~\ref{PhaseDiagram}, the effect occurs universally in all density regimes. The plausibility for such a new qualitative effect can be understood from the fact that the (bare) dissipation rate $\kappa_\mathbf{q} \sim \mathbf{q}^2$ (see below), vanishes in the vicinity of the dark state at $\mathbf{q}=0$: In consequence, there will always be a momentum scale where even an arbitrarily weak interaction energy $Un$ becomes comparable. In the low density limit, it is possible to describe the phenomenon analytically, in this way getting insights into the origin of the additional phase with translation symmetry breaking. To this end, we work with the closed subset of seven correlation functions defined above, which however are time and space dependent. Working in a linear response strategy, we linearize around the homogeneous steady state solution to study its stability. Upon Fourier transform, we obtain a $7\times7$ matrix evolution equation.

\begin{figure}[tbp]
\begin{center}
\includegraphics[width=.9\columnwidth]{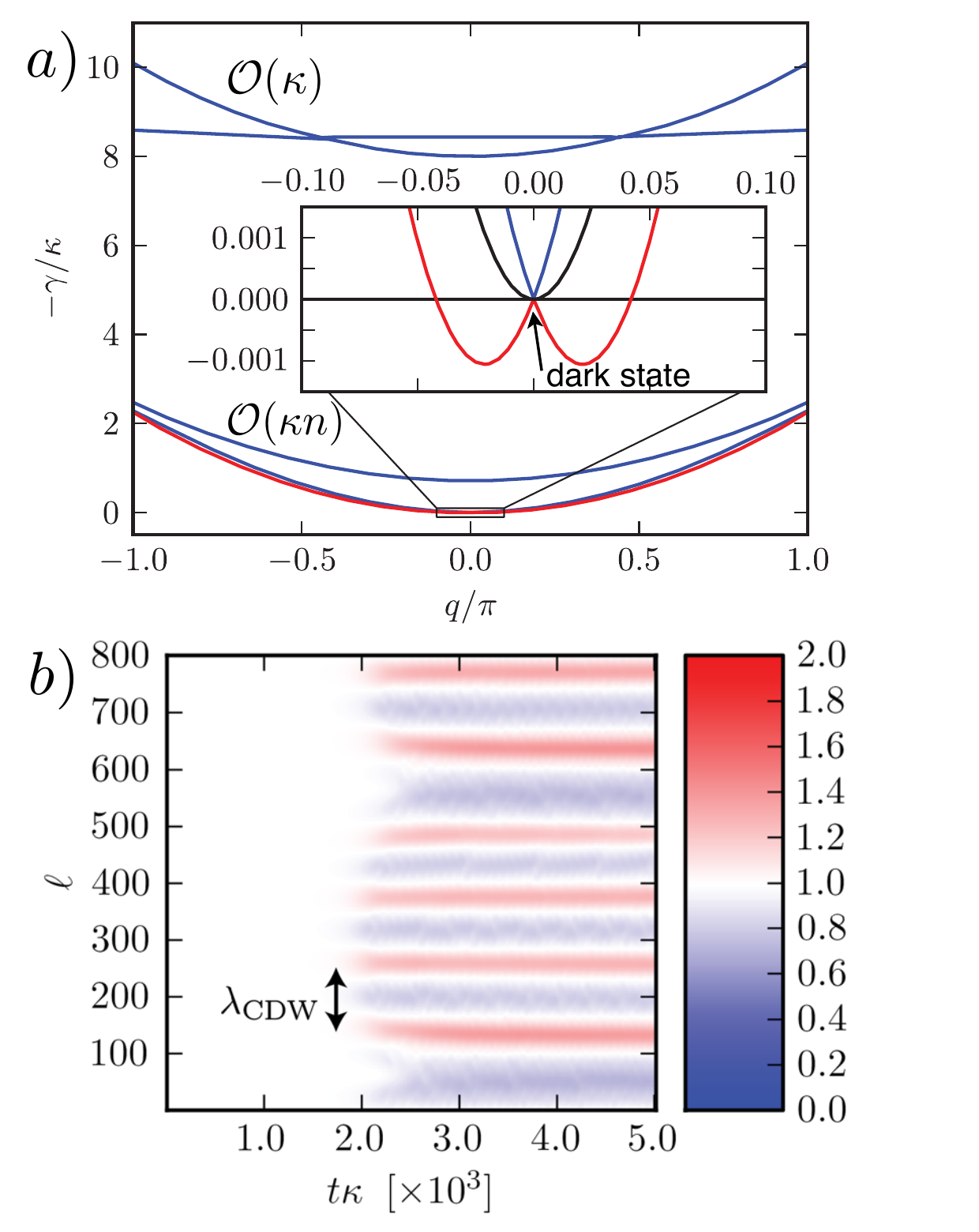}
\end{center}
\caption{(Color online) Dynamical instability: a) Damping spectrum as a function of quasimomentum from linear response around a homogeneous state. There are rapidly damping branches evolving at $\mathcal O (\kappa), \mathcal O (\kappa n)$, as well as two slowly evolving branches associated to single particle excitation damping. Around the dark state at $q=0$, a continuum of unstable modes appears.  b) Numerical evolution of the nonlinear system of correlation functions in the low density limit for 800 lattice sites. The color code represents the density profile, demonstrating an incommensurate charge density wave stationary phase with characteristic wavelength $\lambda_\text{CDW}$. Figure adapted from \citet{Diehl10a} and \citet{Tomadin11}. }
\label{DynamicalInstability}
\end{figure}

We linearize in time the EOM of Eq.~(\ref{eq:redmasterequation}), writing the generic connected correlation function as $\langle \hat{\cal O}_{\ell} \rangle(t) = \langle \hat{\cal O}_{\ell} \rangle_{0} + \delta \langle \hat{\cal O}_{\ell} \rangle(t)$, where $\langle \hat{\cal O}_{\ell} \rangle_{0}$ is evaluated on the homogeneous steady state of the system.
The EOM for the time and space dependent fluctuations $\delta \langle \hat{\cal O}_{\ell} \rangle(t) =\delta \Phi_{\ell}(t)$  is then Fourier transformed, resulting in a $7\times7$ matrix evolution equation $\partial_{t} \delta \Phi_{\textbf{q}} = M_{\textbf{q}} \delta \Phi_{\textbf{q}}$ for the correlation functions $\Phi_{\textbf{q}} = (\langle \delta b\rangle_{\textbf{q}}, \langle \delta b^{\dag}\rangle_{-\textbf{q}},$ $ \langle \delta b^{\dag} \delta b \rangle_{\textbf{q}}, \langle \delta b^{2} \rangle_{\textbf{q}}, \langle \delta b^{\dag 2} \rangle_{-\textbf{q}}, \langle \delta b^{\dag} \delta b^{2} \rangle_{\textbf{q}}, \langle \delta b^{\dag 2} \delta b \rangle_{-\textbf{q}} )$
(We note that the fluctuation $\delta \langle \delta b \rangle_{\textbf{q}}$ ($\delta \langle \delta b^{\dag} \rangle_{\textbf{q}}$) coincides with the fluctuation of the order parameter $\delta \psi_{\textbf{q}}$ ($\delta \psi_{-\textbf{q}}^{\ast}$), since the average of $\delta b_{\textbf{q}}$ on the initial state vanishes by construction.):

\begin{eqnarray}
\left (
\begin{array}{c}
\partial_{t} \delta\Psi_{1,\textbf{q}}(t) \\
\partial_{t} \delta\Psi_{2,\textbf{q}}(t)
\end{array}
\right )
=
\left (
\begin{array}{cc}
M_{11,\textbf{q}} & M_{12,\textbf{q}} \\
M_{21,\textbf{q}} & M_{22,\textbf{q}}
\end{array}
\right )
\left (
\begin{array}{c}
\delta\Psi_{1,\textbf{q}} \\
\delta\Psi_{2,\textbf{q}}
\end{array}
\right ),
\end{eqnarray}
where we have separated a slowly evolving sector describing the single particle fluctuations and containing the dark state $\delta\Psi_{1,\textbf{q}} = ( \delta \psi_{\textbf{q}}, \delta \psi^*_{-\textbf{q}})$, and a sector $\Psi_{2,\textbf{q}} =(\langle \delta b^{\dag} \delta b \rangle_{\textbf{q}}, \langle \delta b^{2} \rangle_{\textbf{q}}, \langle \delta b^{\dag 2} \rangle_{-\textbf{q}}, \langle \delta b^{\dag} \delta b^{2} \rangle_{\textbf{q}}, \langle \delta b^{\dag 2} \delta b \rangle_{-\textbf{q}} )$, whose evolution is seen to be lower bounded by the scale $\kappa n$. This matrix is easily diagonalized numerically, with the result for the imaginary part of the different branches, describing the damping, shown in Fig. \ref{DynamicalInstability}a. A separation of scales for the lower branches $\Psi_1$ and the higher ones $\Psi_2$ is clearly visible for low momenta $\mathbf{q}\to 0$, suggesting to integrate out the fast modes by adiabatic elimination $\partial_t\delta \Psi_2 \equiv0$. This results in a renormalization of the single particle complex excitation spectrum via the terms involving fractions,
\begin{eqnarray}
\partial_{t}
\left (
\begin{array}{c}
 \delta \psi_{\mathbf{q}} \\
\delta \psi^{\ast}_{-\mathbf{q}}
\end{array}
\right )
=
\left(
\begin{array}{cc}
Un + \epsilon_{\textbf{q}} - i \kappa_{\textbf{q}}  &  Un + \frac{9 Un}{4\kappa z}  \kappa_{\textbf{q}}   \\
 - Un -  \frac{9 Un}{4\kappa z} \kappa_{\textbf{q}} &  - Un -  \epsilon_{\textbf{q}}  - i \kappa_{\textbf{q}}
\end{array}
\right)
\left (
\begin{array}{c}
 \delta \psi_{\mathbf{q}} \\
\delta \psi^{\ast}_{-\mathbf{q}}
\end{array}
\right ),
\end{eqnarray}
where $\epsilon_{\textbf{q}}=J\textbf{q}^{2}$ is the kinetic energy and $\kappa_{\textbf{q}}=2(2n+1)\kappa \textbf{q}^{2}$ the bare dissipative spectrum for low momenta. The low-momentum spectrum of this matrix reads
\begin{eqnarray}
\gamma_\textbf{q} \simeq \mathrm i c |\textbf{q}|  + \kappa_{\textbf{q}}, \quad c=\sqrt{2 U n[J - 9 U n / (2z) ]},
\end{eqnarray}
with $c$ the speed of sound. The quadratic $\textbf{q}$-dependence present without renormalization correction is modified by a nonanalytic linear contribution, which dominates at small momenta and reproduces the shape of the unstable modes obtained via diagonalization in Fig.~\ref{DynamicalInstability}.
For a hopping amplitude smaller than the critical value $J_{\rm c} = 9Un / (2z)$, the speed of sound becomes imaginary, rendering the system unstable. The linear slope of the stability border for small $J$ and $U$ is clearly visible from the numerical results in Fig.~\ref{PhaseDiagram}.

Beyond the unstable point, the linearization strategy around the homogeneous state fails in describing the true steady state of the system. In order to extract the correct stationary state in this regime, we resort to a numerical treatment of a large system in the low density limit, where the nonlinearities are fully taken into account. The result is displayed in  Fig.~\ref{DynamicalInstability}b, revealing that the stationary state exhibits charge density wave order with characteristic wavelength $\lambda_{\text{CDW}}$ which is set by the inverse of most unstable momentum mode. Generically, it is incommensurate with the lattice spacing. The scale characterizing the instability is thus transmuted into a physical length scale. The phenomenon is found universally for different system sizes, ruling out the possibility of a mere finite size effect.

At this point, three comments are in order. First, the subtle renormalization effect is not captured by a Gross-Pitaevski type approximation scheme and relies on a suitable treatment of the higher order correlation functions. Second, the new phase emerges at weak coupling already, and for small enough $J$ the homogeneous dissipative condensate is unstable towards the pattern formation at arbitrarily weak interaction. In this weak coupling regime, our approximation scheme is very well controlled. Third, the effect relies on the existence of a continuum of modes, and thus has a truly many-body origin. In summary, the phase with simultaneous spontaneous breaking of phase rotation and lattice translation symmetry is understood as a fluctuation induced beyond (standard) mean field many-body phenomenon, which seems quite unique to the dissipative setting. The full phase diagram discussed here is shown in Fig. \ref{PhaseDiagram}.

\subsection{Dissipative D-Wave Paired States for Fermi-Hubbard Quantum Simulation}
\label{sec:DWave}

\emph{Motivation} -- One of the big experimental challenges in the field of cold atoms is the quantum simulation of the ground state of the Fermi-Hubbard model (FHM) describing two-component fermions interacting locally and \emph{repulsively} on the lattice, whose filling is controlled by a chemical potential. The particular interest in this model roots in the fact that it is believed to be a minimal model for the description of cuprate high-temperature superconductors. The model has challenged theorists for almost thirty years by now, and has proven to be hard to analyze with both advanced analytical approaches and numerical techniques. In particular, from the theory point of view, so far the d-wave ordered nature of the ground state away from half filling, which is observed experimentally, has only the status of a conjecture. Together with the uncertainty whether the model actually faithfully captures the microscopic physics of the cuprates, this situation calls for a quantum simulation of the FHM ground state in a cold atom context, taking advantage of precise microscopic control in such systems.

This goal still remains very challenging, due to tough requirements on the temperature in these systems. In fact, the d-wave gap in the cuprates, setting the temperature scale to be reached, is only $\sim 0.01 T_{\text{F}}$ ($T_{\text{F}}$ the Fermi temperature), and therefore still more than an order of magnitude away from what can currently be reached in the lab. Despite impressive progress in this direction \citep{PhysRevLett.89.220407,PhysRevLett.94.080403,Chin06,Jordens08,Schneider05122008,PhysRevLett.104.180401,EsslingerRev10}, where quantum degeneracy is reached on the lattice, new cooling strategies are needed to achieve this goal. The roadmap using dissipation state engineering is the following: (i) We dissipatively produce a low entropy state that is ``close'' (in a sense specified below) to the expected ground state of the Fermi-Hubbard model away from half filling. (ii) We then construct a suitable adiabatic passage, that consists in slowly switching off the Liouville dynamics while ramping up the Hubbard Hamiltonian.

Here, we will present a mechanism which allows to engineer fermionic paired states of arbitrary symmetry, exemplified here for the case of d-wave symmetry, which is based on dissipative dynamics alone and works in the absence of any attractive conservative forces. 
The mechanism is based on an interplay of the above mechanism of quasilocal phase locking, and Pauli blocking, thus crucially relying on Fermi statistics. A suitable mean field theory, valid for the long-time evolution, has a natural interpretation in terms of damping of fermionic quasiparticles and simplifies the microscopically quartic (interacting) Liouville operator into a quadratic one. We then discuss possible implementations and present numerical results for a suitable adiabatic passage.

\emph{The state to be prepared} -- We target BCS-type states, which represent the conceptually simplest many-body wave functions describing a condensate of $N$ paired spin-1/2 fermionic particles. Working on a bipartite square lattice, and assuming singlet pairs with zero
center-of-mass momentum, we have
\begin{eqnarray}\label{BCSDarkState}
&&|\mathrm{BCS}_{N}\rangle\sim(d^{\dag})^{N/2}|\text{vac}\rangle, \\\nonumber
&& d^{\dag} =\sum_{\mathbf{q} }\varphi_{\mathbf{q}}c_{\mathbf{q},\uparrow}^{\dag}c_{-\mathbf{q},\downarrow}^{\dag}  =\sum_{i,j}\varphi_{ij}c_{i,\uparrow}^{\dagger}c_{j,\downarrow}^{\dagger},
\end{eqnarray}
where $c_{\mathbf{q},\sigma}^{\dagger}$ ($c_{i,\sigma }^{\dagger}$)  denotes the creation operator for fermions with quasimomentum $\mathbf{q}$ (on lattice site $i$) and spin $\sigma=\uparrow,\downarrow$, and $\varphi_{\mathbf{q}}$ ($\varphi_{ij}$) the momentum (relative position) wave function of the pairs. We now specialize to a state close to the conjectured FHM ground state, in what concerns (i) the symmetries and (ii) the ground state energy. For the above pair creation operator $d^\dag$, the pair wave function
\begin{eqnarray}\label{pairWF}
\varphi_{\mathbf{q}}=\cos q_{x}-\cos q_{y} \text{ or } \varphi_{ij}= \tfrac{1}{2}\sum_{\lambda=x,y}\rho_\lambda(\delta_{i,j+\mathbf{e}_{\lambda}}+  \delta_{i,j-\mathbf{e}_{\lambda}})
\end{eqnarray}
with $\rho_x =-\rho_y=1$ ensures the symmetry properties of pairing in the singlet channel and the d-wave transformation law $\varphi_{q_{x},q_{y}}=-\varphi_{-q_{y},q_{x}}=\varphi_{-q_{x},-q_{y}}$ under spatial rotations. The wave function corresponds to the limit of small pairs (see Fig.~\ref{DWaveState}a), and phase coherence is granted by the delocalization of these molecular objects. Pairs with such a short internal coherence length appear in the cuprates in the regime where strong correlations set in upon approaching half filling. No quantitative statement can, of course, be made on the energetic proximity of this wavefunction to the true FHM ground state. However, the fact that the pairing occurs \emph{off-site} avoids excessive double occupancy (which is energetically unfavorable for the strong repulsive onsite interactions), and makes this state an interesting candidate for quantum simulation.

\subsubsection{Dissipative Pairing Mechanism}
\label{sec:DissPair}

We now construct a \emph{parent Liouvillian}, which has the above d-wave state $|\text{d}\rangle$ as a dark state. In other words, we will construct a set of (non-hermitian) jump operators with the property $J_i^\alpha|\mathrm{BCS}_{N}\rangle=0$, where $i=1, ... , M \, (\alpha= x,y,z)$ represents a position (spin) index ($M$ is the number of sites in the lattice). Due to the product form of the dark state wavefunction, a key sufficient condition to fulfill this task is to find a set of normal ordered jump operators $J_i^\alpha$, which commute with the generator of the dark state,
\begin{eqnarray}
[J_i^\alpha , d^\dag] =0 \,\, \forall i,\alpha.
\end{eqnarray}
The appearance of both indices reflects the need to fix the properties of the state in both position and spin space. From a practical point of view, we require the jump operators to be quasilocal, number conserving (i.e. $[J_i^\alpha , \hat N] =0$) and to act on \emph{single particles} only, restricting their class to quasilocal phase rotation invariant fermion bilinears. The above condition is very general and  thus applicable to wider classes of paired, or even more generally, product states. One example discussed in the next section is p-wave paired states for spinless fermions.

\begin{figure}[tbp]
\begin{center}
\includegraphics[width=1.0\columnwidth]{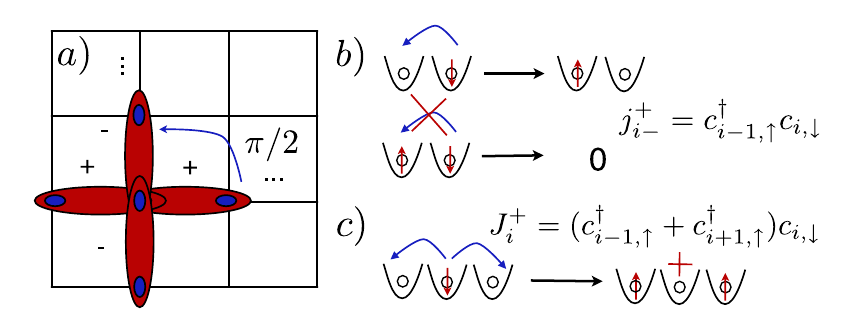}
\end{center}
\caption{(Color online) D-wave state and action of the jump operators. a) Symmetries of the state: an offsite fermion singlet pair exhibits a characteristic sign change under spatial rotations, and is delocalized over the whole lattice. b,c) The dissipative pairing mechanism combines (b) Pauli blocking and (c) delocalization via phase locking. b) The action of Lindblad operators using Pauli blocking for a N\'eel state (see text). c) The d-wave state results as a delocalization of these pairs away from half filling (shown is a cut along one lattice axis). Figure adapted from  \cite{Diehl10b}.}
\label{DWaveState}
\end{figure}

We now turn to the construction of the Lindblad operators for the d-wave BCS state. To this end, we follow the physical picture that d-wave superconductivity (or superfluidity) on a lattice can be viewed as delocalized antiferromagnetic order, obtained when moving away from half filling \citep{Anderson87,Zhang88,Gros88,Altman02,Paramekanti04}. Therefore, we will first construct the parent Liouvillian for a N\'{e}el state at half filling, which is the conceptually simplest (product) wavefunction representing antiferromagnetism, and then generalize to the BCS state.  There are two N\'{e}el states at half filling, related by a global spin flip, $|\text{N}+\rangle=\prod_{i\in A}c_{i+\mathbf{e}_{x},\uparrow}^{\dag}c_{i,\downarrow}^{\dag}|\text{vac}\rangle$, $|\text{N}-\rangle=\prod_{i\in A}c_{i+\mathbf{e}_{x},\downarrow }^{\dag}c_{i,\uparrow}^{\dag}|\text{vac}\rangle$ with $A$ a sublattice in a two-dimensional bipartite lattice. For later convenience we introduce \textquotedblleft N\'{e}el unit cell operators\textquotedblright\ $\hat{S}_{i,\nu}^{a}=c_{i+\mathbf{e}_{\nu}}^{\dag}\sigma^{a}c_{i}^{\dag}$ ($a=\pm,\mathbf{e}_\nu=\{\pm\mathbf{e}_x,\pm\mathbf{e}_y\},$ and two-component spinor $c_i = (c_{i,\uparrow}, c_{i,\downarrow})$), such that the state can be written in eight different forms, $|\text{N}\pm\rangle=\prod_{i\in A}\hat{S}_{i,\nu}^{\pm}|\text{vac}\rangle=(-1)^{M/2}\prod_{i\in B}\hat{S}_{i,-\nu}^{\mp}|\text{vac}\rangle$. We then see that the Lindblad operators must obey $[j_{i,\nu}^{a},\hat{S}_{j,\mu} ^{b}]=0$ for all $i,j$ located on the same sublattice $A$ or $B$, which holds for the set
\begin{eqnarray}\label{AFjump}
j_{i,\nu}^{a}=c_{i+\mathbf{e}_{\nu}}^{\dag} \sigma^{a}c_{i}, \, i \in A\,\text{or}\, B.
\end{eqnarray}
The presence of fermionic statistics is essential for the action of the operators $j_{i,\nu}^{a}$, as illustrated in Fig.~\ref{DWaveState}b: they generate spin
flipping transport according to e.g. $j_{i,\nu}^{+}=c_{i+\mathbf{e}_{\nu},\uparrow}^{\dag}c_{i,\downarrow}$, not possible when the N\'{e}el order is already present. It is then easy to prove the uniqueness of the N\'{e}el steady state up to double degeneracy: The steady state must fulfill the quasilocal condition that for any site occupied by a certain spin, its neighboring sites must be filled by opposite spins. For half filling, the only states with this property are $|\text{N}\pm\rangle$. The residual twofold degeneracy can be lifted by adding a single operator $j_{i}=c_{i+\mathbf{e}_{\nu}}^{\dag}(\mathbf{1}+\sigma^{z})c_{i}$ on an arbitrary site $i$.

The Lindblad operators for the d-wave BCS state can now be constructed along a similar strategy. First we rewrite the d-wave generator in terms of antiferromagnetic unit cell operators $\hat {S}_{i}^{a}$,
\begin{align}\label{BCSWF}
d^{\dag}   =  \tfrac{\mathrm i}{2}\sum_{i}(c_{i+\mathbf{e}_{x}}^{\dag} -c_{i+\mathbf{e}_{y}}^{\dag
})\sigma^{y}c_{i}^{\dag}= \tfrac{a}{2}\sum_{i}\hat{D}_{i}^{a},\quad
\hat{D}_{i}^{a}   =\sum_{\nu}\rho_\nu\hat{S}_{i,\nu}^{a},
\end{align}
where $\rho_{\pm x}=-\rho_{\pm y}=1$, and the quasilocal d-wave pair $\hat D_{i}^{a}$ may be seen as the ``d-wave unit cell operators". This form makes the picture of d-wave superconductivity as delocalized antiferromagnetic order transparent, and we note the freedom $a=\pm$ in writing the state. The condition
$[J_{i}^{\alpha},\sum_{j}\hat{D}_{j}^{b}]=0$ ($\alpha=(a,z)\, \text{or} \,(x,y,z))$ is fulfilled by
\begin{eqnarray} \label{BCSLindblads1}
J_{i}^{\alpha}=\sum_{\nu}\rho_\nu j_{i,\nu}%
^{a},
\quad j_{i,\nu}^{\alpha}=c_{i+\mathbf{e}_{\nu}}^{\dag }\sigma^{\alpha}c_{i}, 
\end{eqnarray}
which is our main result. Coherence is created by these operators via phase locking between adjacent cloverleaves of sites.

The uniqueness of this state as a stationary state for the Lindblad operators \eqref{BCSLindblads1} is less obvious then in the antiferromagnetic case and we argue based on symmetry arguments. Uniqueness is equivalent to the uniqueness of the ground state of the associated hermitian Hamiltonian $H=V\sum_{i,\alpha=\pm,z}J_{i}^{\alpha \dag}J_{i}^{\alpha}$ for $V>0$. The state generated by \eqref{BCSLindblads1} shares  the Hamiltonian symmetries of global phase and spin rotations, and translation invariance. Assuming that no other symmetries exist, we then expect the ground state to be unique. The full set $\{J_i^\alpha\}$ is necessary for uniqueness: Omitting e.g. $\{J_i^z\}$ generates an additional discrete symmetry in $H$ resulting in two-fold ground state degeneracy. We confirmed these results with small scale numerical simulations for periodic boundary conditions, cf. Fig.~\ref{FidelityEntropy}. We note that the above construction method allows us to find ``parent'' Lindblad operators for a much wider class of BCS-type states, see \cite{Yi11}.

\subsubsection{Dissipative Gap}
\label{sec:DissGap}

A remarkable feature of the dissipative dynamics defined with the set of operators \eqref{BCSLindblads1} is the emergence of a ``dissipative gap'' in the long time evolution of the master equation. Such a dissipative gap is a minimal damping rate which crucially remains finite in the thermodynamic limit. The phenomenon is a dissipative counterpart of a coherent gap suppressing single particle fermion excitations in a BCS superfluid, where it is a characteristic feature of the low energy effective theory.
\begin{figure}[tbp]
\begin{center}
\includegraphics[width=.9\columnwidth]{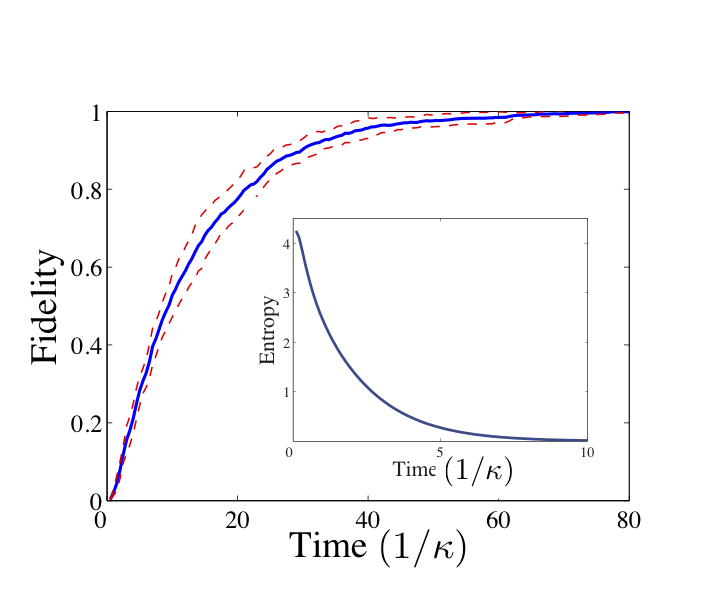}
\end{center}
\caption{(Color online) Uniqueness of the d-wave dark state for the master equation with Lindblad operators from Eq.~\eqref{BCSLindblads1}: Fidelity to the d-wave BCS state, $\langle {\rm BCS}_N |\rho |{\rm BCS}_N\rangle$ for 4 atoms on a 4$\times$4 grid, showing exponential convergence from a completely mixed state to a pure state. Dashed lines denote sampling errors. (Inset): Entropy evolution for four atoms on a 4x1 lattice. Figure adapted from \cite{Diehl10b}.}
\label{FidelityEntropy}
\end{figure}

The dissipative gap can be established in a mean field theory which is controlled by the proximity to the exactly known stationary dark state. For this purpose it is convenient to give up exact particle number conservation and to work with fixed phase coherent states $|\mathrm{BCS}_{\theta}\rangle=\mathcal{N}^{-1/2}\exp(e^{\mathrm{i}\theta}d^{\dag})|\text{vac}\rangle$ instead of the fixed number states $|\mathrm{BCS}_{N}\rangle$ \citep{leggettbook}, where $\mathcal{N} =\prod_{\mathbf{q}}(1+\varphi_{\mathbf{q}}^{2})$ ensures the normalization. The equivalence of these approaches in the thermodynamic limit is granted by the fact that the relative number fluctuations in BCS coherent states scale $\sim 1/\sqrt{N}$, where $N$ is the number of degrees of freedom in the system. The density matrix for the coherent states factorizes in momentum space $\exp (e^{\mathrm{i}\theta}d^{\dag})|\text{vac}\rangle=\prod_{\mathbf{q}}(1+e^{\mathrm{i}\theta}\varphi_{\mathbf{q}}c_{\mathbf{q},\uparrow}^{\dag}c_{-\mathbf{q},\downarrow}^{\dag})|\text{vac}\rangle$. At late times, we can make use of this factorization property and expand the state around $|\mathrm{BCS}_{\theta}\rangle$, implemented with the ansatz $\rho=\prod_{\mathbf{q}}\rho_{\mathbf{q}}$, where $\rho_{\mathbf{q}}$ contains the mode pair $\pm(\mathbf{q},\sigma)$ necessary to describe pairing. We then find a linearized evolution equation for the density operator,
\begin{eqnarray}
\mathcal L [\rho] = \sum_{\mathbf{q},\sigma}\kappa_\mathbf{q} [\gamma_{\mathbf{q},\sigma} \rho \gamma^\dag_{\mathbf{q},\sigma} - \tfrac{1}{2}\{\gamma^\dag_{\mathbf{q},\sigma}\gamma_{\mathbf{q},\sigma},\rho\}],
\end{eqnarray}
with quasiparticle Lindblad operators and momentum dependent damping rate given by
\begin{eqnarray}
\gamma_{\mathbf{q},\sigma}&=&(1+\varphi_{\mathbf{q}}^{2})^{-1/2}\,(c_{-\mathbf{q},\sigma}+s_\sigma\varphi_{\mathbf{q}}c_{\mathbf{q},-\sigma
}^{\dag}),\\\nonumber
\kappa_{\mathbf{q}}&=&\kappa\, \tilde n \,(1+\varphi_{\mathbf{q}}^{2})\geq \kappa\,\tilde n,
\end{eqnarray}
with $s_\uparrow = -1,s_\downarrow =1$, the wavefunction specified in Eq. \eqref{pairWF}, and the value $\tilde n= 2\int\tfrac{d\mathbf{q}}{(2\pi)^2}\tfrac{|\varphi_\mathbf{q}|^2}{1 +|\varphi_\mathbf{q}|^2} \approx 0.72$ dictated by the presence of nonzero mean fields resulting from a coupling to other momentum modes, and the proximity to the final state.

The linearized Lindblad operators have analogous properties to quasiparticle operators familiar from interaction pairing problems: (i) They annihilate the (unique) steady state $\gamma_{\mathbf{q},\sigma}|\mathrm{BCS}_{\theta}\rangle=0$; (ii) they obey the Dirac algebra $\{\gamma_{\mathbf{q},\sigma},\gamma_{\mathbf{q}^{\prime},\sigma^{\prime}}^{\dag}\}=\delta_{\mathbf{q},\mathbf{q}^{\prime}}\delta_{\sigma,\sigma^{\prime}}$ and zero otherwise; and (iii) in consequence are related to the original fermions via a canonical transformation.

Physically, the dissipative gap $\kappa\,\tilde n$ implies an exponential approach to the steady d-wave BCS state for long times. This is easily seen in a quantum trajectory representation of the master equation, where the time evolution of the system is described by a stochastic system wavefunction $|\psi(t)\rangle$ undergoing a time evolution with non-hermitian ``effective'' Hamiltonian $|\psi(t)\rangle =e^{-iH_{\mathrm{eff}}t}|\psi(0)\rangle/\left\Vert \ldots\right\Vert $ ($H_{\text{eff}}= H - \mathrm i \kappa \sum_{i,\alpha} J_{i}^{\alpha\,\dag}J_{i}^{\alpha}$ here) punctuated with rate $\kappa\left\Vert j_{\ell}|\psi(t)\rangle\right\Vert ^{2}$ by quantum jumps $|\psi(t)\rangle\rightarrow j_{\ell}|\psi (t)\rangle/\left\Vert \ldots\right\Vert $ such that $\rho(t)=\langle|\psi(t)\rangle\langle\psi(t)|\rangle_{\mathrm{stoch}}$ (see, e.g., \cite{zollerbook}). We thus see that (i) the BCS state is a dark state of the dissipative dynamics in that $j_{\ell}|\mathrm{BCS}_{N}\rangle=0$ implies that no quantum jump will ever occur, i.e. the state remains in $|\mathrm{BCS}_{N}\rangle$, and (ii) states near $|\mathrm{BCS}_{N}\rangle$ decay exponentially with rate lower-bounded by the dissipative gap.

This dissipatively gapped behavior strongly contrasts the bosonic case, where the dissipation is gapless as we have seen above, in the sense that $\kappa_\mathbf{q} \sim \mathbf{q}^2$ for $\mathbf{q}\to0$. One crucial difference between the bosonic and fermionic evolutions is then the fact that many-body observables involving a continuum of modes behave polynomially in the boson case, due to the slow decay in the vicinity of the dark state. For fermions instead, the dark state property is not encoded in a zero of the decay rate, but rather in the annihilation property of the linearized Lindblad operators on a nontrivial BCS vacuum. In this case, even many-body observables will relax exponentially. More generally, the generation of a finite gap scale at long times makes the fermionic dissipatively induced phases potentially more stable than the bosonic ones, as one may compare competing energy or rate scales to that finite scale.

This convergence to a unique pure state is illustrated in Fig.~\ref{FidelityEntropy} using numerical quantum trajectory simulations for small systems. We show the fidelity of the BCS state for a small 2D grid as a function of time, computed for the full density matrix via the quantum trajectories method. The inset shows the entropy evolution for a small 1D system (where one direction of the d-wave cloverleaf is simply omitted).
\begin{figure}[tbp]
\begin{center}
\hspace{-0.9cm}\includegraphics[width=1.1\columnwidth]{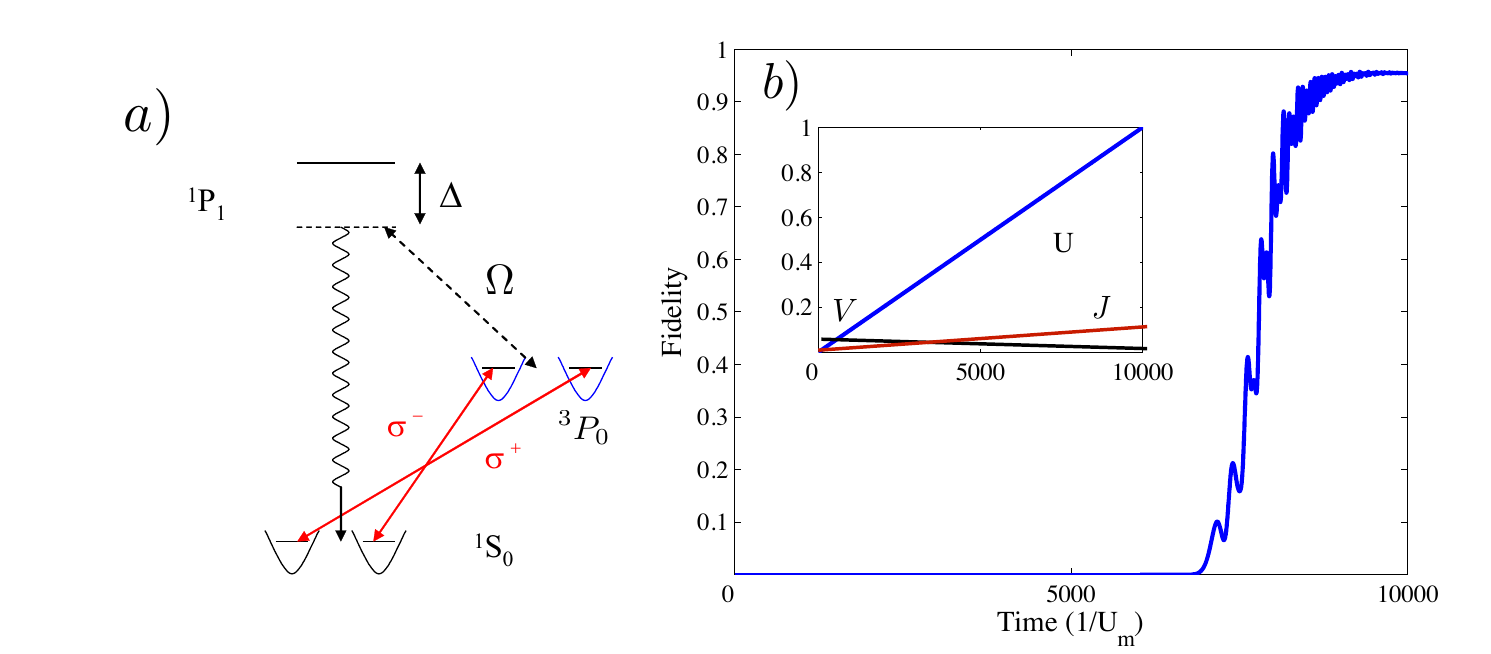}
\end{center}
\caption{ (Color online) a) Level scheme for physical implementation. The spin flip operation is implemented via off-resonant coherent coupling to the $^3$P$_0$ manifold with circularly polarized light (red arrows). The long lived $^3$P$_0$ states are coupled to the $^1$P$_1$ level in a two-photon process, from which spontaneous emission into a cavity is induced, leading back to the $^1 S_0$ manifold encoding the physical fermionic states. b) Adiabatic passage connecting the mean field d-wave state with the ground state of the FHM on a $2\times6$
ladder with 4 atoms with parent Hamiltonian $H_{\mathrm{p}}$ (see text). Evolution of fidelity of the instantaneous system state with respect to the final ground state of the FHM is calculated. (inset): Parameters hopping $J$, onsite interaction $U$ of the Fermi-Hubbard Hamiltonian $H = -
J\sum_{\langle i,j\rangle,\sigma}c^{\dag}_{i\sigma}c_{j\sigma} + U\sum_{i}c^{\dag}_{i\uparrow}c_{i\uparrow}c^{\dag}_{i\downarrow}c_{i\downarrow}$, and the parent Hamiltonian strength $V$, as a function of time in units of the maximal final interaction strength $U_m$. Figure adapted from \cite{Yi11}.}
\label{AdiabaticPassage}
\end{figure}

\subsubsection{State Preparation}
\label{sec:StatePrep}

\emph{Implementation with alkaline earth-like atoms} -- The conceptually simple quasilocal and number-conserving form of $J_{i}^{\alpha}$ raises the possibility to realize dissipation induced pairing via reservoir engineering with cold atoms. We illustrate this in 1D, taking the example of $J_{i}^+ = (c_{i+1,\uparrow}^{\dag}+ c_{i- 1,\uparrow}^{\dag})c_{\downarrow}$. Implementation requires (i) a spin flip, (ii) a spatial redistribution of the atom onto sites neighbouring the central one, and (iii) a dissipative process which preserves the coherence over several lattice sites. These ingredients can be met using alkaline earth-like atoms \citep{Ye27062008,PhysRevLett.99.123001,PhysRevLett.101.170504,Gorshkov10} with nuclear spin (e.g., $I=1/2$ for $^{171}$Yb), and a long-lived metastable $^{3}$P$_{0}$ manifold as a physical basis, see \cite{AJD11} for a recent review. In this setting, one can construct a stroboscopic implementation, where the action of each $J_{i}^{\alpha}$ is realized successively. The level scheme and the spin flip process are described in Fig. \ref{AdiabaticPassage}a. There we concentrate on the spatial redistribution of the atoms using the fact that the $^{3}$P$_{0}$ states can be trapped independently of the ground $^{1}$S$_{0}$ manifold. The $^{3}$P$_{0}$ state is trapped in a lattice of three times the period as that for the $^{1}$S$_{0}$ state, defining blocks of three sites in the original lattice. Using this, any $\downarrow$ atom in $^{1}$S$_{0}$ on the central site is excited to the $\uparrow$ state of the $^{3}$P$_{0}$ manifold. By adding an additional potential, the traps for $^{3}$P$_{0}$ are coherently divided so that atoms confined in them overlap the right and left sites of the original block. Decay is induced by coupling atoms in the $^{3}$P$_{0}$ state off-resonantly to the $^{1}$P$_{1}$ state, as depicted in Fig.~\ref{AdiabaticPassage}a, with coupling strength $\Omega$, and detuning $\Delta$. By coupling the $^{1}$S$_{0}$--$^{1}$P$_{1}$ transition to a cavity mode with linewidth $\Gamma$ and vacuum Rabi frequency $g$, the decay is coherent over the triple of sites. In the limit $\Delta\gg\Omega$ and $\Gamma\gg\frac{\Omega g}{\Delta}$, an effective decay rate $\Gamma_{\mathrm{eff}}=\frac{\Omega^{2}g^{2}}{\Delta^{2}\Gamma}\sim9$kHz results for typical parameters. Fermi statistics will be respected in this process, as long as the atoms remain in the lowest band. This operation can be performed in parallel for different triples, and needs to be repeated with the superlattice shifted for other central sites. Similar operations combined with rotations of the nuclear spin before and after these operations allows implementation of $J_{i}^{-}$ and $J_{i}^{z}$. In 2D 3x3 plaquettes are defined by the appropriate superlattice potential for the $^{3}$P$_{0}$ level, and the adiabatic manipulation of the potential has to be adjusted to ensure the correct relative phases for atoms transported in orthogonal directions. Such a digital or stroboscopic scheme is rather demanding in the context of cold atoms, and most of the complication comes from the need to fix the spin quantum number. Below, we discuss spinless fermions and see that there, an ``analog'' implementation along the lines of Sect.~\ref{sec:Concepts} with continuous driving and dissipation is possible.

\emph{Adiabatic Passage} -- To reach the ground state of the FHM in small scale numerical simulations, we found it efficient to introduce in addition to the parent Liouvillian a parent Hamiltonian $H_{\mathrm{p}}=V\sum_{i,\alpha}J^{\alpha\,\dag}_{i} J_{i}^\alpha$, which has the above d-wave state as the exact unique (fixed number) ground state for $V>0$, and which could be obtained by replacing the decay step into the cavity by induced interactions between atoms. The result of the numerical calculation is reported in Fig.~\ref{AdiabaticPassage}b, where convergence to the FHM ground state is clearly seen. In a large system, one should additionally be able to take advantage of the fact that (i) in the initial stages the system is protected by a gap $\sim 0.72V$, and (ii) the d-wave state has identical symmetry and similar energy to the conjectured Fermi-Hubbard ground state away from half filling. Thus, a d-wave superfluid gap protection is present through the whole passage path, since no phase transition is crossed.

\subsection{Dissipative Topological States of Fermions}
\label{sec:Topological}

\emph{Motivation} -- Topological phases of matter exhibit ordering phenomena beyond the Landau paradigm, where order is described by local order parameters. Instead, these phases are characterized by nonlocal order parameters, the topological invariants \citep{HasanKane,RevModPhys.83.1057}.  Observable physical manifestations of topological order emerge when these systems are subject to boundary conditions in space, such as the appearance of Majorana modes localized to suitably designed edges in certain one- or two-dimensional superfluids  \citep{Kitaev00,ReadGreen}. These modes are robust against large classes of environmental perturbations and imperfections. This gives them a potentially high practical relevance, and they are discussed as candidates for providing the building blocks for topologically protected quantum memories and computations \citep{nayak-rmp-80-1083}.

So far, the concept of topological order and its physical consequences have been discussed mainly in a Hamiltonian ground state context. Motivated by the prospects of combining topological protection with a targeted dissipative engineering of the corresponding states, in  \cite{Diehl11,Bardyn12} we have shown how such concepts and phenomena manifest themselves in systems governed by driven-dissipative Lindblad dynamics. Here we will give a brief review of these results, focusing on the simplest paradigmatic model discussed in  \cite{Diehl11}, a dissipative quantum wire of spinless atomic fermions. This model is the counterpart of Kitaev's quantum wire, which provides a minimal one-dimensional model for topological order, and hosts Majorana edge modes in a finite wire geometry. In particular, we establish dissipative Majorana modes, and discuss their interpretation in terms of a nonlocal decoherence free subspace. We give an argument for the nonabelian exchange statistics, and sketch the construction of a topological invariant for density matrices corresponding to mixed states pinpointing the topological origin of the edge modes. We also highlight a phase transition induced by ``loss of topology'' which has no Hamiltonian counterpart. Beyond these theoretical findings, we argue that due to the spinless nature of the atomic constituents, an implementation along the lines of Sect. \ref{sec:Concepts} is possible. Remarkably, all that needs to be done is to replace the bosonic operators in Eq. \eqref{eq:jump} by spinless fermionic ones, and to put proper boundaries using the new experimental tools offered by singe-site addressability \citep{bakr-nature-462-74, bakr-science-329-547,sherson-nature-467-68,weitenberg-nature-471-319}. Together with practical preparation protocols and detection schemes \citep{Kraus12}, this makes dissipative state engineering an attractive route for realizing Majorana physics in the lab.

\emph{Topological quantum wire in Hamiltonian setting} -- Before embarking the construction of a dissipative quantum wire, we first recapitulate briefly Kitaev's Hamiltonian scenario. We discuss spinless fermions $a_{i},a_{i}^{\dag}$ on a finite chain of $N$ sites $i$ described by a quadratic Hamiltonian $H=\sum_{i} \left[  \left( -Ja_{i}^{\dag}a_{i+1} - \Delta a_{i}a_{i+1} + \text{h.c.} \right) -\mu a_{i}^{\dag}a_{i} \right]$,  with hopping amplitude $J$, a pairing term with order parameter $\Delta$, and a chemical potential $\mu$. The topologically non-trivial phase of the model is best illustrated for parameters $J=|\Delta|$ and $\mu=0$, where the Hamiltonian simplifies to
\begin{equation}
H=  2J\sum_{i=1}^{N-1}  \left( \tilde{a}_{i}^{\dag} \tilde{a}_{i} - \frac{1}{2} \right) = \mathrm{i}J\sum_{i=1}^{N-1}c_{2i}\,c_{2i+1}. \label{HKit}
\end{equation}
Here we write the Hamiltonian in a  complex Bogoliubov basis defined with quasilocal fermionic quasiparticle operators $\tilde{a}_{i}$, and in terms of Majorana operators $c_{i}$, which are given by the quadrature components of the original complex fermion operators $a_{i}=\frac{1}{2}\left( c_{2i} - \mathrm{i } c_{2i-1} \right)  $, respectively. With these preparations, we collect some key properties of this model: The \emph{bulk properties} are most clearly revealed in the complex Bogoliubov basis, where the Hamiltonian is diagonal: The ground state is determined by the condition $\tilde{a}_{i}|G\rangle=0$ for all $i$, and the bulk describes a fermionic BCS-type $p$-wave superfluid with a bulk spectral gap, which for the above parameter choice equals the constant dispersion $\epsilon_{k}=2J$. The Majorana representation instead gives rise to a picture of the bulk in terms of pairing of Majoranas from different physical sites. In view of the \emph{edge physics}, the absence of the term $2 ( \tilde{a}_{N}^{\dag}\tilde{a}_{N} -1/2) =\mathrm{i}c_{2N}c_{1}$ for a finite wire indicates the existence of a two-dimensional zero energy fermionic subspace spanned by $|\alpha\rangle\in\{ |0\rangle,|1\rangle=\tilde{a}_{N}^{\dag}|0\rangle\} $, which is highly non-local in terms of the complex fermions. In contrast, in the Majorana basis the situation is described in terms of two Majorana edge modes $\gamma_{L}=c_{1}$ ($\gamma_{R}=c_{2N}$), which are completely localized on the leftmost (rightmost) Majorana site $1$ ($2N$), describing ``half'' a fermion each. These edge modes remain exponentially localized in the whole parameter regime $-2J<\mu<2J$, however leaking more and more strongly into the wire when approaching the critical values. Their existence is robust against perturbations such as disorder, which can be traced back to the bulk gap in connection with their topological origin \citep{Kitaev00}.

\subsubsection{Dissipative Topological Quantum Wire}
\label{sec:DissTopWire}

\begin{figure}[tbp]
\begin{center}
\includegraphics[width=.9\columnwidth]{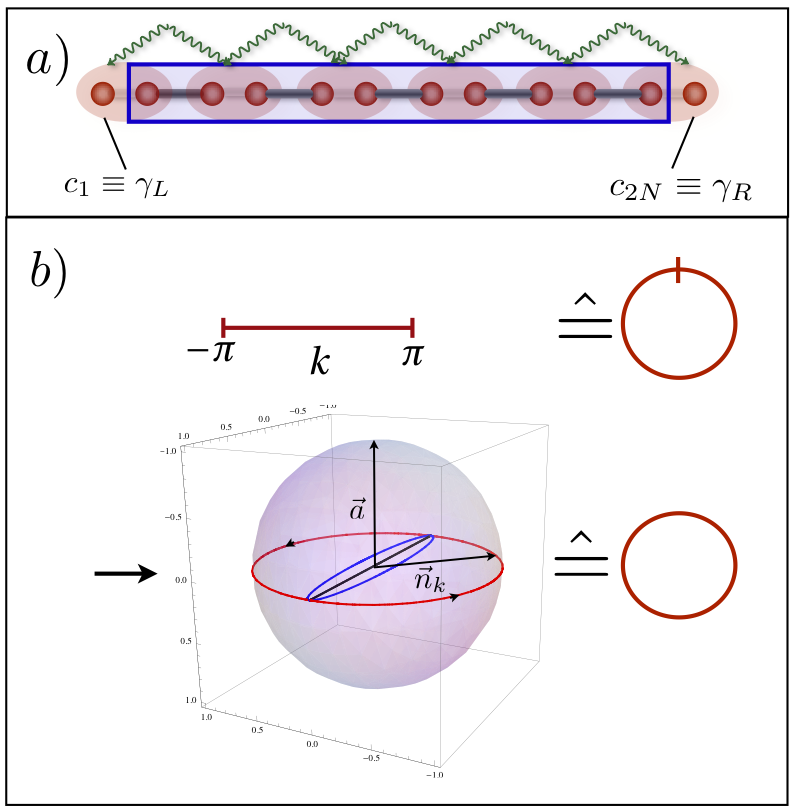}
\end{center}
\caption{(Color online) a) The Lindblad operators act on each link of the finite wire, in this way isolating the edge mode subspace described by $\gamma_L,\gamma_R$, which together define the Hilbert space of one complex fermion (see text). The bulk (blue shaded) is cooled to a p-wave superfluid, with pairing links between different physical sites established dissipatively. b) Visualization of the winding number $\nu$ for chirally symmetric mixed states, characterizing the mapping from the Brillouin zone $\simeq S^1$ to the vector $\vec n_k$, which due to chiral symmetry is constrained to a great circle $\simeq S^1$. For pure states, it is furthermore pinned to unit length (large circle). Tuning the Liouville parameters can destroy the purity and deforms the circle to an ellipse (blue), while the topological invariant remains well defined. A phase transition occurs when the ellipse shrinks to a line (dark line). The values of $\theta$ are: $\pi/4$ (large circle); $1.9 \pi /4$ (ellipse); $\pi/2$ (line). Figure adapted from \cite{Diehl11}. }
\label{WindingNumber}
\end{figure}

\emph{ i) Bulk properties} -- In view of constructing an open system analog of the above scenario, we consider a purely dissipative ($H=0$) Lindblad master equation of the form of Eq.~\eqref{eq:master_equation} for spinless fermions in a chain with $N$ sites and rate $\kappa$. We choose the Lindblad jump operators $j_{i}$ as the above Bogoliubov quasiparticle operators, with the explicit form
\begin{equation}\label{equ:jump}
j_{i} \equiv\tilde{a}_{i} =\frac{1}{2}(a_{i}+a_{i}^{\dag }-a_{i+1}+a_{i+1}^{\dag}), ~\, ~ (i=1,\ldots,N-1).
\end{equation}
These Lindblad operators are quasi-local superpositions of annihilation and creation operators, leading to a Liouville operator which is quadratic in the fermions, and act on the links of each pair of lattice sites (see Fig. \ref{WindingNumber}a). We indicate below how such a setting emerges naturally in the long-time evolution of a microscopically number conserving (quartic) Liouville dynamics, relying on a mean-field theory as discussed in Sect. \ref{sec:DissGap}, and taking advantage of the quasilocal nature of the target Lindblad operators. Crucially, the ground state condition $\tilde{a}_{i}|G\rangle=0$ now plays the role of a dark state condition. Since the operators $j_{i}$ obey the Dirac algebra, in a translation invariant setting this dark state is unique and pure. In particular, the bulk of the system cools to the $p$-wave superfluid ground state of the Hamiltonian (\ref{HKit}).
The approach to this steady state is governed by the damping spectrum of the Liouvillian $\mathcal{L}$. In analogy to the Hamiltonian gap in Kitaev's model, diagonality of $\mathcal{L}$ in the $\tilde{a}_{i} $ now implies a flat damping spectrum $\kappa_{k}=\kappa$, and in particular the existence of a dissipative gap.

\emph{ii) Edge modes as nonlocal decoherence free subspace} -- For a finite wire we find dissipative zero modes related to the absence of the Lindblad operator $\tilde{a}_{N}$. More precisely, there exists a subspace spanned by the edge-localized Majorana modes $\tilde{a}_{N}=\frac{1}{2} \left(  \mathrm{i}\gamma_{L}+\gamma_{R}\right)  $, with the above Fock basis $|\alpha\rangle\in\{ |0\rangle,|1\rangle\}$, which is decoupled from dissipation, i.e. $\partial_{t}\rho_{\alpha\beta}(t)=0$ with $\rho_{\alpha\beta}\equiv\langle\alpha|\rho|\beta\rangle$.
 These dissipative edge modes are readily revealed in solutions of the master equation defined with jump operators Eq.~(\ref{equ:jump}). The fact that the master equation is quadratic in the fermion operators implies solutions in terms of Gaussian density operators $\rho(t) \sim\exp\left[-\frac{\mathrm{i}}{4} c^{T}G(t)c \right]  $. Here we have defined a column vector $c$ of the $2N$ Majorana operators, and $G$ is a real antisymmetric matrix related to the correlation matrix $\Gamma_{ab}(t)=\frac{\mathrm{i}}{2}\langle[c_{a}, c_{b}]\rangle= \mathrm{i }[\tanh(\mathrm{i }G/2)]_{ab}$, which equally is real and antisymmetric. Writing the Lindblad operators in the Majorana basis, $j_{i} =l_{i}^{T}c, j^{\dag}_{i} = c^{T} l_{i}^{*}$, such that the Liouvillian parameters are encoded in a hermitian $2N\times2N$ matrix $M = \sum_{i} l_{i}\otimes l_{i}^{\dag}$, this covariance matrix  obeys the dissipation-fluctuation equation \citep{Prosen08a,Prosen08b,Eisert10}, $\partial_{t}\Gamma=- \{X,\Gamma\} + Y$, with real matrices $X=2 \mathrm{Re} M = X^{T}$ and $Y = 4 \mathrm{Im} M = -Y^{T}$. Physically, the matrix $X$ describes damping, while the matrix $Y$ is related to fluctuations in the stationary state, determined by $\{X,\bar{\Gamma}\} =  Y$. Note that $Y$ corresponds to the first quantized description of the effective Hamiltonian associated to the master equation: Due to Fermi statistics, only the antisymmetric part of $M$ contains nontrivial information, and thus $H_\text{eff} = \sum_i j_i^\dag j_i =  \tfrac{\mathrm i }{4}c^T Y c$. Writing $\Gamma= \bar{\Gamma}+\delta\Gamma$, the approach to steady state is governed by $\partial_{t} \delta\Gamma= - \{X,\delta\Gamma\}$, i.e., the eigenvalues of the positive semi-definite matrix $X$ give the damping spectrum. The ``dark'' nonlocal subspace of edge modes, decoupled from dissipation, is thus associated with the subspace of zero eigenvalues of the damping matrix $X$. We refer to \cite{Bardyn12} for a more comprehensive discussion of the roles of $X$ and $Y$.

\emph{iii) Bulk-edge dynamics and dissipative isolation} -- In a spectral decomposition $X = \sum_{r} \lambda_{r} |r \rangle\langle r|$, and identifying by greek subscripts the zero eigenvalues subspace, we can write
\begin{align}
\label{Partition}\partial_{t}\left(
\begin{array}
[c]{cc}
\Gamma_{\alpha\beta} & \Gamma_{\alpha s}\\
\vspace{0.1cm} \Gamma_{r\beta} & \Gamma_{rs}
\end{array}
\right)   &  = \left(
\begin{array}
[c]{cc}
0 & - ( \Gamma\lambda)_{\alpha s}\\
\vspace{0.1cm} - (\lambda\Gamma)_{r\beta} & ( - \{\lambda,\Gamma\} +  Y)_{rs}
\end{array}
\right)  .
\end{align}
While the bulk ($rs$ sector) damps out to the steady state by dissipative evolution, the density matrix in the edge mode subspace ($\alpha\beta$ sector) does not evolve and therefore preserves its initial correlations. The coupling density matrix elements (mixed sectors) damp out according to $\Gamma_{r\beta}(t)= e^{- \lambda_{r} t} \Gamma_{r\beta}(0)$. In the presence of a dissipative gap as in the example above, this fadeout of correlations is exponentially fast, leading to a dynamical decoupling of the edge subspace and the bulk. 

In summary, we arrive at the physical picture that dissipative evolution cools the bulk into a p-wave superfluid, and thereby isolates the edge mode subspace, $\rho(t\rightarrow\infty)\rightarrow\rho_{\text{edge}}\otimes\rho_{\text{bulk}}$, providing a highly nonlocal decoherence free subspace \citep{Lidar98}.

So far, we did not yet address the preparation of the edge mode subspace. Generically, when starting from a wire geometry, the initial edge mode subspace is strongly mixed. Since its correlations are preserved during dissipative evolution, it thus will be useless e.g. as a building block for a qubit (Note that this property is also shared with a Hamiltonian setting, where the equilibrium density matrix $\rho_\text{eq} \sim e^{-H/(k_\text{B}T)}$, $k_\text{B}$ the Boltzmann constant, is purified by lowering the temperature. The subspace of this density matrix associated to the zero modes of $H$ is not purified by such cooling.). Therefore, in \cite{Kraus12} we discuss a scheme where the starting point is a ring geometry, where the stationary state is unique and has even parity, since it corresponds to a paired state of fermions. The ring is then adiabatically ``cut'' by removing dissipative links quasi-locally. In this way, it is possible to obtain a pure Majorana subspace with non-local edge-edge correlations.

\subsubsection{Nonabelian Character of Dissipative Majorana Modes}
\label{sec:Nonabelian}

There is a simple and general argument for the nonabelian exchange statistics of dissipative Majorana modes, highlighting the universality of this property that holds beyond the Hamiltonian setting. Consider the time evolution of the density matrix in a co-moving basis $|a(t) \rangle= U(t) | a(0) \rangle$ which follows the decoherence free subspace of edge modes, i.e.~preserves $\dot{\rho}_{\alpha\beta}= 0$. Demanding normalization of the instantaneous basis for all times, $\langle b(t) |a(t) \rangle= \delta_{ab}$, this yields
\begin{equation}
\frac{d}{dt} \rho= -\mathrm{i }[A , \rho] + \sum_{a,b} |a\rangle\dot{\rho}_{ab} \langle b |,
\end{equation}
with the hermitean connection operator $A =\mathrm{i }\dot{U}^{\dagger} U$ and $\dot{\rho}_{ab} \equiv\langle a(t) | \partial_{t} \rho| b(t) \rangle$ the time evolution in the instantaneous basis. The Heisenberg commutator reflects the emergence of a gauge structure \citep{Berry84,Simon83,WilczekZee84,Pachos99,Carollo2003a} in the density matrix formalism, which appears \emph{independently} of what kind of dynamics -- unitary or dissipative -- generates the physical time evolution, represented by the second contribution to the above equation. 
We note that an adiabaticity condition $\dot{\theta}/\kappa_{0}\ll1$ on the rate of parameter changes vs. the bulk dissipative gap has to be accommodated in order to keep the protected subspace. Since the subspace has no intrinsic evolution, this  provides a natural separation of time scales  which prevents the decoherence-free subspace from being left, a phenomenon sometimes referred to as the Quantum Zeno effect \citep{Beige00}.

Starting from this understanding, one can now construct adiabatic local parameter changes in the Liouvillian at the edges of a chain to perform elementary dissipative Majorana moves. Applying such procedure sequentially, and operating on a T-junction in full analogy to the proposal by \cite{Alicea11} for Hamiltonian ground states in order to exchange the two modes while permanently keeping them sufficiently far apart from each other, the unitary braiding matrix describing the process is $B_{ij}=\exp{\left(  \frac{\pi}{4} \gamma_{i}\gamma_{j} \right)  }$ for two Majorana modes $i,j$. This  demonstrates non-abelian statistics since $[B_{ij},B_{jk}]\neq0$ for $i\neq j$. Here we use that the above general considerations are not restricted to a single quantum wire but apply to more general quantum wire networks.

\subsubsection{Topological Order in Density Matrices}
\label{sec:TopOrder}

\emph{Density matrix topological invariant} -- In numerical calculations we have verified that the Majorana modes are robust under wide classes of translation-invariance breaking perturbations such as random local variations of the Lindblad operators of Eq.~\eqref{equ:jump}, suggesting a topological origin. Indeed, we can connect the existence of the edge modes to topological order in the bulk of the stationary state. This is achieved by constructing a topological invariant for the distinction of topologically inequivalent states. This classification is formulated in terms of the density matrix alone and does not rely on the existence of a Hamiltonian or on the purity of the state, in contrast to existing constructions.

As shown in \cite{Diehl11}, the topological information of the stationary state of a Gaussian translationally invariant Liouville evolution is encoded in the even occupation subspace of each momentum mode pair $\pm k$, $\rho_{2k} \propto \frac{1}{2}(\mathbf{1}+\vec{n}_{k}\vec{\sigma})$, where $\vec{\sigma}$ is the vector of Pauli matrices and $\vec{n}_{k}$ is a real three-component vector $0\leq|\vec{n}_{k}|\leq1$. The special case of pure states corresponds to $\rho_{k}^{2}=\rho_{k}$, i.e. $|\vec{n}_{k}|=1$ for all $k$. In the more general case, once the vector $\vec{n}_{k}$ is nonzero for all $k$, a normalized vector $\hat{\vec{n}}_{k}=|\vec{n}_{k}|^{-1}\vec{n}_{k}$ can be introduced. This then defines a mapping $S^{1}\rightarrow S^{2}$, where $S^{1}$ is the circle defined with the Brillouin zone $-\pi\leq k\leq\pi$ with identified end points $k=\pm\pi$ as usual, and the unit sphere $S^{2}$ is given by the end points of $\hat{\vec{n}}_{k}$, as illustrated in Fig. \ref{WindingNumber}. This mapping, however, is generically topologically trivial, with corresponding homotopy group $\pi_{1}(S^{2})=0$, since a circle can always be continuously shrunk into a point on the sphere. In order to introduce a nontrivial topology, we therefore need an additional constraint on $\vec{n}_{k}$ . In our setting, motivated by Kitaev's model Hamiltonian \citep{Kitaev00}, this is provided by the chiral symmetry \citep{Altland97,Ryu10}. In terms of the density matrix, the latter is equivalent to the existence of a $k$-independent unitary matrix $\Sigma$ with $\Sigma^{2}=\mathbf{1}$, which anticommutes with the traceless part of the density matrix ($\vec{n}_{k}\vec{\sigma}$ in our case): $\Sigma \,\vec{n}_{k}\vec{\sigma}\,\Sigma=-\vec{n}_{k}\vec{\sigma}$. This condition can be turned into a geometric one, by representing the matrix $\Sigma$ in terms of a constant unit vector $\vec{a}$, $\Sigma=\vec{a}\vec{\sigma}$. The chiral symmetry condition then translates into an orthogonality condition $\vec{n}_{k}\vec{a}=0$ for all $k$. The end point of $\hat{\vec{n}}_{k}$ is now pinned to a great circle $S^{1}$ on the sphere such that the vector $\hat{\vec{n}}_{k}$ defines a mapping $S^{1}\rightarrow S^{1}$ from the Brillouin zone into a circle, see Fig. \ref{WindingNumber}b. The corresponding homotopy group is now nontrivial, $\pi_{1}(S^{1})=\mathbf{Z}$, and such mappings are divided into different topological classes distinguished by an integer topological invariant, the winding number, with the explicit form
\begin{align}
\label{winding}\nu=\frac{1}{2\pi}\int_{-\pi}^{\pi}dk\,\vec{a}\cdot(\hat
{\vec{n}}_{k}\times\partial_{k}\hat{\vec{n}}_{k})\in\mathbf{Z}.
\end{align}
Geometrically, $\nu$ counts the number of times the unit vector $\hat{\vec{n} }_{k}$ winds around the origin when $k$ goes across the Brillouin zone. Crucially, the resulting topological distinction of different density matrices for translationally invariant, chirally symmetric Gaussian systems works without restriction on the purity of the state. Using bulk-edge correspondence established for Hamiltonian settings \citep{PhysRevLett.71.3697,Kitaev06}, a nonzero value of the invariant would imply the existence of edge modes as found above. However, in a general dissipative setting it is possible to break this bulk-edge correspondence. For a discussion of this subject, and interesting consequences of it, we refer to \cite{Bardyn12}.

\emph{Phase transition by ``loss of topology''} -- In Fig.~\ref{WindingNumber}b we illustrate a situation described by a one-parameter deformation of the vector $\vec{n}_{k}(\theta)$. This is induced by a corresponding deformation on the Lindblad operators according to  $j_i(\theta) = \tfrac{1}{\sqrt{2}} (  \sin{\theta} \, ( a^{\dagger}_i -  a_{i+1} ) + \cos{\theta} \, (  a_i +a^{\dagger}_{i+1}  ) )$, where Eq. \eqref{equ:jump} is reproduced for $\theta =\pi/4$. For this deformation, the purity is not conserved while preserving the chiral symmetry, reflected in the fact that the vector in general lies on an ellipsis $0\leq |\vec{n}_k| \leq 1$. Topological order is meaningfully defined as long as the first inequality is strict as discussed above, i.e. as long as there is a ``purity gap''. However, at the points $\theta=\theta_{s} = \pi s/2$ ($s$ integer), not only the direction of $\vec{a}$ but also the topological invariant is not defined, since $\vec{n}_{k}$, aligned in the $y$-direction for all $k$, has zeroes and the purity gap closes: $\vec{n}_{k=0,\pi}=0$, meaning physically that these modes are in a completely mixed state. The "loss" of topology at $\theta=\theta_{s}$ can be viewed as a non-equilibrium topological phase transition\citep{Rudner09,Lindner11,Kitagawa10} as a result of changing the Liouville parameters: $\theta=\theta_{s}$ also implies a closing of the dissipative gap in the damping spectrum, which leads to critical behavior manifesting itself via diverging time scales, resulting e.g in an algebraic approach to steady state (as opposed to exponential behavior away from criticality) \citep{diehl-natphys-4-878,kraus-pra-78-042307,verstraete-nphys-5-633,Diehl10a,Eisert10}. We emphasize that the symmetry pattern of the steady state is identical on both sides of the transition, ruling out a conventional Landau-Ginzburg type transition and underpinning the topological nature of the transition.

\subsubsection{Physical Implementation}
\label{sec:PImplementation}

As mentioned above, a physical implementation of this scenario is provided by a microscopically number conserving Liouville dynamics as discussed in Sect. \ref{sec:DissBEC}, with jump operators of the form Eq. \eqref{eq:jump}, where boson operators are replaced by spinless fermionic ones. We note that in this implementation setting, the role of the bath is played by the bosonic atoms from a surrounding BEC, and originates microscopically from standard contact density-density interactions, thus imposing a natural parity conservation for the fermionic system constituents due to fermionic superselection rules. This contrasts potential solid state realizations, where the environmental degrees of freedom are fermionic as well. Explicitly, we choose
\begin{align}
\label{Jump}
J_{i}  &  = \frac{1}{4} (a_{i}^{\dag}+ a_{i+1}^{\dag}) (a_{i}- a_{i+1} ) = C_{i}^{\dag}A_{i}.
\end{align}
From a formal point of view, the sequence of annihilation ($A_i = \tfrac{1}{2} (a_{i}- a_{i+1} ) $) and creation ($C^\dag_i= \tfrac{1}{2}(a_{i}^{\dag}+ a_{i+1}^{\dag})$) part, gives rise to dissipative pairing of spinless fermions in the absence of any conservative forces, in complete analogy to the discussion for the spinful case in Sect. \ref{sec:DissPair}. In the present case, the mean field construction outlined above can be simplified. It can be shown \citep{Diehl11} that in the long-time and thermodynamic limit, the following general relation between fixed number ($J_{i}$) and fixed phase ($j_{i}$) Lindblad operators holds,
\begin{align}
\label{Equiv}J_{i} = C_{i}^{\dag}A_{i} \Leftrightarrow j_{i} = C_{i}^{\dag}+ A_{i}.
\end{align}
The relation to the Majorana operators is now apparent: It is precisely Kitaev's quasiparticle operators which are obtained as effective Lindblad operators in the late time evolution, $j_{i} = \tilde a_{i}$. The role of phase fluctuations remains to be investigated. The explicit mean field calculation shows that a master equation with jump operators Eq. (\ref{equ:jump}) is produced, with effective dissipative rate $\kappa= \tilde\kappa/8$. An analysis of the leading imperfections shows that they preserve the chiral symmetry, and so keep the system in the above described topological class. We furthermore emphasize that the practical simplifications in view of engineering such dissipative dynamics in the lab compared to the stroboscopic implementation of Sect. \ref{sec:StatePrep} is mainly due to the fact that the spin quantum numbers do not have to be fixed in the present case.

Recently, we have also investigated two-dimensional dissipatively induced fermionic paired states with $p_x + \mathrm i p_y$ order parameter \citep{Bardyn12}. Intriguingly, in such systems we established a mechanism that guarantees the existence of a single localized Majorana mode at the core of a dissipative vortex in a phase with vanishing bulk topological invariant. This phenomenon ultimately relies on a violation of the bulk-edge correspondence which is unique to the dissipative dynamics and has no Hamiltonian counterpart. The Majorana modes could be generated dynamically with the implementation strategy outlined here by additionally imprinting optical angular momentum onto the matter system \citep{Brachmann11}, potentially circumventing the need of single-site addressability.


\section{Outlook}
\label{sec:Conclusions_Outlook}




In the present work we have summarized recent advances in digital quantum
simulation and engineering of open many-body systems with atoms and ions, where our main emphasis
has been on presenting new concepts and tools. We conclude our discussion with few remarks on open theoretical and experimental problems and challenges.

With regard to the digital quantum simulation approach discussed in Section~\ref{sec:Digital_QS}, the described experiments realized with trapped ions \citep{lanyon_science-334-57,barreiro-nature-470-486} demonstrate in principle the feasibility of the digital simulation approach for the study of open many-particle quantum systems. They have been carried out in setups of linear ion chains and are, in their present form, not immediately scalable to large systems. However, similar protocols can be realized in scalable and two-dimensional ion-trap architectures, whose development is currently at the center of an intense research effort \citep{blakestad-prl-102-153002,home-science-325-1227,hensinger-apl-88-034101,schmied-prl-102-233002,clark-j-appl-phys-105-013114}.

In view of the big challenge of scaling up the simulations to larger systems, the Rydberg-based simulator architecture with cold atoms in optical lattices \citep{weimer-nphys-6-382} provides an \emph{a priori} scalable simulation platform. Especially in view of the recent experimental achievement of the first entangling Rydberg gates \citep{isenhower-prl-104-010503,wilk-prl-104-010502} and single-site addressability \citep{bakr-nature-462-74,sherson-nature-467-68,bakr-science-329-547,weitenberg-nature-471-319}, it seems to be a promising route towards large-system digital quantum simulators with control over some tens to hundred qubits (spins). 
This would outperform state-of-the-art classical numerical simulation techniques. However, it remains to be seen if neutral atoms or other competing simulation platforms will be able to achieve the remarkable fidelities of quantum gate operations demonstrated with ions \citep{lanyon_science-334-57,barreiro-nature-470-486}. In fact, the concepts discussed here for trapped ions and Rydberg atoms can be readily adapted to other physical simulation platforms ranging from optical, atomic and molecular systems to solid-state devices \citep{ladd-nature-464-45,obrien-science-318-1567,clarke-nature-453-1031,wrachtrup-jpcm-18-S807,hanson-rmp-79-1217,vandersypen-rmp-76-1037}.

From a fundamental point of view, it will be most interesting to connect the driven-dissipative ensembles discussed in Section~\ref{sec:Analog_QS_DissDyn} to other fields, such as non-equilibrium statistical mechanics. For example, as known from classical problems in this context (see, e.g., \cite{altlandbook}), strong non-equilibrium drive can give rise to new universality classes beyond those known in thermodynamic equilibrium. It seems plausible that similar phenomena could be present in our systems as well, possibly enriched by their intrinsic quantum mechanical character. More broadly speaking, the goal is the identification of universal hallmark signatures for the intrinsic non-equilibrium nature of these systems.

From a practical perspective, the recent experiments with atomic ensembles \citep{PhysRevLett.107.080503} and trapped ions \citep{barreiro-nature-470-486,lanyon_science-334-57} suggest that a strong dissipative drive can protect against ubiquitous unwanted decoherence mechanisms -- while a system with dominant unitary dynamics alone is sensitive to decoherence. This sparks the more general question if systematic criteria for the stability of many-body states under competing unitary and dissipative dynamics can be established, starting from the promising results on the existence of dissipative gaps described above. A general scenario of Òdissipative protectionÓ clearly would give a high practical relevance to dissipative quantum state engineering. Ultimately, if these questions can be answered positively, it will be intriguing to investigate whether the robustness benefits of dissipative quantum computation \citep{verstraete-nphys-5-633} and memories \citep{pastawski-pra-83-012304}, as well as topological quantum computation \citep{nayak-rmp-80-1083}, can be sensibly combined in one unified setting. Clearly, answering such questions also requires the development of new theoretical tools. A promising avenue is provided by a Keldysh functional integral approach \citep{KL2009}, within which the powerful toolbox of advanced field theoretical methods could be leveraged over to driven-dissipative many-body systems.

Furthermore, in view of quantum engineering, it is an important goal to extend the scope of many-body physics with driven-dissipative ensembles to new physical platforms. This concerns not only trapped ion systems, but also microcavity arrays, which have a strong potential of being developed into true many-body scenarios in the future  \citep{Mariantoni11}. Each of these systems will also add new theoretical challenges, such as the intrinsic non-number conserving nature of systems whose basic constituents are photons. In addition, it will be intriguing to explore the theoretical crosslinks between analog and digital quantum simulations in a many-body context.

Finally, proper quantitative assessment of errors poses a non-trivial task and remains a challenge for future work, although first steps in this direction have been taken in~\cite{lanyon_science-334-57}. In contrast to quantum computing, quantum simulation is usually not interested in obtaining the many-body wave function in a faithful way, but rather aims at an accurate prediction of low order correlation functions, as is relevant, for example, for phase diagrams in equilibrium physics. Thus it is generally argued that quantum simulation is more robust against errors and imperfections than quantum computing, and from an experimental point of view the realization of a large-scale quantum simulator is expected to be a more realistic short-term goal than building a fault-tolerant quantum computer. However, one of the outstanding problems is to investigate the role of errors in an interplay between theory and experiment. Along a similar line, questions of validation and verification of quantum simulators need to be addressed in these future studies. In the context of digital quantum simulation, the good news is that -- if the gate fidelities and system sizes can be further increased -- the gate-based approach can incorporate quantum error correction protocols. These might prove essential for fault-tolerant quantum simulation, in particular for future large-scale quantum simulations of complex many-body models.

In the field of quantum information processing, it is one of the grand challenges and visions to build in the laboratory
a quantum device which performs tasks not achievable on a classical level. A next generation quantum simulation experiment involving (experimentally proven) large-scale entanglement may be the first laboratory demonstration that fulfills this promise in a convincing way. This would be an exciting and big step forward towards the realization of Feynman's 30-years-old dream of building a programmable quantum simulator, which might not only provide us with answers to long-standing open questions, but also allow us to explore new realms of physics, such as many-body quantum dynamics beyond thermodynamic equilibrium.

\section{Acknowledgments}

 We acknowledge support by the Austrian Science Fund (FWF) through SFB FOQUS and the START grant Y 581-N16 (S. D.), the European Commission (AQUTE, NAMEQUAM), the Institut fuer Quanteninformation GmbH and the DARPA OLE program. M. M. acknowledges support by the CAM research consortium QUITEMAD S2009-ESP-1594, European Commission PICC: FP7 2007-2013, Grant No. 249958, and the Spanish MICINN grant FIS2009-10061.








\end{document}